\documentclass[hyper,a4paper]{JHEP}
   
\usepackage{epsfig}
\usepackage{latexsym}
\usepackage{graphicx}
\usepackage{amsmath}
\usepackage{amsfonts}   % if you want the fonts
\usepackage{amssymb}    % if you want extra symbols
\usepackage{float}
\usepackage{bm}
\def\ee{e^+e^-\to h^0Z}

\def\lsim{\raise0.3ex\hbox{$\;<$\kern-0.75em\raise-1.1ex\hbox{$\sim\;$}}}
\def\gsim{\raise0.3ex\hbox{$\;>$\kern-0.75em\raise-1.1ex\hbox{$\sim\;$}}}
\newcommand{\captions}{\sf\caption}
\def    \be            {\begin{equation}}
\def    \ee            {\end{equation}}
\def    \bea           {\begin{eqnarray}}
\def    \eea           {\end{eqnarray}}

\def \snuc{\tilde{\nu}^c}
\def \snu{\tilde{\nu}}
\def\sw2{sin^2 \theta_w}

\def\a^tau{\alpha_{\tau}}

\def\beq{\begin{equation}}
\def\eeq{\end{equation}}
\def\beqa{\begin{eqnarray}}
\def\eeqa{\end{eqnarray}}

\newcommand{\gev}{\,\textrm{GeV}}

\newcommand{\kev}{\,\textrm{keV}}

\newcommand{\neutralino}{{\tilde{\chi}^0}}
\newcommand{\mneutralino}{{m_{\tilde{\chi}^0}}}

\newcommand{\newc}{\newcommand}
\newc\BR{BR}
\newc{\akappa}{A_{\kappa} }

\newc\deltagmtwo{\delta (g-2)_{\mu}} 
\newc\deltaamu{\Delta a_{\mu}}

\def\anti{\overline}

\newc{\haa}{BR\(h_1\to a_1 a_1\)}
\newc{\abb}{BR\(a_1\to b\anti{b}\)}
\newc{\hbb}{BR\(h_1\to b\anti{b}\)}
\newc{\abund}{\Omega h^2}

\newc\bsgamma{b\rightarrow s \gamma }
\newc\bxsgamma{\overline{B}\rightarrow X_{s}\gamma}
\newc\brbsgamma{\BR(\overline{B}\rightarrow X_s\gamma)}

\title{The Higgs sector of the $\mu\nu$SSM and collider physics}
\author{} 
 \author{Javier Fidalgo\\
         Departamento de F\'{\i}sica Te\'{o}rica UAM
         and Instituto de F\'{\i}sica Te\'{o}rica UAM/CSIC,\\
         Universidad Aut\'{o}noma de Madrid (UAM), Cantoblanco,
         28049 Madrid, Spain\\
         E-mail: \email{javier.fidalgo@uam.es}}
 \author{Daniel E. L\'{o}pez-Fogliani\\
        Laboratoire de Physique Th\'eorique,\\ 
 Universit\'e  Paris-Sud, F-91405 Orsay, France \\
         E-mail: \email{daniel.lopez@th.u-psud.fr}}
 \author{Carlos Mu\~noz\\
         Departamento de F\'{\i}sica Te\'{o}rica UAM
         and Instituto de F\'{\i}sica Te\'{o}rica UAM/CSIC,\\
         Universidad Aut\'{o}noma de Madrid (UAM), Cantoblanco,
         28049 Madrid, Spain\\
         E-mail: \email{carlos.munnoz@uam.es}}
 \author{Roberto Ruiz de Austri\\ Instituto de F\'{\i}sica Corpuscular UV/CSIC, Universidad de Valencia,\\ Edificio Institutos de Paterna, Apt. 22085, E-46071 Valencia, Spain\\
         E-mail: \email{rruiz@ific.uv.es}}

\abstract{\small The $\mu\nu$SSM is a supersymmetric standard model that accounts for light neutrino masses and solves the $\mu$ problem of the MSSM by simply using right-handed neutrino superfields. 
Since this mechanism breaks $R$-parity, a
peculiar structure for the mass matrices is generated. 
The neutral Higgses are mixed with the right- and left-handed sneutrinos producing 8$\times$8 neutral scalar mass matrices. 
We analyse the Higgs sector of the $\mu\nu$SSM
in detail, with special emphasis in possible signals at colliders.
After studying in general the decays of the Higges, 
we focus on those processes that are genuine of the $\mu\nu$SSM, and could serve to distinguish it from other supersymmetric models.
In particular, we present viable benchmark points for LHC searches.
For example, we find decays of a MSSM-like Higgs into two lightest neutralinos, with the latter decaying inside the detector leading to displaced vertices, and producing final states with 4 and 8 $b$-jets
plus missing energy. 
Final states with leptons and missing energy are also found.
}

\preprint{
\rightline{FTUAM-11-52, IFT-UAM/CSIC-11-51, LPT-Orsay 11/62, July 2011}
}

\begin{document}
%%%%%%%%%%%%%%%%%%%%%%%%%%%%%%%%%%%%%%%%%%%%%%%%%%
%                                                %
%    BEGINNING OF TEXT                           %
%                                                %
%%%%%%%%%%%%%%%%%%%%%%%%%%%%%%%%%%%%%%%%%%%%%%%%%%

%%%%%%%%%%%%%%%%%%%%%%%%%%%%%%%%%%%%%%%%%%%%%%%
\section{Introduction}
%%%%%%%%%%%%%%%%%%%%%%%%%%%%%%%%%%%%%%%%%%%%%%%

% The minimal supersymmetric extension of the standard model (SM), the well known minimal supersymmetric standard model (MSSM) \cite{mssm}, is  an attractive candidate for physics beyond the SM.
% It solves many theoretical puzzles affecting  the SM and one expects to find its signatures in the forthcoming large 
% hadron collider (LHC). 
% 
% 
% In the proceeding sections  , are  given in the Appendices.

The $\mu$ from $\nu$ Supersymmetric Standard 
Model ($\mu\nu$SSM) \cite{MuNuSSM,unpublished,MuNuSSMreview}, uses right-handed neutrino superfield(s) to generate light neutrino masses and to solve the $\mu$-problem \cite{muproblem} of the Minimal Supersymmetric Standard Model (MSSM) \cite{mssm}. Thus the $\mu\nu$SSM is a minimal model in the sense that no extra singlet superfield has to be added to the spectrum for solving the $\mu$ problem, as it is e.g. the case of 
the Next-to-MSSM (NMSSM) \cite{ana}. The spectrum and the vacua of the $\mu\nu$SSM were studied in \cite{MuNuSSM2, MuNuSSM SCPV}.

The superpotential of the $\mu\nu$SSM contains, in addition to the
usual Yukawas for quarks and charged leptons,
Yukawas for neutrinos
$Y_\nu\ \hat H_u\,  \hat L \, \hat \nu^c$, terms of the type
$\lambda\ \hat \nu^c \hat H_d\hat H_u$ 
producing an effective  $\mu$ term through electroweak (EW)-scale right-handed sneutrino
vacuum expectation values (VEVs),
and also terms of the type $\kappa\ \hat \nu^c \hat \nu^c \hat \nu^c$  
avoiding the existence of a Goldstone boson and
generating EW-scale
effective Majorana masses for neutrinos, i.e. giving rise to an EW-scale seesaw.
Notice that, since only dimensionless trilinear couplings are present in the superpotential of the model, the EW scale arises through the soft supersymmetry (SUSY)-breaking terms in the scalar potential. Thus all known particle physics phenomenology can be reproduced in the $\mu\nu$SSM with only one scale.
For example, {\it ad hoc} high-energy scales in order to generate a GUT-scale seesaw are not needed.
With the EW-scale seesaw of this model, neutrino Yukawa couplings $Y_\nu$ of the order of $10^{-6}$ (like the electron Yukawa coupling) are sufficient to reproduce the correct neutrino masses.
The neutrino sector was studied in detail in \cite{MuNuSSM2, MuNuSSM SCPV, Ghosh:2008yh, Ghosh2010}, obtaining that current neutrino data (the measured mass differences and mixing angles) can be easily reproduced.
%, essentially because the see-saw scale is the electroweak one %and R-parity is broken \cite{MuNuSSM SCPV}.

The above terms in the superpotential produce the explicit breaking
of $R$-parity in this model.
The size of the breaking can be 
%The breaking of R-parity in the $\mu\nu$SSM can be 
easily understood realizing that in the limit where 
%neutrino Yukawa couplings 
$Y_{\nu}$ are vanishing, the 
$\hat \nu^c$ are 
ordinary singlet superfields like the $\hat S$ of the NMSSM, 
without any connection with neutrinos,
and 
%{\bf this model would be like the
%NMSSM 
%(although having three singlets instead of one) 
%with three singlets},
%where 
$R$-parity is therefore conserved.
Once $Y_{\nu}$ are switched on, 
the 
%singlets
$\hat \nu^c$ become right-handed neutrinos, and, as a consequence, $R$-parity
is broken. 
% {\bf This breaking has to be small because} of the
% EW-scale seesaw implying that {\bf the values of $Y_{\nu}$ 
%neutrino Yukawa couplings 
% cannot be larger than $10^{-6}$} (like the electron Yukawa) to reproduce the correct neutrino masses ($\lsim 10^{-2}$ eV).
Thus the breaking is small because
the EW-scale seesaw implies small values for $Y_{\nu}$.

% Actually, the explicit breaking of R-parity in this model 
% by the above terms produces the mixing of neutralinos with
% left- and right-handed neutrinos, and as a consequence a generalized matrix of the
% seesaw type that gives rise at tree level to three
% light eigenvalues corresponding to neutrino 
% masses \cite{MuNuSSM}.

%  Since the right-handed neutrino supermultiplets generate the $\mu$-term through the vacuum expectation values of the right-handed sneutrinos, R-parity could not be an exact symmetry of the theory. Then, to prevent fast proton decay an additional symmetry as Baryon-triality could be invoked \cite{Baryon triality}.
 
Concerning cosmological issues, dark matter and baryon asymmetry have been analysed in the model.
Since the lightest supersymmetric particle (LSP) is not stable when $R$-parity is broken,
%In this context
the neutralino \cite{reviewmio}
%\cite{reviewmio} 
or the right-handed sneutrino \cite{sneutrino}, 
%\cite{sneutrino},
with very short lifetimes,
are no longer candidates for the dark matter of the Universe.
Nevertheless, the
gravitino, present in the local SUSY version of the model, 
could be a good dark matter candidate as discussed 
in \cite{gravitino}, where 
its possible detection through the observation of a monochromatic gamma-ray line in the Fermi satellite was also studied.
In \cite{EWbaryo}, the generation of the baryon asymmetry of the universe was analysed in detail in the context of the $\mu\nu$SSM, with the interesting result that electroweak baryogenesis can be realized.

% The smallness of neutrino masses could be understood through the presence of an approximate symmetry, due to the fact that if the neutrino Yukawas are set to zero, R-parity is restored as an exact symmetry of the model and neutrino masses would vanish. 
% Respect to the cosmological issues of the model, because the presence of a $Z_3$ symmetry in the superpotential, one could expects a cosmological domain-walls problem. In a similar way as in the NMSSM, the domain-walls problem could be solved through the presence of non-renormalizable terms, without reintroducing hierarchy problems \cite{nowall}. 

Summarizing, the $\mu \nu$SSM is a very well motivated and attractive model, and, as a consequence, a complete study of possible signals at colliders is required. In this respect, there are two main features that could help to distinguish the $\mu \nu$SSM from other SUSY models. On the one hand, since the LSP is no longer stable due to the breaking of $R$-parity, not all SUSY chains must yield missing energy events.
In 
\cite{Ghosh:2008yh,Hirsch0,Porod} the decays of the lightest neutralino were discussed, as well as the correlations of the decay branching ratios with the neutrino mixing angles.
On the other hand, the breaking of $R$-parity also generates a peculiar structure for the mass matrices. In particular, the presence of right and left-handed sneutrino VEVs leads to mixing of the neutral Higgses with the sneutrinos producing 8$\times$8
neutral scalar mass matrices. This extended Higgs sector could be very helpful for testing the $\mu \nu$SSM.

In this work we will continue the analysis of the Higgs sector of the model started in \cite{MuNuSSM2}, putting special emphasis in possible signals at colliders \cite{Hirsch0,indios3}. In Section \ref{The model} we will briefly review the $\mu \nu$SSM, describing the superpotential and deriving the neutral scalar potential. 
In Section \ref{Higgs sector and decays} we will analyze the Higgs sector. In particular, we will study first the Higgs mixings, and second the possible Higgs decays taking place once a Higgs particle is produced at colliders. Finally, we will 
discuss the LEP constraints. For that we will
compute the couplings of the Higgses with the $Z$ boson, and the sum rules. 
% In Section \ref{Production mechanisms at colliders} we will briefly discuss the production mechanisms of Higgses at lepton and hadron colliders. 
In Section \ref{numericalresults} we will
concentrate on Higgs decays
that are genuine of this model, and could therefore serve to 
distinguish it from other SUSY models in certain regions of the parameter space. We will present a sample of numerical examples of viable benchmark points for LHC searches. For that, we will focus first our attention on the decays of a MSSM-like Higgs with a sizeable branching ratio into two lightest neutralinos.
These neutralinos could decay inside the detector leading to displaced vertices. This fact could be used to distinguish the $\mu\nu$SSM from R-parity conserving models.
Also, the product of the decays can be used to distinguish it from other R-parity breaking models.
Higgs-to-Higgs cascade decays will also be studied, and we will discuss an interesting benchmark point with
% another genuine signal of the $\mu\nu$SSM.
similar signals to the NMSSM that could also serve to distinguish the $\mu \nu$SSM from other R-parity breaking models.
For completeness, we will discuss in Section~\ref{gravitino} the possibility that gravitino dark matter in this model might alter the collider phenomenology through the decay channel neutralino to gravitino-photon.
We will see that this branching ratio turns out to be negligible.
% We have also emphasized that in the $\mu \nu$SSM the gravitino could be a viable dark matter candidate, accessible to indirect detection experiments, and without altering the collider phenomenology described along this paper. In particular, the 
% branching ratio of neutralino to gravitino-photon turns out to be negligible.
% In section blablabla while brief comments, on the role of the gravitino at colliders,  on some special limits of the model (related with the issue of missing energy at colliders), are left for Section~\ref{comments}.}
Finally, the conclusions are left for Section~\ref{Conclusions}.

%%%%%%%%%%%%%%%%%%%%%%%%%%%%%%%%%%%%%%%%%%%%%%%
\section{The $\mu \nu$SSM}
\label{The model}
%%%%%%%%%%%%%%%%%%%%%%%%%%%%%%%%%%%%%%%%%%%%%%%

The  superpotential of the $\mu \nu$SSM is given 
by \cite{MuNuSSM}\footnote{Although we assume three families of right-handed neutrinos motivated by the Standard Model generation replication pattern, a different number of right-handed neutrinos can be used.}:
\begin{align}\label{superpotential}
W  &= 
\ \epsilon_{ab} \left(
Y_{u_{ij}} \, \hat H_u^b\, \hat Q^a_i \, \hat u_j^c +
Y_{d_{ij}} \, \hat H_d^a\, \hat Q^b_i \, \hat d_j^c +
Y_{e_{ij}} \, \hat H_d^a\, \hat L^b_i \, \hat e_j^c +
Y_{\nu_{ij}} \, \hat H_u^b\, \hat L^a_i \, \hat \nu^c_j 
\right)
\nonumber\\ 
&- \epsilon{_{ab}} \lambda_{i} \, \hat \nu^c_i\,\hat H_d^a \hat H_u^b
+
\frac{1}{3}
\kappa{_{ijk}} 
\hat \nu^c_i\hat \nu^c_j\hat \nu^c_k \ ,
%\label{superpotential}
\end{align}
where we take $\hat H_d^T=(\hat H_d^0, \hat H_d^-)$, $\hat H_u^T=(\hat
H_u^+, \hat H_u^0)$, $\hat Q_i^T=(\hat u_i, \hat d_i)$, $\hat
L_i^T=(\hat \nu_i, \hat e_i)$, 
$i,j,k=1,2,3$ are family indices, $a,b=1,2$ are $SU(2)_L$ indices with
$\epsilon_{12}=1$, and
$Y$, $\lambda$, $\kappa$ are dimensionless
matrices, a vector, and a totally symmetric tensor, respectively. 
In the following the summation convention on repeated indices is implied.

Working in the framework of supergravity, the Lagrangian
$\mathcal{L}_{\text{soft}}$
is given by: 
%mediated SUSY breaking, we write 
% the soft terms appearing in the Lagrangian, $\mathcal{L}_{\text{soft}}$, as:
%
\begin{eqnarray}
-\mathcal{L}_{\text{soft}} & =&
 m_{\tilde{Q}_{ij} }^2\, \tilde{Q^a_i}^* \, \tilde{Q^a_j}
+m_{\tilde{u}_{ij}^c}^{2} 
\, \tilde{u^c_i}^* \, \tilde u^c_j
+m_{\tilde{d}_{ij}^c}^2 \, \tilde{d^c_i}^* \, \tilde d^c_j
+m_{\tilde{L}_{ij} }^2 \, \tilde{L^a_i}^* \, \tilde{L^a_j}
+m_{\tilde{e}_{ij} ^c}^2 \, \tilde{e^c_i}^* \, \tilde e^c_j
\nonumber \\
&+ &
m_{H_d}^2 \,{H^a_d}^*\,H^a_d + m_{H_u}^2 \,{H^a_u}^* H^a_u +
m_{\tilde{\nu}_{ij}^c}^2 \,\tilde{{\nu}^c_i}^* \tilde\nu^c_j 
\nonumber \\
&+&
\epsilon_{ab} \left[
(A_uY_u)_{ij} \, H_u^b\, \tilde Q^a_i \, \tilde u_j^c +
(A_dY_d)_{ij} \, H_d^a\, \tilde Q^b_i \, \tilde d_j^c +
(A_eY_e)_{ij} \, H_d^a\, \tilde L^b_i \, \tilde e_j^c 
\right.
\nonumber \\
&+&
\left.
(A_{\nu}Y_{\nu})_{ij} \, H_u^b\, \tilde L^a_i \, \tilde \nu^c_j 
+ \text{c.c.}
\right] 
\nonumber \\
&+&
\left[-\epsilon_{ab} (A_{\lambda}\lambda)_{i} \, \tilde \nu^c_i\, H_d^a  H_u^b
+
\frac{1}{3}
(A_{\kappa}\kappa)_{ijk} \, 
\tilde \nu^c_i \tilde \nu^c_j \tilde \nu^c_k\
+ \text{c.c.} \right]
\nonumber \\
&-&  \frac{1}{2}\, \left(M_3\, \tilde\lambda_3\, \tilde\lambda_3+M_2\,
  \tilde\lambda_2\, \tilde
\lambda_2
+M_1\, \tilde\lambda_1 \, \tilde\lambda_1 + \text{c.c.} \right) \,.
\label{2:Vsoft}
\end{eqnarray}
%
%one obtains from (\ref{2:Vsoft}) 
%the neutral scalars develop in general
%the VEVs:
%
%
%\begin{equation}\label{2:vevs}
%\langle H_d^0 \rangle = v_d \, , \quad
%\langle H_u^0 \rangle = v_u \, , \quad
%\langle \tilde \nu_i \rangle = \nu_i \, , \quad
%\langle \tilde \nu_i^c \rangle = \nu_i^c \,,
%\end{equation}
% 

In addition to terms from $\mathcal{L}_{\text{soft}}$, the tree-level neutral scalar potential
receives the usual $D$ and $F$ term contributions and is 
given by \cite{MuNuSSM,MuNuSSM2}:
\begin{equation}
V^0 = V_{\text{soft}} + V_D  +  V_F\ , 
\label{finalpotential}
\end{equation}
where
\begin{eqnarray}
V_{\text{soft}} &=& 
 m_{H_d}^{2}H^0_{d}H^{0*}_{d}+m_{H_u}^{2}H^0_{u}H_{u}^{0*}+
m_{\tilde{L}_{ij} }^2 \, \tilde{\nu}_i \, \tilde{\nu}_j^* +
m_{\tilde{\nu}^c_{ij}}^{2}\tilde{\nu}^c_{i}\tilde{\nu}^{c*}_{j}
\nonumber\\
&+&
\left(
a_{\nu_{ij}}H^0_{u}\tilde{\nu}_{i}\tilde{\nu}^{c}_{j}
%m_{\tilde{\nu}_{i}}^{2}\tilde{\nu}_{i}\tilde{\nu}_{i}^{*}
-a_{\lambda_{i}}\tilde{\nu}^c_{i}H^0_{d}H^0_{u}
+
\frac{1}{3} {a_{\kappa_{ijk}}\tilde{\nu}^c_{i}\tilde{\nu}^c_{j}\tilde{\nu}^c_{k}}
+
\text{c.c.} \right)\ ,
\label{akappa}
\end{eqnarray}
with
$a_{\nu_{ij}}\equiv (A_\nu Y_\nu)_{ij}$, 
$a_{\lambda_i}\equiv (A_\lambda\lambda)_i$,  
$a_{\kappa_{ijk}}\equiv (A_\kappa \kappa)_{ijk}$,
%Besides the potential receives the $D$ and $F$-term
%contributions
\begin{equation}
V_D  =
\frac{G^2}{8}\left(\tilde{\nu}_{i}\tilde{\nu}_{i}^* 
                   + H^0_{d}H_{d}^{0*}-H^0_{u}H_{u}^{0*}\right)^{2},
\end{equation}
with $G^2\equiv g_{1}^{2}+g_{2}^{2}$, and
\begin{eqnarray}
V_{F}  &=&
 \lambda_{j}\lambda_{j}^{*}H^0_{d}H_{d}^{0*}H^0_{u}H_{u}^{0*}
 +\lambda_{i}\lambda_{j}^{*}H^0_{d}H_{d}^{0*}\tilde{\nu}^c_{i}
                                                   \tilde{\nu}^{c*}_{j}
 +\lambda_{i}\lambda_{j}^*H^0_{u}H_{u}^{0*}\tilde{\nu}^c_{i}
                                                   \tilde{\nu}^{c*}_{j}
 +\kappa_{ijk}\kappa_{ljm}^{*}\tilde{\nu}^c_{i}\tilde{\nu}^{c*}_{l}
                                   \tilde{\nu}^c_{k}\tilde{\nu}^{c*}_{m}
\nonumber\\
 &-& (\kappa_{ijk}\lambda_{j}^*H_{d}^{0*}H_{u}^{0*}
                                      \tilde{\nu}^c_{i}\tilde{\nu}^c_{k}
 -Y_{\nu_{ij}}\kappa_{ljk}^{*}H^0_{u}\tilde{\nu}_{i}\tilde{\nu}^{c*}_{l}
                                                     \tilde{\nu}^{c*}_{k}
 +Y_{\nu_{ij}}\lambda_{j}^{*}H_{d}^{0*}H_{u}^{0*}H^0_{u}\tilde{\nu}_{i} 
\nonumber \\
 &+& Y_{\nu_{ij}}^{*}\lambda_{k}H^0_{d}\tilde{\nu}^c_{k} \tilde{\nu}_{i}^{*}
                                  \tilde{\nu}^{c*}_{j} 
 + \text{c.c.}) 
\nonumber \\
 &+& Y_{\nu_{ij}}Y_{\nu_{ik}}^*H^0_{u}H_{u}^{0*} \tilde{\nu}^c_{j}
                                                \tilde{\nu}^{c*}_{k}
 +Y_{\nu_{ij}}Y_{\nu_{lk}}^{*}\tilde{\nu}_{i}\tilde{\nu}_{l}^{*}
                                    \tilde{\nu}^c_{j}\tilde{\nu}^{c*}_{k}
 +Y_{\nu_{ji}}Y_{\nu_{ki}}^{*}H^0_{u}H_{u}^{0*}\tilde{\nu}_{j}\tilde{\nu}_{k}^* \, .
\end{eqnarray}

Once the EW symmetry is spontaneously broken, the neutral scalars develop in general the following VEVs:
\begin{equation}\label{2:vevs}
\langle H_d^0 \rangle = v_d \, , \quad
\langle H_u^0 \rangle = v_u \, , \quad
\langle \tilde \nu_i \rangle = \nu_i \, , \quad
\langle \tilde \nu_i^c \rangle = \nu_i^c \,.
%\label{esperados}
\end{equation}
In the following we will assume for simplicity that all parameters in the potential are real, as well as the VEVs, i.e. that CP is conserved\footnote{For an analysis of spontaneous CP violation, see \cite{MuNuSSM SCPV}.}.
As a consequence, the neutral CP-even scalars are not mixed with the neutral CP-odd scalars\footnote{
In general, all neutral scalars are mixed, and since all of them get VEVs, we will call them Higgses throughout this work. 
To be more precise, we will use the term 
`Higgses' for the mass eigenstates, and `Higgs doublets' for 
$H_d^T=(H_d^0, H_d^-)$, $H_u^T=(
H_u^+, H_u^0)$.}. 

Let us now discuss in the next section the neutral Higgs sector of the model and possible signals at colliders.

% [HAY QUE PONER ESTO? We want to stress that the presence in the superpotential of terms with $Y_{\nu_{ij}}$ at the same time of those  with $\lambda_i$, $\kappa{_{ijk}}$ produces the explicit breaking of R-parity. Then, all the processes that are breaking R-parity must contain at least one neutrino Yukawa parameter in addition to at least one $\kappa$ or $\lambda$ parameter.]

% Regarding the mass eigenstates, as was explained in \cite{MuNuSSM2}, all  the fermions with the same electric charge mix together, as also all the scalars with the same electric charge do. Since we are assuming CP conservation, the neutral CP-even scalars are not mixed with the neutral CP-odd scalars.

%For further purpose we also 
%define as family indices $l,m=1,2,3$, and a sum will be implicit for 
%repeated indices.

% Since R-parity is not a symmetry of the theory neutrinos right-handed mix with MSSM neutralinos and neutrinos left-handed. Then we will call all this particles as neutralinos. Then the lightess neutralinos are mostly neutrinos left-handed particles.  Also sneutrinos right-handed mix with the doublet-Higgses and sneutrinos left-handed we call all this particles Higgses. In this case is not clear the composition of the lightest particles.

\section{Higgs sector and decays}\label{Higgs sector and decays}
In this section we will analyse the mixings in the scalar (Higgs) sector of the $\mu \nu$SSM, and we will also study the possible decay modes of Higgses once they are generated at colliders. We will focus our attention on the novelties that this extended Higgs sector introduces compared to Higgs sectors of other models, like e.g. the one of the NMSSM. Finally, we will discuss the LEP constraints in the context of this model.

\subsection{Higgs sector mixings}\label{Higgs sector mixings}

The presence of right and left-handed sneutrino VEVs in the 
$\mu \nu$SSM leads to mixing of the
neutral components of the Higgs doublets with the sneutrinos producing the 
$8\times 8$ neutral scalar mass matrices for the 
CP-even 
and CP-odd states \cite{MuNuSSM2} that can be found in
Appendix \ref{Appendix Mass matrices}, where we have defined as usual 
\begin{eqnarray}
&& H^0_u = \frac{h_u+\,iP_u}{\sqrt{2}}+v_u\ , \,\,\,\,\, H_d^0=\frac{h_d+\,iP_d}{\sqrt{2}}+v_d,
\nonumber\\ 
&& \widetilde{\nu}^c_i=\frac{
     (\widetilde{\nu}^c_i)^R+\, i(\widetilde{\nu}_i^c)^I}{\sqrt{2}}+\nu_i^c\ ,  
\,\,\,\,\, \widetilde{\nu}_i=\frac{(\widetilde{\nu}_i)^R  +\, i (\widetilde{\nu}_i)^I}{\sqrt{2}}
  + \nu_i\ .
\end{eqnarray}
Note that after rotating away the CP-odd would be Goldstone boson,
we are left with seven states.
It is also worth noticing here that in the CP-even sector
the $5\times 5$ Higgs doublets--right handed sneutrino submatrix is basically decoupled from the 
$3\times 3$ left handed sneutrino submatrix, since the mixing occurs only through terms proportional
to $\nu_i$ or $Y_{\nu_{ij}}$ in
(\ref{Adl}), (\ref{Aul}) and (\ref{Alr}). 
As discussed in \cite{MuNuSSM}, because of the contribution of the small couplings $Y_{\nu}\sim 10^{-6,-7}$ to the minimization conditions for the left-handed sneutrinos, their VEVs turn out to be small $\nu \sim 10^{-4,-5}$~GeV.
Then, all terms containing $Y_{\nu}$ or $\nu$ are negligible compared to the rest of terms that are of the order of the EW scale.
The same decoupling between Higgs doublets-right handed sneutrinos and left-handed sneutrinos is true for the CP-odd sector.

On the contrary, the mixing between
Higgs doublets and right-handed sneutrinos is not necessarily small.
In the CP-even sector this is given 
by (\ref{Adr}) and (\ref{Aur}):
\begin{eqnarray}
M_{h_{d}(\widetilde{\nu}_i^c)^R }^{2}  &= & -a_{\lambda_{i}}v_{u}+2\lambda_{i}\lambda_{j}v_{d}\nu_j^c -2\lambda_{k} \kappa_{ijk}v_{u}\nu_j^c-Y_{\nu_{ji}}\lambda_{k}\nu_{j}\nu_k^c -Y_{\nu_{jk}}\lambda_{i}\nu_{j}\nu_k^c\ , \\
M_{h_{u}(\widetilde{\nu}_i^c)^R }^{2} &= & -a_{\lambda_{i}}v_{d}+a_{\nu_{ji}}\nu_{j}+2\lambda_{i}\lambda_{j}v_{u}\nu^c_{j}-2\lambda_{k}\kappa_{ilk}v_{d}\nu^c_{l}+2Y_{\nu_{jk}}\kappa_{ilk}\nu_{j}\nu^c_{l}
+2Y_{\nu_{jk}}Y_{\nu_{ji}}v_{u}\nu^c_{k}\ . \nonumber \\
\end{eqnarray}
Neglecting  terms proportional to $Y_{\nu_{ij}}$, $\nu_{i}$, using $a_{\lambda_i}= (A_\lambda\lambda)_i$, and defining $\mu \equiv \lambda_j \nu_j^c$,  one can write the above equations as
\begin{align}
M_{h_{d}(\widetilde{\nu}_i^c)^R }^{2} \approx 2\lambda_{i} \mu v_{d}
-2\lambda_{k} \kappa_{ijk}v_u\nu_j^c
-v_u (A_\lambda\lambda)_i
%-(2\lambda_{k} \kappa_{ijk}\nu_j^c+a_{\lambda_{i}})v_{u}
\ ,
\label{mixturehds1}\\
%\end{align}
%
%\begin{align}
M_{h_{u}(\widetilde{\nu}_i^c)^R }^{2} \approx 2\lambda_{i}\mu v_{u}
-2\lambda_{k}\kappa_{ilk}v_d\nu^c_{l}-v_d 
(A_\lambda\lambda)_i
%-(2\lambda_{k}\kappa_{ilk}\nu^c_{l}+a_{\lambda_{i}})v_{d}
\ .
\label{mixturehus1}
\end{align}

Let us now discuss how to suppress these mixings. This can be used to have very light $\tilde \nu^c$-like Higgses avoiding collider constraints,
but also, as we will discuss below, to have a doublet-like Higgs as the lightest one being as heavy as possible. The simplest possibility to suppress the mixings is that Eqs. (\ref{mixturehds1}) and (\ref{mixturehus1}) vanish.
Clearly, this can be obtained with $\lambda_i \to 0$.
%(recall that $a_\lambda\equiv \lambda A_\lambda$). 
%Thus in this situation it is possible to have very light $\tilde{\nu}^c$-like Higgses that could escape collider constraints.
Another possibility is that the sum of the three terms in the above equations vanishes.
% \begin{align}
% 0 \approx 2\lambda_{i} \mu v_{d}
% -2\lambda_{k} \kappa_{ijk}v_u\nu_j^c
% -v_u (A_\lambda\lambda)_i
%-(2\lambda_{k} \kappa_{ijk}\nu_j^c+a_{\lambda_{i}})v_{u}
% \ ,
% \label{mixturehds}\\
% \end{align}
% \begin{align}
% 0 \approx 2\lambda_{i}\mu v_{u}
% -2\lambda_{k}\kappa_{ilk}v_d\nu^c_{l}-v_d 
% (A_\lambda\lambda)_i
%-(2\lambda_{k}\kappa_{ilk}\nu^c_{l}+a_{\lambda_{i}})v_{d}
% \ .
% \label{mixturehus}
% \end{align}
%
To simplify this analysis let us start with only one generation of right-handed neutrinos. 
Then, 
%Eqs. (\ref{mixturehds}) and (\ref{mixturehus}) 
%can be written as
\begin{align}
0 \approx 2\lambda \mu v_{u}-2\lambda \kappa v_d \nu^c - v_d 
A_\lambda\lambda\ , \\
0 \approx 2\lambda \mu v_{d}-2\lambda \kappa v_u \nu^c - 
v_{u}A_\lambda\lambda\ ,
\end{align}
and after a rotation in the mass matrix we obtain the condition \cite{MuNuSSM2}
\begin{align}
A_{\lambda}=
\frac{2 \mu}{\sin{2 \beta}}
%2 \mu \sin^{-1}{(2 \beta)}
-2\kappa \nu^c  \ ,
\end{align}
similar to the one of the NMSSM (with $\nu^c \rightarrow S$) \cite{NMSSMmezclas}.
Following the same arguments as above, in the CP-odd sector,
and after a rotation in the mass-squared matrix to isolate the Goldstone boson, we obtain the condition,
\begin{align}
\lambda (A_\lambda-2 \kappa \nu^c)v=0\ ,
\end{align}
implying $\lambda \to 0$ or $A_\lambda=2 \kappa \nu^c$.
The generalization of these results to three generations of right-handed neutrinos is straightforward. 
In addition to the solution $\lambda_i \to 0$, we obtain
\begin{align}
A_{\lambda_i} &= \frac{2 \mu}{\sin{2 \beta}}-\frac{2}{\lambda_i}\sum_{j,k} \kappa_{ijk} \lambda_j \nu_k^c\ ,
\label{abc}\\
%\end{align}
%\begin{align}
A_{\lambda_i} &= \frac{2}{\lambda_i} \sum_{j,k}\kappa_{ijk}\lambda_j \nu_k^c\ ,
\label{abcd}
\end{align}
for the CP-even and CP-odd sectors, respectively.

Nevertheless, although the above solutions
for the decoupling of Higgs doublets and right-handed sneutrinos can be used in general, they are
sufficient but not necessary conditions.
% that can be used in the general case for the decoupling of the Higgs doublets and the right-handed sneutrinos. 
As was shown in \cite{MuNuSSM2}, there are regions of the parameter space where the off-diagonal mixing terms of the neutral scalar mass matrices are smaller than the diagonal terms, and then quite pure singlets can also be obtained. Actually, we will use this mechanism in Section \ref{numericalresults} in order to search for interesting signals at colliders.

Let us finally emphasize that some of these conditions can be applied not only to obtain a very light $\tilde{\nu}^c$-like lightest Higgs, as discussed above, but also to have the lightest scalar as heavy as possible\footnote{Notice that the upper bound on the lightest Higgs boson mass for the $\mu \nu$SSM turns out to be similar to the one of the NMSSM \cite{MuNuSSM2}.}. Clearly this lightest scalar, for being as heavy as possible, must be Higgs doublet-like, since the
right- and left-handed sneutrinos can be as heavy as we want. 
%In order to have what we want 
Thus to have it as heavy as possible the contamination with right-handed sneutrinos should be small. For this to happen the right-handed sneutrinos must be very heavy and/or the mixing should be small. Notice however that for the latter we cannot use one of the conditions discussed above, $\lambda_i \to 0$, since $\lambda_i$ must be as large as possible to saturate the upper bound on the lightest Higgs boson mass \cite{MuNuSSM2}.
% and conditions (\ref{abc}, \ref{abcd}) are more appropriated. 

% \textbf{We want to notice that these are sufficient but not necessary conditions that can be used in the general case for the decoupling of the Higgs doublets and the right-handed sneutrinos. Nevertheless, for the results of the computation presented in Section \ref{numericalresults}, to attain the decoupling, we will use the degeneracy of parameters for having very pure singlets as we will explain below.}

% Summarizing, we have discussed in this subsection the mixing in the Higgs sector of the 
% $\mu\nu$SSM. In particular, we have learnt how to suppress the mixings between right-handed sneutrinos and Higgs doublets \textbf{for having} very light $\tilde{\nu^c}$-like Higgses avoiding collider constraints,
% or \textbf{for having} a doublet-like Higgs as the lightest one being as heavy as possible.

\subsection{Decays}\label{Decays}

Here we will study possible decay modes of the Higgses in the 
$\mu \nu$SSM, pointing out novel features with respect to the NMSSM/MSSM.
%an overview of the decays that would take place in the Higgs sector of the $\mu \nu$SSM in order to point out the novel features respect to models with simpler Higgs sectors as the NMSSM. 
The presence of new fields extending the Higgs sector, and the fact that $R$-parity is not a symmetry of the model, give rise to new decays, thus 
%respect to the MSSM and NMSSM case, 
changing substantially the phenomenology.

\begin{figure}[t]
  \begin{center} 
\hspace*{-10mm}
%    \begin{tabular}{cc}
	\epsfig{file=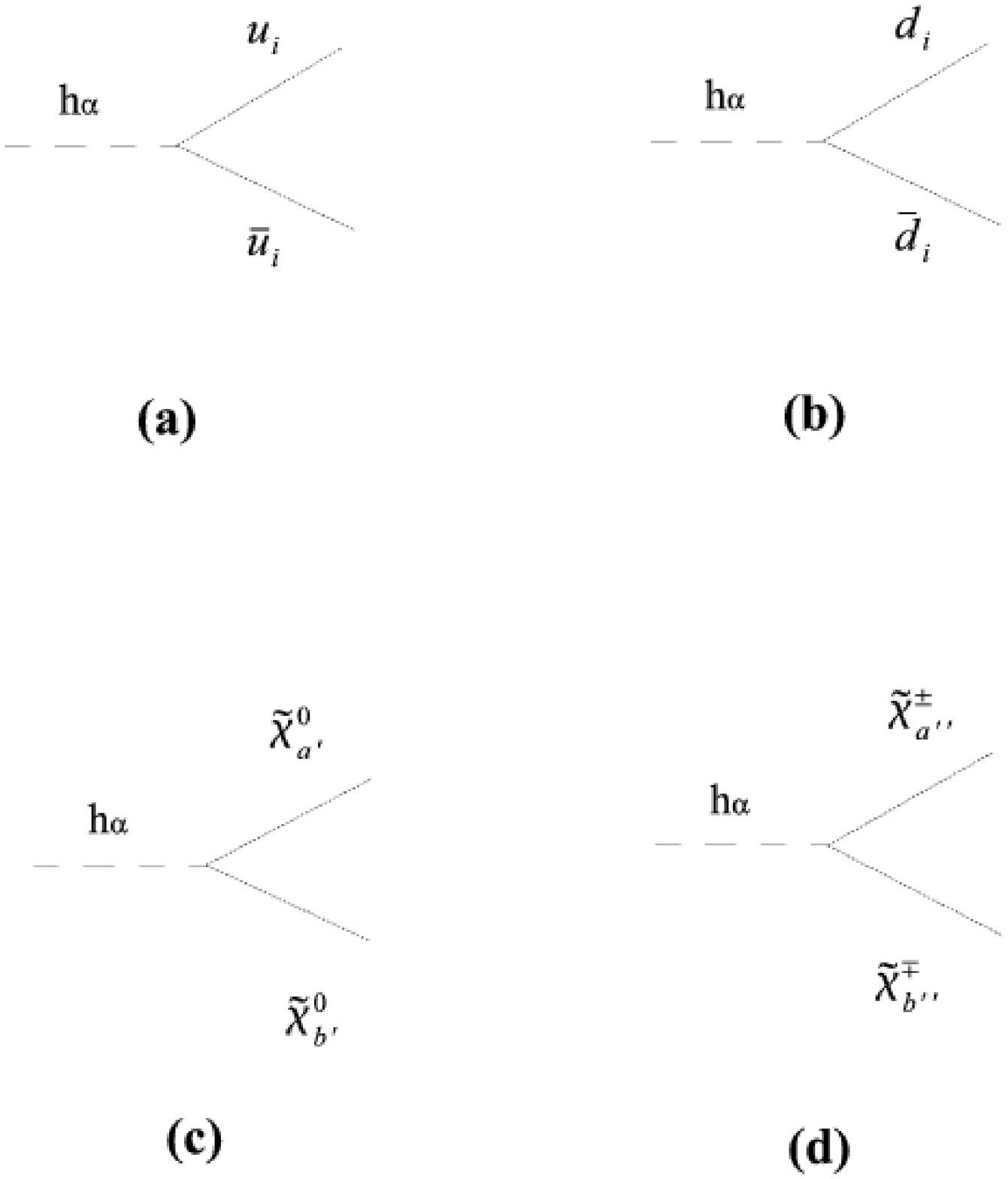,width=140mm,angle=0,clip=} 
% & \hspace*{-10mm} \epsfig{file=figures/DibujoDiagTreeSToSS.ps,width=90mm,angle=0,clip=}
%    \end{tabular}
\vspace*{2cm}
\captions{Feynman diagrams of Higgs decay to fermions, with $\alpha=1,...,8$, $i=1,2,3$, $a', b'=1,...,10$, and $a'', b''=1,...,5$. Replacing the decaying $h_\alpha$ by a pseudoscalar $P_{\alpha'}$, with 
$\alpha'=1,...,7$,
all the Feynman diagrams are valid.} 
    \label{fig:fermions}
  \end{center}
\end{figure}

\begin{figure}[t]
  \begin{center} 
\hspace*{-10mm}
%    \begin{tabular}{cc}
	\epsfig{file=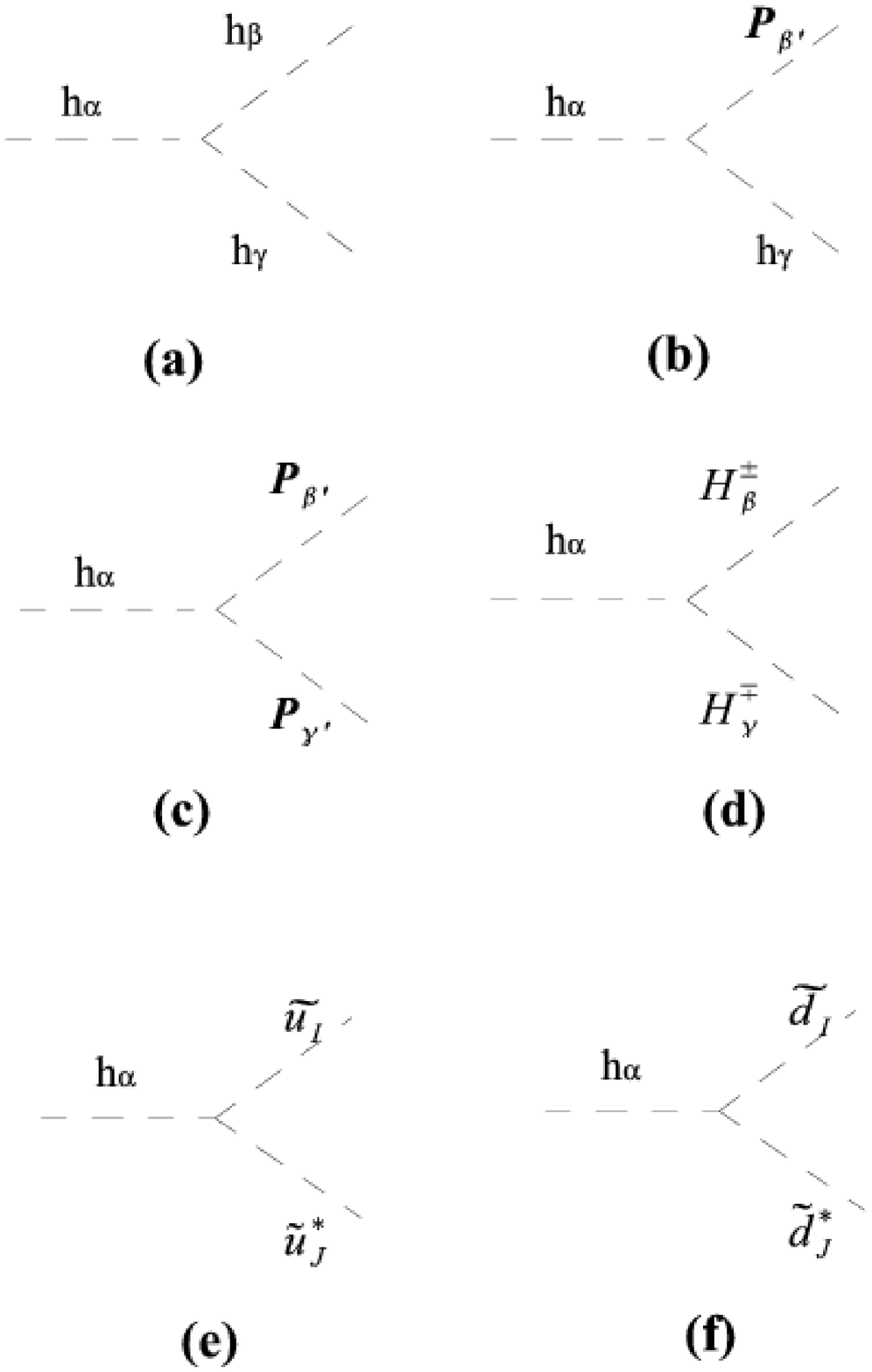,width=140mm,angle=0,clip=}

% & \hspace*{-10mm} \epsfig{file=figures/DibujoDiagTreeSToSS.ps,width=90mm,angle=0,clip=}
%    \end{tabular}
\captions{Feynman diagrams of Higgs decays to scalars (under Lorentz), with $I,J=1..6$.  Replacing the decaying $h_\alpha$ by a pseudoscalar $P_{\alpha'}$, all the Feynman diagrams are valid.
The index convention is like in Fig.~1.} 
    \label{fig:scalars}
  \end{center}
\end{figure}

\begin{figure}[t]
  \begin{center} 
\hspace*{-10mm}
%    \begin{tabular}{cc}
	\epsfig{file=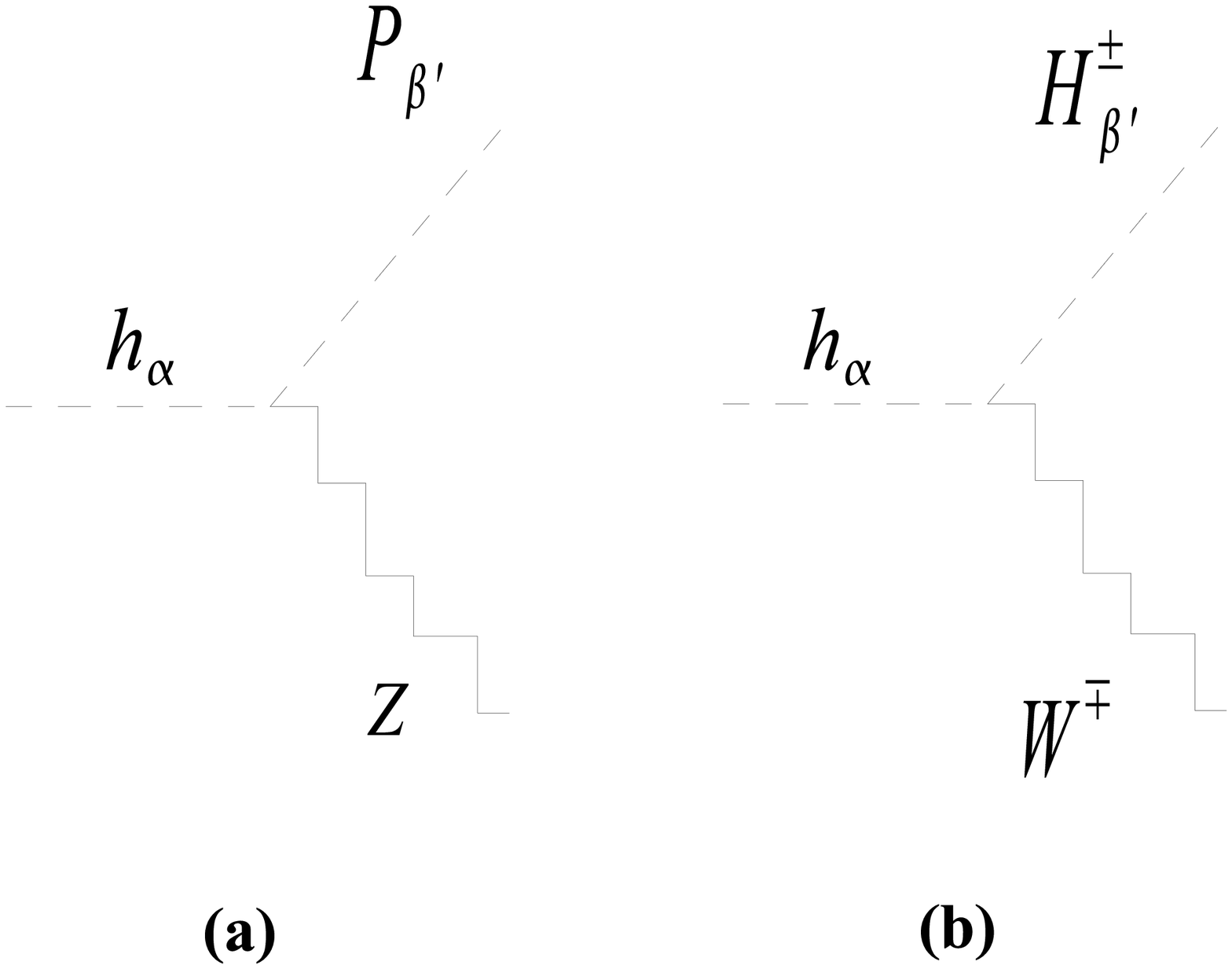,height=90.5mm,angle=0,clip=} 
% & \hspace*{-10mm} \epsfig{file=figures/DibujoDiagTreeSToSS.ps,width=90mm,angle=0,clip=}
%    \end{tabular}
\captions{Feynman diagrams of Higgs decays  to scalars and vectors (under Lorentz). Replacing the decaying $h_\alpha$ by a pseudoscalar $P_{\alpha'}$, all the Feynman diagrams are valid. The index convention is like in Fig.~1.} 
    \label{fig:scalarsvector}
  \end{center}
\end{figure}

\begin{figure}[t]
  \begin{center} 
\hspace*{-10mm}
%    \begin{tabular}{cc}
	\epsfig{file=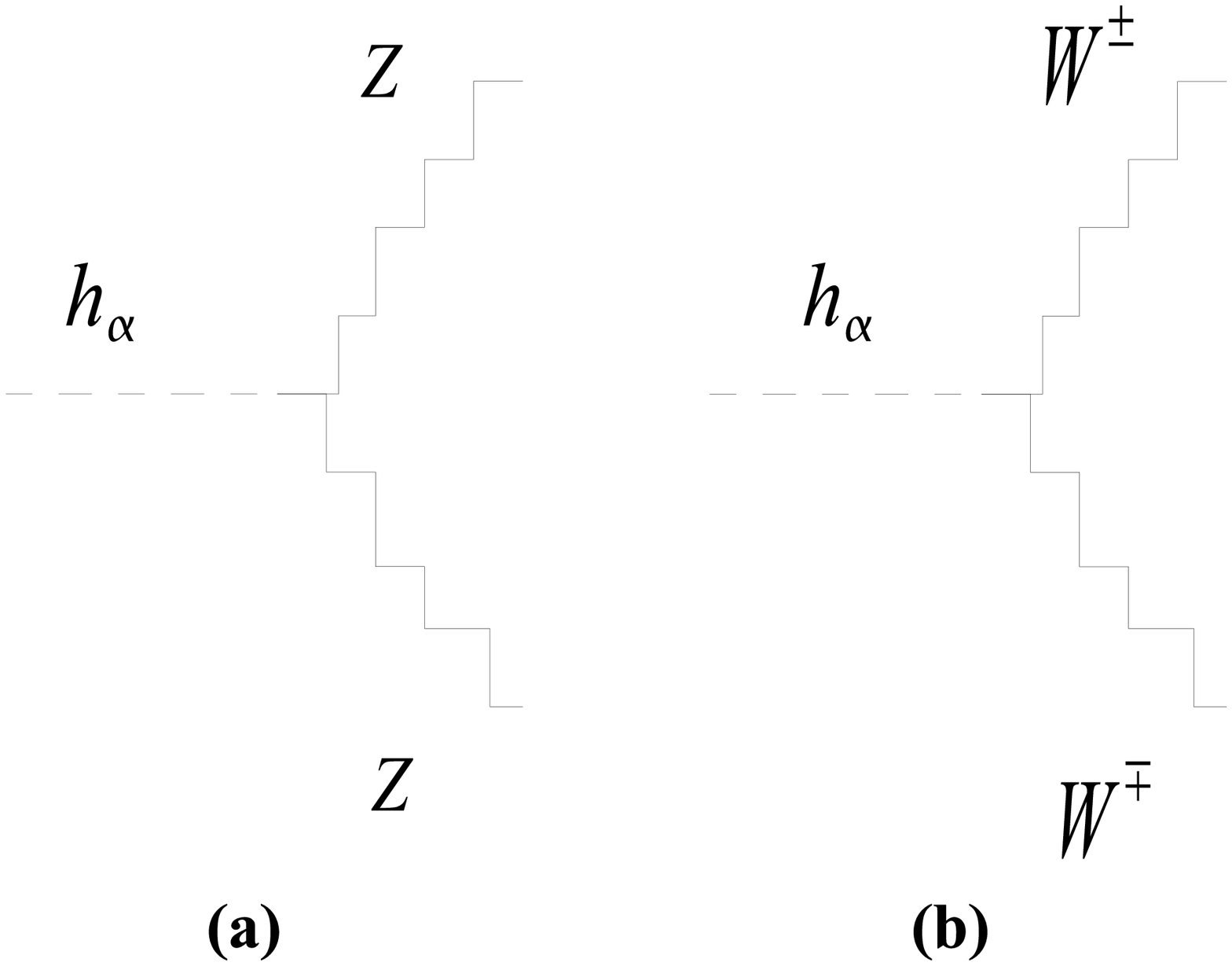,height=90.5mm,angle=0,clip=} 
% & \hspace*{-10mm} \epsfig{file=figures/DibujoDiagTreeSToSS.ps,width=90mm,angle=0,clip=}
%    \end{tabular}
    \captions{Feynman diagrams of Higgs decays to vectors (under Lorentz). The index convention is like in Fig.~1.} 
    \label{fig:vectors}
  \end{center}
\end{figure}

First of all, the Higgs-to-Higgs cascade decays can be more complicated since more Higgses are present in this model compared to the NMSSM. As discussed above, in the $\mu \nu$SSM there are eight CP-even and seven CP-odd Higgses, while in the NMSSM there are three CP-even and two CP-odd Higgses. The relevant couplings  for Higgs-to-Higgs decays in the $\mu \nu$SSM are written in Appendix \ref{Appendix Higgs sector couplings}, and the Feynman diagrams of all possible tree-level decays of the Higgses are given in Figs.~\ref{fig:fermions}-\ref{fig:vectors}.
In particular, for a CP-even (CP-odd) decaying scalar we can see in Fig.~\ref{fig:scalars} that the Feynman diagrams {\bf a} and {\bf c} ({\bf b}) are crucial to understand new decays with respect to the NMSSM ones.
Note that the Feynman diagram(s) {\bf b} ({\bf a} and {\bf c}) in the figure is (are) present only if a source of CP violation is taken into account\footnote{In order to reduce the number of Feynman diagrams shown in the figure, we allow an abuse of notation in the diagrams, since if CP is violated, CP-even and CP-odd Higgses mix together and the notation ceases to make sense.}. 
%For the CP-odd scalars (the pseudoscalars), N3 and N4 are the crucial ones, while N2 and N5 are only present if there is a CP violating source. 
%If we take the Higgs-strahlung process (Z to Z* Higgs), in this model the Higgs could decays through these Feynman diagrams.

Let us assume that we have enough energy to generate only one CP-even Higgs at a collider, i.e., only one Higgs, $h_1$, has mass below the threshold energy.
Then the following decay is possible:
\bea
  h_1  \to 2 \text{ Standard Model fermions} 
%\nonumber
\ . 
\eea
In case that the second lightest Higgs, $h_2$,
%are below the threshold energy 
can be generated, the following cascade decay is possible if kinematically allowed:
\bea
h_2 \to 2h_1  \to 4 \text{ Standard Model fermions} 
%\nonumber
\ . 
\eea
If the third lightest Higgs, $h_3$, 
%below the threshold, 
can be generated, then if kinematically allowed we have the possibility
\bea
h_3 \to 2h_2 \to 4h_1 \to 8 \text{ Standard Model fermions} 
%\nonumber
\ . 
\eea
The situation turns out to be more complicated if we take into account the decays to scalars that are not the ones immediately below in mass. Also we have the possibility of having light pseudoscalars entering in the game. In the $\mu\nu$SSM we have three/two (six/five including left-handed sneutrinos) pseudoscalars more than in the MSSM/NMSSM case,  and they could be very light. Thus we may need to include the following decays (if kinematically allowed) into the cascades:
\bea
h_{\alpha} \to h_{\beta}  h_{\gamma} \ , \qquad \hspace*{2cm} h_{\alpha} \to P_{\beta'}  P_{\gamma'} \ , \qquad \hspace*{2cm} P_{\alpha'} \to P_{\beta'}  h_{\gamma} \label{decayeq} 
%\nonumber 
\ ,
 \eea
where $\alpha, \ \beta, \ \gamma=1,...,8$ and $\alpha', \ \beta', \ \gamma'=1,...,7$.
% If this is the case we could have the following decays,
% \bea
% k_{a} \to k_{a-1} k_{a-1} \to k_{a-2} k_{a-2} k_{a-2} k_{a-2} \to ...\to  2^i  \text{ standard model fermions} \label{decayeq} \\ \nonumber 
% \eea
% where $k_a$ contains all the pseudoscalars and scalars that are below the threshold ordered from lightest to heaviest. Including slepton left-handed $a$ could have the maximun value of 15 (excluding left-handed sleptons because are almost decouple the maximun value is 9).
% %. The first one must be a scalar.

In benchmark point 7 of Section \ref{numericalresults} we will study an example where these types of Higg-to-Higgs cascade decays are 
present.
%relevant to distinguish the $\mu \nu$SSM from other SUSY models. 
Working with a MSSM-like CP even Higgs, $h_{\text{MSSM}}$, it will
decay into $b \bar b$ or through the cascades typical of the NMSSM, 
$h_{MSSM} \to 2P \to 2 b 2 \bar b$, in most of the cases. Nevertheless we will see that 
the following cascade is also possible:
$h_{MSSM} \to 2h \to 4P \to 4b 4\bar b$.
In benchmark point 8 we will see that $h_{MSSM}$ 
can decay with the following relevant cascades: $h_{MSSM}\to 2 h_1 \to 4P_{1,2}\to 4\tau^+4\tau^-$, $h_{MSSM}\to 2P_3 \to 2b 2 \bar b$, 
because for the singlet-like pseudoscalars $P_{1,2}$ the decay into $b \bar b$ is kinematically forbidden, whereas for $P_3$ it is allowed. These cascades are
genuine of the $\mu \nu$SSM.

\begin{figure}[h!]
 \begin{center}
\hspace*{-8mm}
    \begin{tabular}{cc}
      \epsfig{file=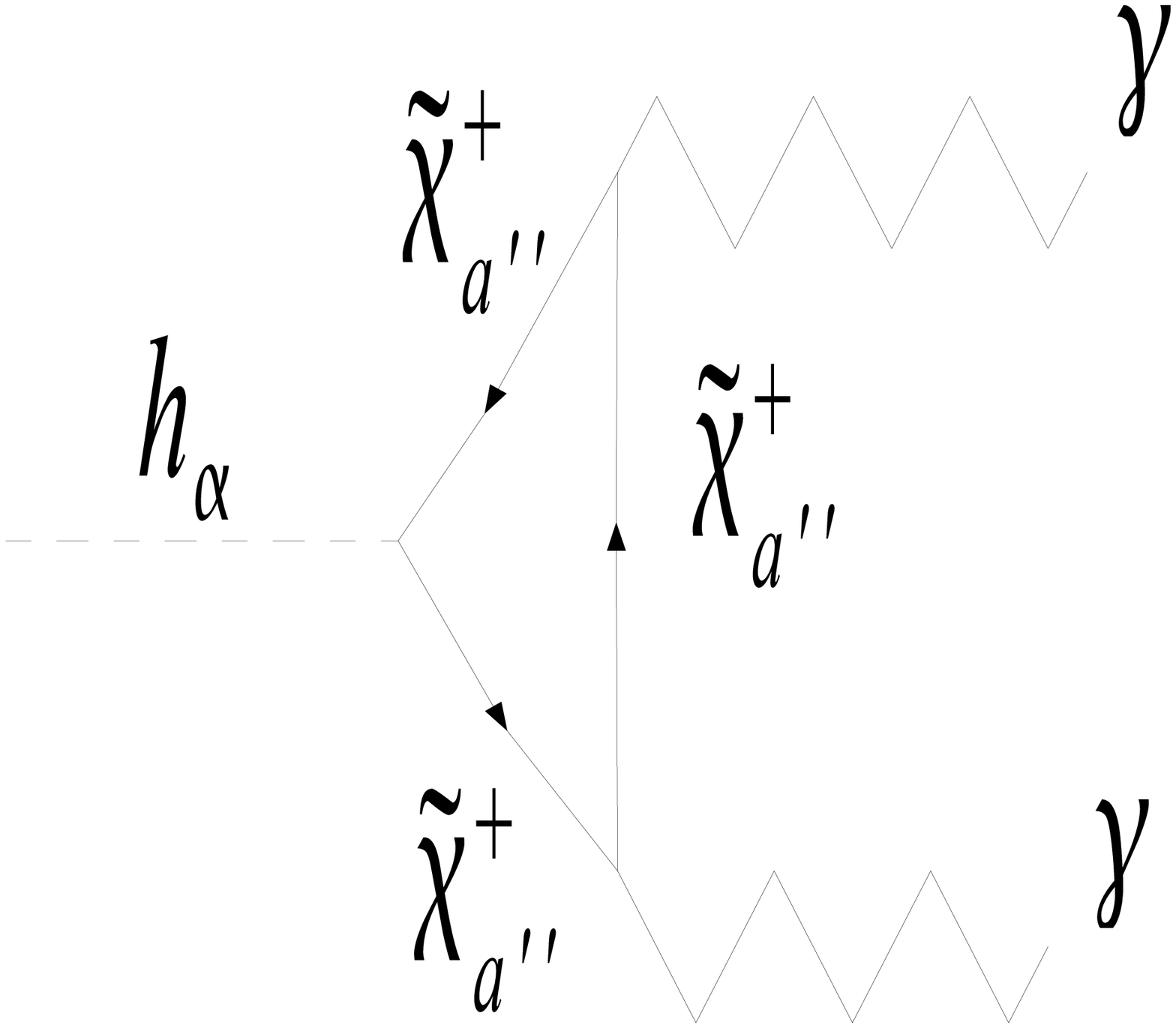,height=5.3cm,angle=-0}
%         \hspace*{0mm}&\hspace*{-3mm}
%         \epsfig{file=figures/BolaCharginoA.eps,height=2.5cm,angle=-0}
     %   \hspace*{0mm}&\hspace*{-3mm}
% \epsfig{file=figures/HiggsDecayToNeutrinos.eps,height=2.5cm,angle=-0}
%       \\ & \\
%         \hspace*{-1.1cm} (a) 
%& \hspace*{-1cm} (b)
    \end{tabular}
\captions{
%(a) 
Di-photon Higgs decay. 
Replacing the decaying $h_\alpha$ by a pseudoscalar $P_{\alpha'}$, the Feynman diagram is valid.
%We can observe the 
%proportionality between them. 
%(b) Di-photon CP-odd Higgs decay. 
%Tree-level CP-even Higgs decay to neutrinos.
The index convention is like in Fig.~1.}
    \label{fig: BolaCharginos and HiggsDecayToNeutrinos}
 \end{center}
\end{figure}

Another difference of the $\mu\nu$SSM compared to the NMSSM, that comes from the breaking of $R$-parity, is that a very light lightest Higgs with the decays into $b \bar b$ or $\tau^+ \tau^-$ kinematically forbidden, could decay into two neutrinos $\nu_i \nu_j$ at the tree-level.
%(this process is supressed by the Yukawa coupling of the neutrinos), 
This possibility is included in the Feynman diagram {\bf c} of Fig.~\ref{fig:fermions}, due to the mixing of the MSSM neutralinos and neutrinos.
This decay
takes place due to the presence of the superpotential terms $Y_\nu\hat H_u \hat L \hat \nu^c$. A Higgs with $H_u$ composition can decay in this way
%to $\nu_i \nu_j$, 
because the light neutrinos, that are mainly left-handed, have small right-handed neutrino $\nu^c$ components. 
A Higgs with $\tilde{\nu}$ component can decay to two neutrinos 
%to $\nu_i \nu_j$, 
because the light neutrinos can have respectively $\tilde{H_u}$ and $\nu^c$ components. Of special interest is the fact that a Higgs with $\tilde{\nu^c}$ composition can also decay into $\nu_i \nu_j$ because the light neutrinos can have $\tilde{H_u}$ component, as mentioned above, or 
%because the $\nu^c$ composition of the light neutrinos
through the $\kappa \hat \nu_i^c \hat \nu_j^c \hat \nu_k^c$ terms in the superpotential, taking into account that left- and  right-handed neutrinos mix together.
% and the mixtures in the mass matrices. 
However, since the neutrino Yukawa couplings are small, it is difficult to compete with the usual 1-loop decay into photons through the chargino loop process
(see Fig.~\ref{fig: BolaCharginos and HiggsDecayToNeutrinos}).
%%,\ref{fig: BolaCharginos and HiggsDecayToNeutrinos pseudo}a),
%%and 
%%\ref{fig: BolaCharginos and HiggsDecayToNeutrinos},
%%\ref{fig: BolaCharginos and HiggsDecayToNeutrinos pseudo}b).
%i.e. it is not easy to have a significant reduction for the
%branching ratio to photons by an increase in the annihilation to neutrinos.
%% via a Higgs.
Then, the usual constraints for very light Higgses annihilating to photons \cite{gamma} still apply.

Also we must take into account that, unless they are not kinematically allowed, new decays to leptons are present, as can be deduced from the Feyman diagram {\bf d} of Fig.~\ref{fig:fermions}, since the charged leptons are mixed with the MSSM charginos. Then, a singlet-like Higgs could decay to charged leptons through the $\lambda \hat \nu^c \hat H_u \hat H_d $ terms in the superpotential, due to the chargino composition.
% For very pure singlet-like Higgs the situation could be different from the NMSSM one specially for very light ones. 
The mixing of charged leptons with charginos also affects the loop diagrams describing Higgs decaying into photons 
(Fig. \ref{fig: BolaCharginos and HiggsDecayToNeutrinos})
%,\ref{fig: BolaCharginos and HiggsDecayToNeutrinos pseudo}a),
due to
% in the $\mu \nu$SSM 
the contribution from charged leptons running in the loop, since the charginos are contaminated with them.
%In this case this decay could be in competition with the chargino loop  process, pseudoscalar to two photons (Higgs----bola de chargino---- con dos fotones que salen).
Besides, a Higgs with $\tilde{\nu}$ component can also decay into charged leptons through the
$Y_e \hat H_d \hat L \hat e^c$ term in the superpotential.
Notice that it can also decay into two light neutrinos through the contamination 
with $\tilde{H}_u^0$ and $\nu^c$ in the 
Yukawa term $Y_\nu \hat H_u \hat L \hat \nu^c$.
For example, for the benchmark point 2 shown in Table 2 of Section \ref{numericalresults}, the light singlet-like pseudoscalars $P_{1,2,3}$ decay  mainly into $\tau^+ \tau^-$ because of the small contamination with doublets.

\vspace{0.5cm}
An interesting situation that we will study in detail in Section \ref{numericalresults},
occurs when a MSSM-like CP even Higgs, $h_{\text{MSSM}}$,
has a sizeable branching ratio to two light neutralinos 
$h_{\text{MSSM}} \rightarrow \neutralino \neutralino$.
Since $R$-parity is broken, 
neutralinos can decay into a Higgs and a neutrino inside the detector leading to displaced vertices. 
This possibility is included in the Feynman diagram {\bf c} of Fig.~\ref{fig:fermions}, due to the mixing of the MSSM neutralinos and neutrinos.
Thus 
working with light on-shell singlet-like pseudoscalars,
cascades of the type $h_{MSSM} \to \tilde \chi^0 \tilde \chi^0  \to 2 P
%P_{\tilde \nu^c} 
2\nu \to 2b2\bar b 2\nu$, leading to the final state 4 $b$-jets plus missing energy, will be present.
If the decay of the pseudoscalars into two $b$'s is kinematically forbidden, then they decay into $\tau^+ \tau^-$ generating the following cascade: $h_{MSSM} \to \tilde \chi^0 \tilde \chi^0 \to 2 P 2\nu \to 2 \tau^+ 2 \tau^- 2\nu$.
We will also see that the final state 8 $b$-jets plus missing energy is possible in situations where singlet-like scalars are produced by the decay of the neutralino, and they decay to pseudoscalars as shown 
in (\ref{decayeq}),
$h_{MSSM} \to \tilde \chi^0 \tilde \chi^0  \to 2 h 2\nu \to 4P2\nu \to 4b4\bar b 2\nu$. As mentioned above, in benchmark point 8 of 
Section \ref{numericalresults}, 
for the singlet-like pseudoscalars $P_{1,2}$ the decay into $b \bar b$ is kinematically forbidden, whereas for $P_3$ it is allowed, thus 
the following relevant cascades can be produced: $h_4 \to \tilde \chi_4^0 \tilde \chi_4^0 \to 2  P_{1,2} 2\nu \to 2\tau^+2\tau^- 2\nu $, $h_4 \to \tilde \chi_4^0 \tilde \chi_4^0 \to 2 h_{1,2,3} 2\nu \to 4P_{1,2}2 \nu \to 4\tau^+4\tau^- 2 \nu$, $h_4 \to \tilde \chi_4^0 \tilde \chi_4^0 \to 2 P_{3} 2\nu \to 2 b 2\bar b 2  \nu$.

%In addition, as R-parity is broken, the lightest neutralino can decay to a Higgs and a neutrino. Then, neutralinos can also play a role in the Higgs-to-Higgs cascade decays as we will show in Section \ref{numericalresults}. We would like to remark that in the computation, we have used degenerate parameters for the right-handed neutrino sector and the cascades we obtain are simplified since the singlet-like Higgses are almost degenerated. Breaking the degeneration of parameters, in general, will turn on more complicated cascades that can be important to study in a future work. In the numerical study, we have focused our attention on the case of a SM Higgs with mass closed to $114$ GeV in order to be detectable in the near future. Having a sizeable branching ratio to two lightest neutralinos $\tilde \chi_4^0$ that decay inside the detector leading to displaced vertices, cascades of the type $h_{MSSM} \to \tilde \chi_4^0 \tilde \chi_4^0 \to 2 P_{\tilde \nu^c} 2\nu \to bb\bar b \bar b 2\nu$ are studied. 

Displaced vertices are common signals of $R$-parity violating models and could help to distinguish the $\mu \nu$SSM from the NMSSM. In addition, other R-parity breaking models such as the BRpV \cite{hall} do not have singlets in the spectrum, and, as a consequence, the above decays 
%to be studied in Section \ref{numericalresults} 
can be considered as genuine of the $\mu \nu$SSM. 
%In our model, for the points presented in Section \ref{numericalresults} the lightest neutralino would decay via a two-body process to an on-shell pseudoscalar (decaying to $b \bar b$) and a neutrino. 
%This is a remarkable difference with respect to other R-parity breaking models as
Note e.g. that in the BRpV, if the lightest neutralino is lighter than the gauge bosons, only three-body processes are available for its decay.

\vspace{0.5cm}

\begin{figure}[b!]
 \begin{center}
\hspace*{-8mm}
    \begin{tabular}{ccc}
      \epsfig{file=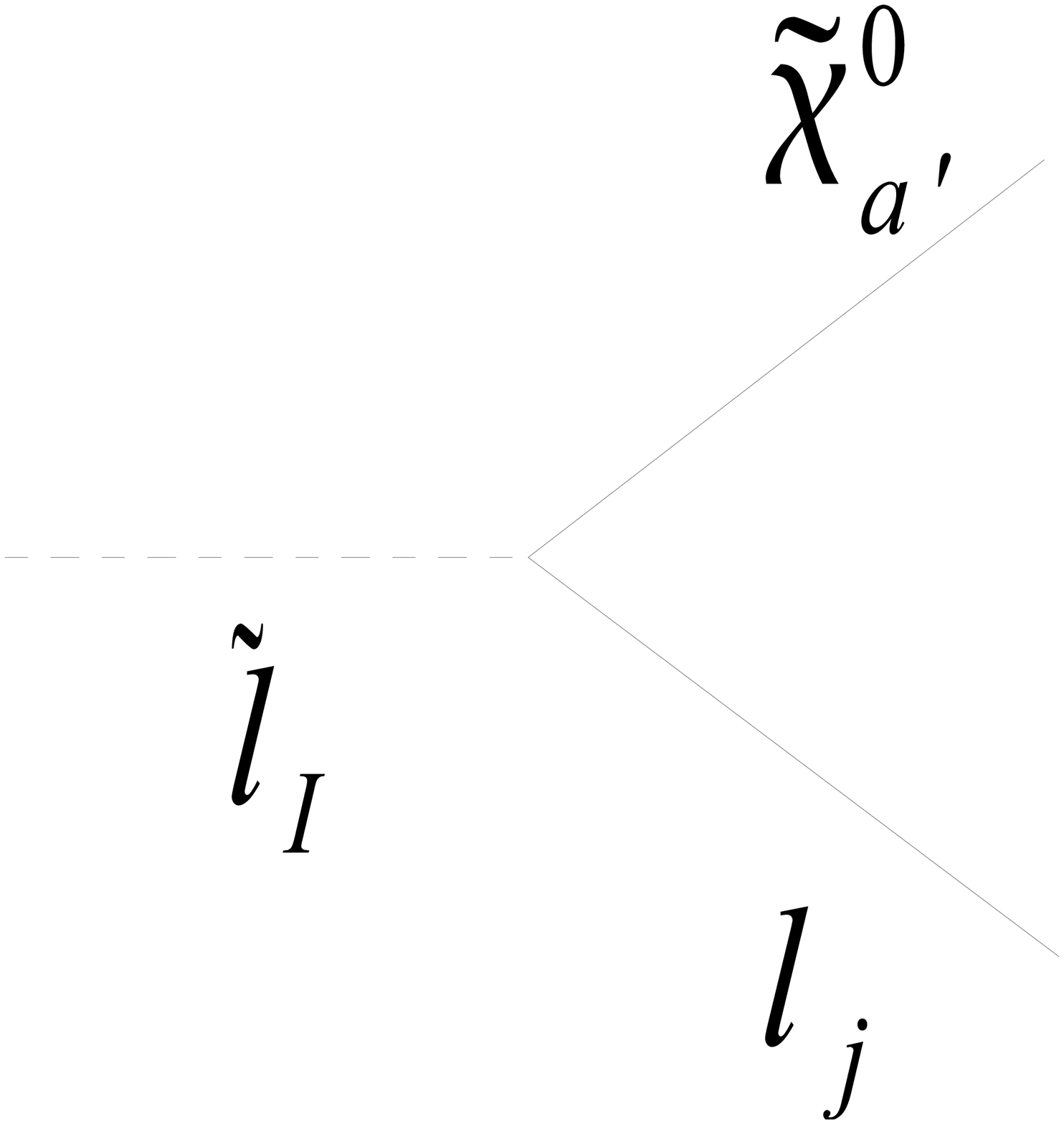,height=4.0cm,angle=-0}
        \hspace*{0mm}&\hspace*{-1mm}
\epsfig{file=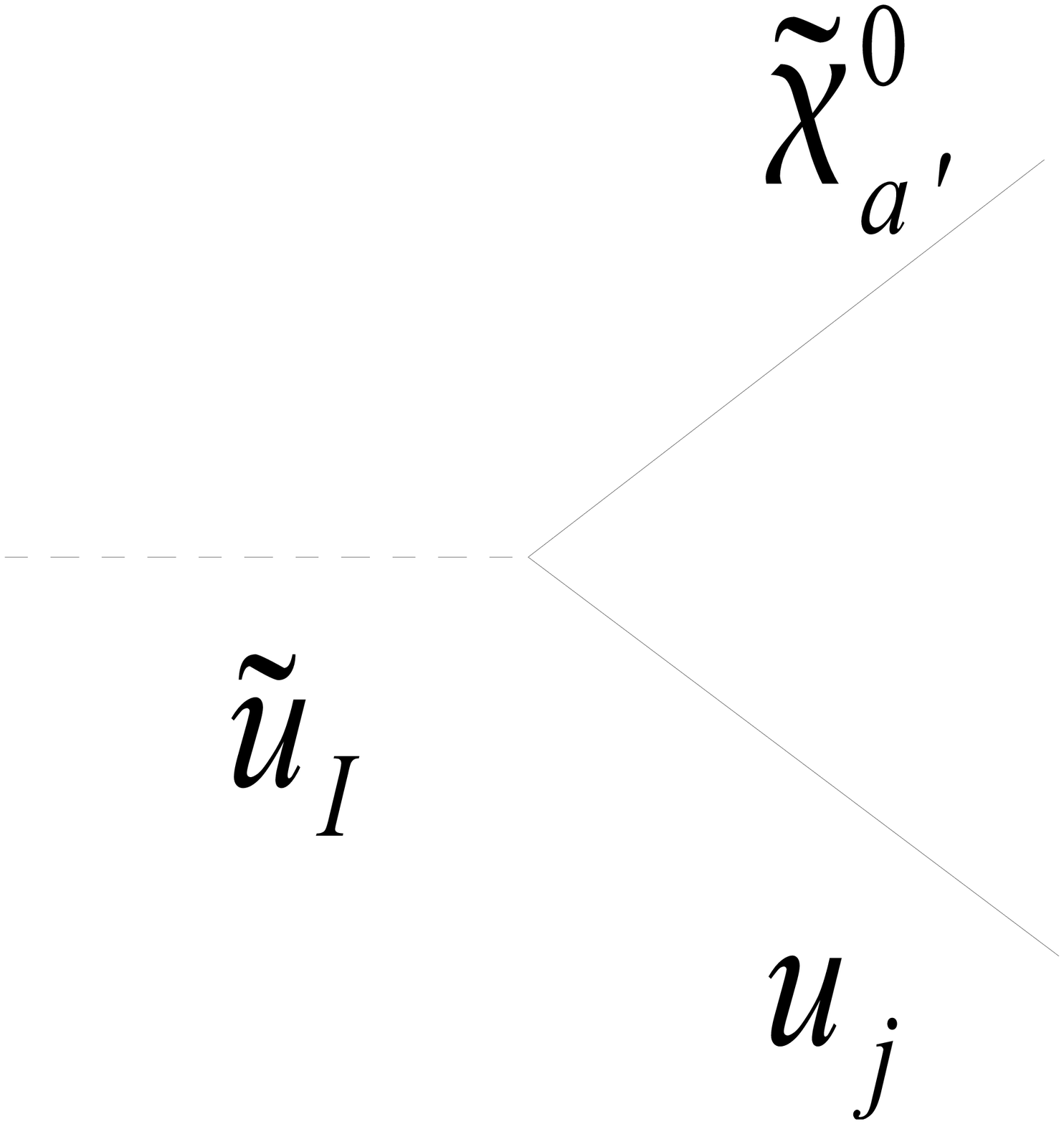,height=4.0cm,angle=-0}
\hspace*{0mm}&\hspace*{-1mm}
\epsfig{file=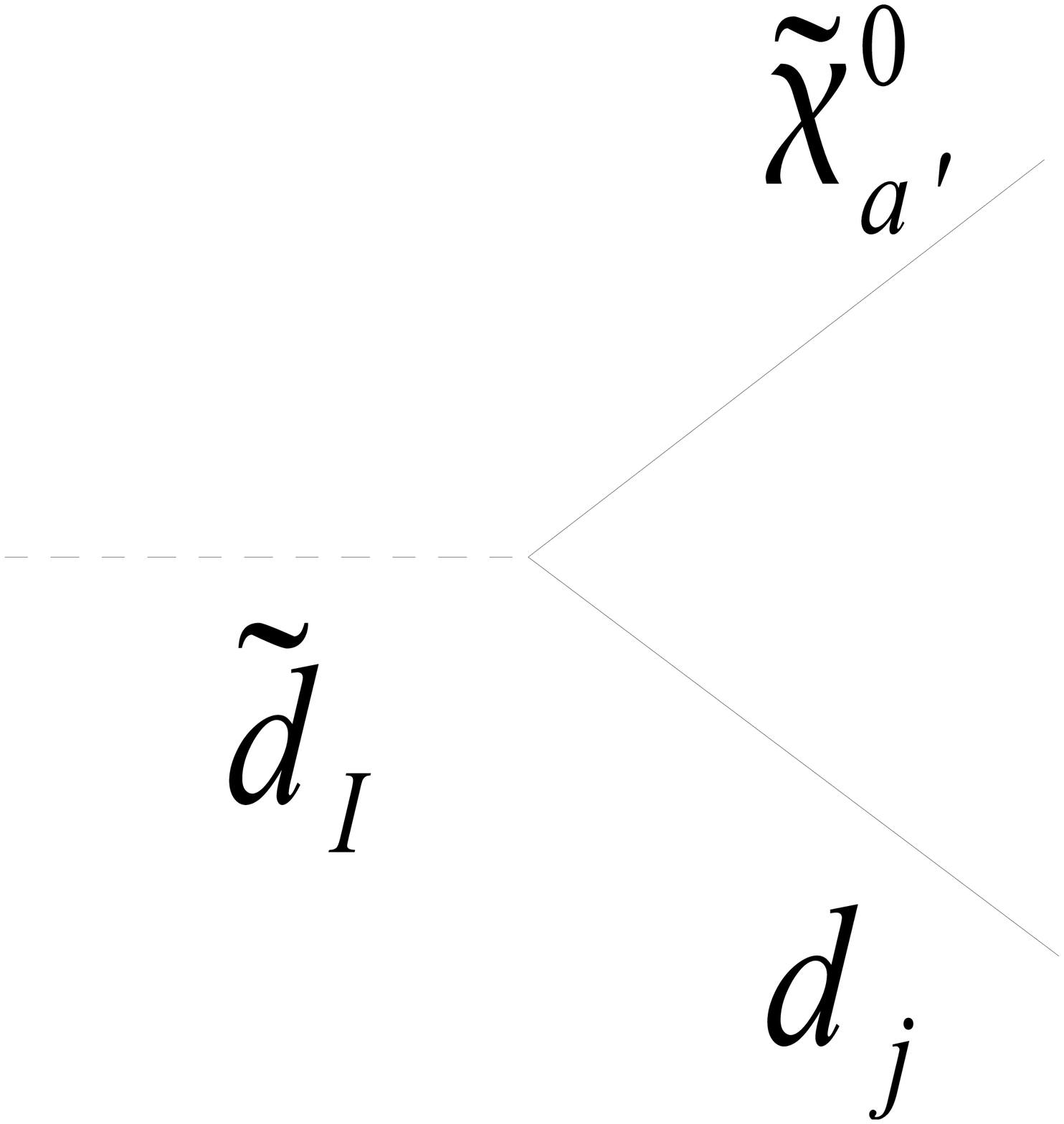,height=4.0cm,angle=-0}
      \\ & \\ & \\
       \hspace*{-1.1cm} (a) & \hspace*{-1cm} (b) & \hspace*{-1cm} (c)
    \end{tabular}
\captions{
(a) Charged slepton decay.
%We can observe the 
%proportionality between them. 
(b) Up squark decay. (c) Down squark decay. 
%The indices take the values $I=1..6$, $j=1..3$, $a'=1..10$.
The index convention is like in Figs.~1 and 2.
}
    \label{fig:sparticlesdecay}
 \end{center}
\end{figure}

Regarding the charged Higgses, as was discussed in \cite{MuNuSSM2}, they are mixed with the sleptons opening the following possibility. 
As usual, a slepton can decay into a neutralino and a lepton as shown in Fig.~\ref{fig:sparticlesdecay}{\bf a}.
In a $R$-parity conserving model, if the neutralino is heavier than the slepton 
%(and then, if the slepton is also lighter that the squarks, it would be stable, NO SE QUE PINTA AQUI EL SQUARK??)
the latter will be stable. However, when $R$-parity is broken, the left-handed neutrinos mix with the neutralinos, and then the slepton decays into a lepton and a light neutrino. 
Since the charged Higgses are mixed with the sleptons, they can also decay in this way. 
% Making the Yukawa of the neutrino zero makes the mixture between MSSM neutralinos plus right-handed neutrinos with the left-handed neutrinos impossible, as also the decay to light neutrinos. 

It is worth noticing here that, similarly to a slepton, a squark can decay into a quark and a light neutrino. This can be deduced from 
Figs.~\ref{fig:sparticlesdecay}{\bf b} and \ref{fig:sparticlesdecay}{\bf c}
using again that neutrinos and neutralinos mix together.
% In this case a squark decays to a quark and a light neutrino, again the reason is because neutrino and neutralino mix together.
Let us also mention that, as usual in $R$-parity breaking models, the squarks or the sleptons can be the LSP\footnote{In the following we will define the LSP as the lightest supersymmetric particle present in Lagrangian when the neutrino Yukawas are set to zero. 
As usual in $R$-parity breaking models, the LSP is not really well defined. For example, 
the lightest scalar with a singlet sneutrino composition can be lighter than the  lightest neutralino. Also the left-handed neutrinos are very light and are mixed with the MSSM neutralinos.}
%Since these Yukawa couplings are not zero, the LSP is not really well defined, and this is an abuse of language that we will allow in such a way to make simpler the discussions.  
%We are going to make an special comment regarding the gravitino in section \ref{gravitino}
without conflict with experimental bounds. Whereas in the MSSM/NMSSM this would imply a stable charged particle incompatible with these bounds, in the $\mu \nu$SSM the LSP decays.

% In particular we want to compare the situation with the one when R-parity is a symmetry, as for instance in the MSSM (similar conclusions apply if we compare with the NMSSM). In that case a squark or slepton () could be the LSP (usually $\tilde{t}$ or $\tilde{\tau}$) but this is excluded because it is charged and stable. In this model since R-parity is not a symmetry of the theory the lightest slepton or squark  will decay also if it is lighter than the MSSM neutralinos. Then, in the $\mu \nu$SSM, the case of a slepton or squark LSP is not ruled out 

In the next subsection we will study the couplings of the Higgses with the $Z$ boson and the sum rules in the $\mu \nu$SSM, discussing also the LEP constraints.

% %%%%%%%%%%%%%%                               %%%%%%%%%%%%%%%%%%%%%%%%%%%
% {\bf DL:} [Escrir sobre todas las posibilidades, escalares o pseudoescalares  $\to$ escalares o pseudoescalares, y asi toda la cascada posible. Terminando en fermiones, o pseudoescalares o lo que sea ...]
%%%%%%%%%%%%%                                   %%%%%%%%%%%%%%%%%%%%%%%%%%%
\subsection{Couplings with the $Z$ boson and sum rules}\label{Coupling with $Z$ bosons and summation rules.}

\begin{figure}[h!]
 \begin{center}
\hspace*{-8mm}
    \begin{tabular}{cc}
%       \epsfig{file=figures/PairProduction2.eps,height=2.5cm,angle=-0}
%         \hspace*{0mm}&\hspace*{-3mm}
\epsfig{file=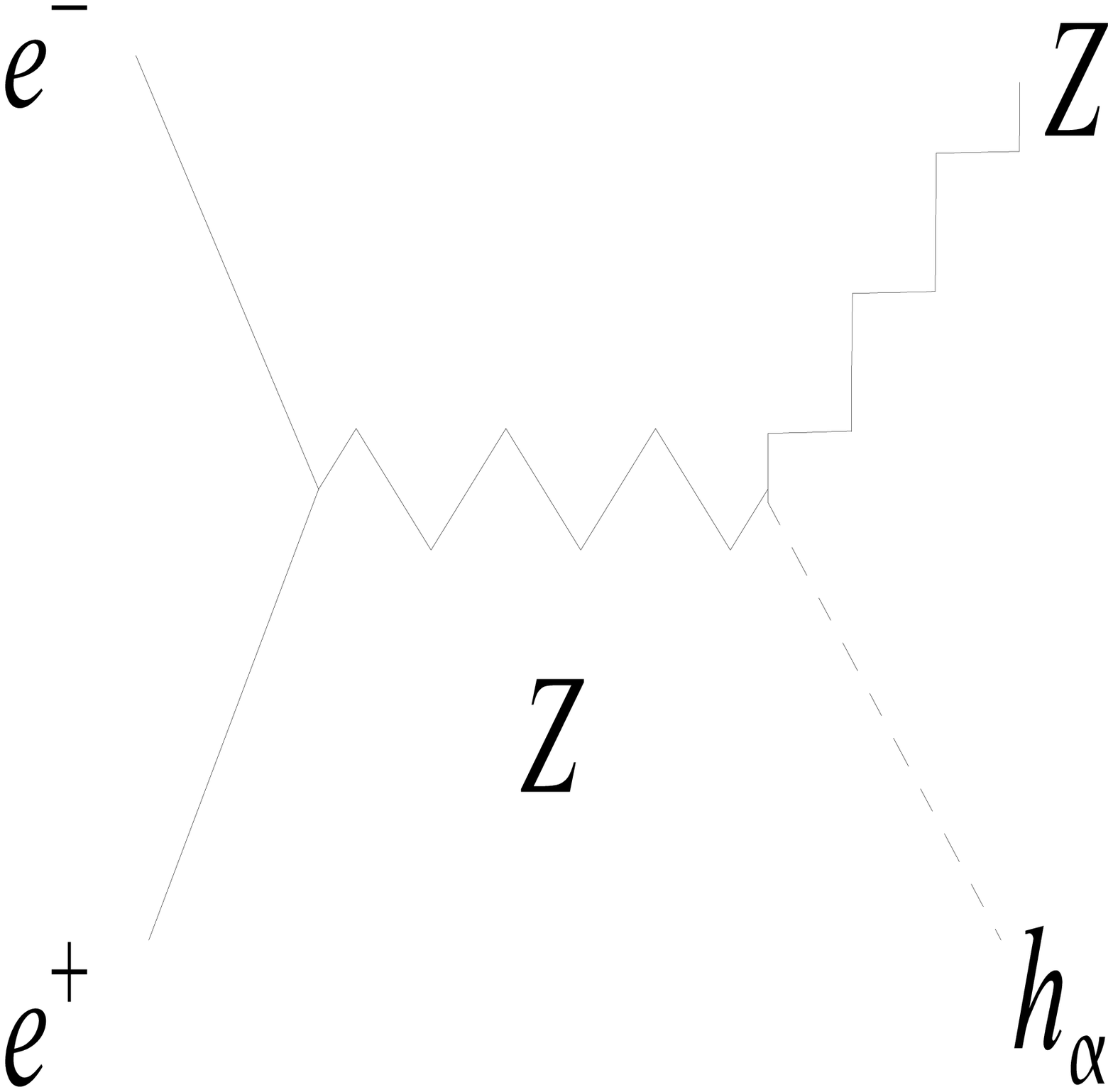,height=5.0cm,angle=-0}
      \\ & \\
%        \hspace*{-1.1cm} (a) & \hspace*{-1cm} (b)
    \end{tabular}
\captions{
% (a) Pair production of two scalar Higgses.
%We can observe the 
%proportionality between them. 
%(b) 
Higgs-strahlung. The index convention is like in Fig.~1.}
    \label{fig: PairProduction2 and HiggsStrahlung0}
 \end{center}
\end{figure}

In the following we will discuss the LEP constraints, especially the ones coming from the Higgs-strahlung process shown in 
Fig. \ref{fig: PairProduction2 and HiggsStrahlung0}. In the previous subsection we have discussed Higgs-to-Higgs decays in the $\mu \nu$SSM (see Eq.(\ref{decayeq})). Thus a CP-even Higgs originated through a Higgs-strahlung could decay in that way. 
%To start the discussion let us analyse the constraints considering the case of the CP-even scalars.

\begin{figure}[t]
  \begin{center} 
\hspace*{-10mm}
%    \begin{tabular}{cc}
	\epsfig{file=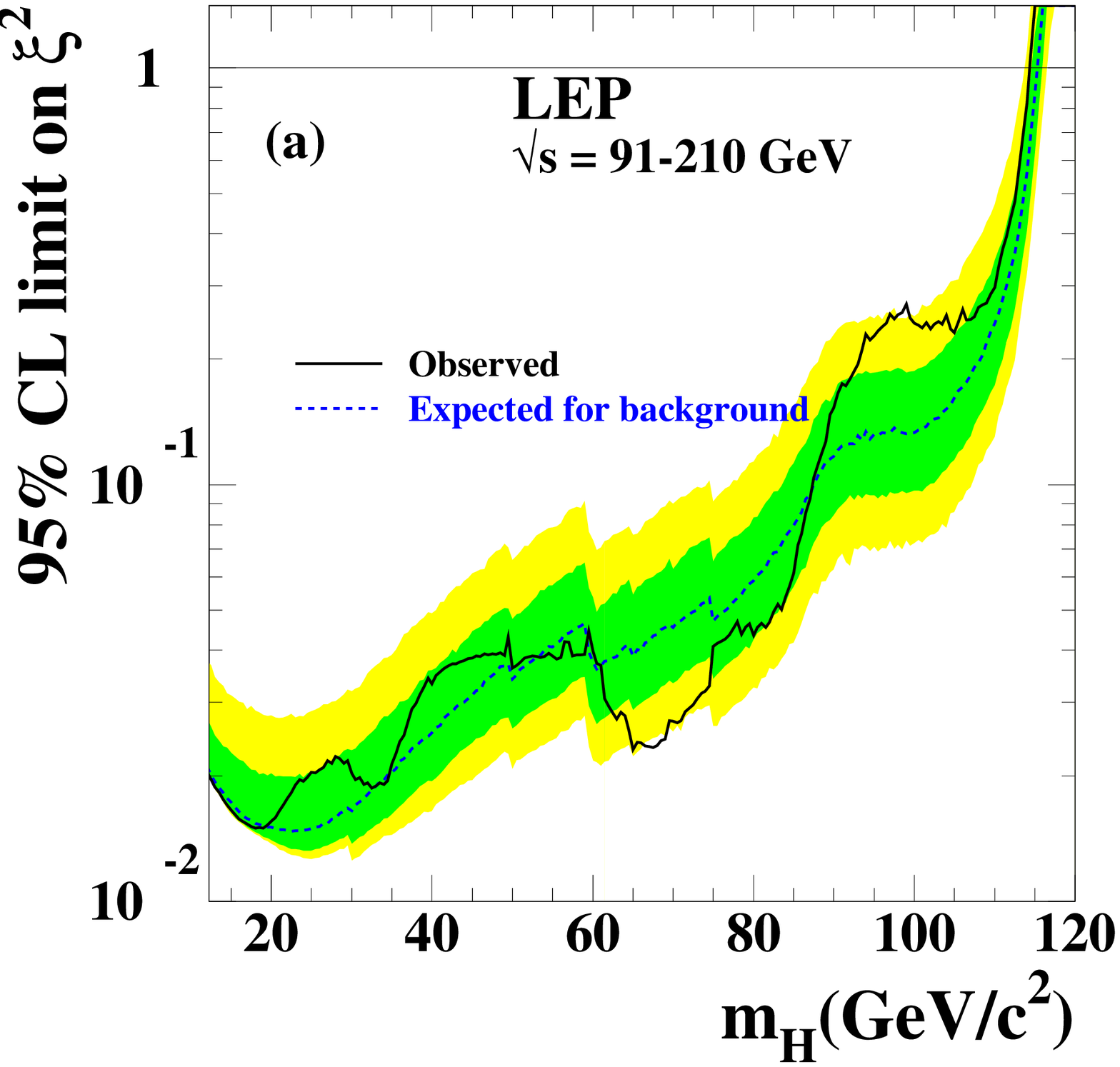,width=60mm,angle=0,clip=}

% & \hspace*{-10mm} \epsfig{file=figures/DibujoDiagTreeSToSS.ps,width=90mm,angle=0,clip=}
%    \end{tabular}
\captions{
The 95\% confidence level upper bound on the ratio 
$\xi^2=(g_{hZZ}/g^{SM}_{hZZ})^2$ from \cite{LEPH}. The dark and light shaded bands around the median expected line correspond to the 68\% and 95\% probability bands. 
The horizontal line corresponds to the Standard Model coupling for Higgs boson decays predicted by the Standard Model.}
%Limits on:
%(a) $\xi^2 Br(H\to b \bar{b})$  (b) $\xi^2 Br(H\to \tau^+ \tau^-)$. 
    \label{fig:1byc}
  \end{center}
\end{figure}

Let us remember that LEP data can be used to set lower bounds on the lightest Higgs boson mass in non-standard models, as shown in Fig.~\ref{fig:1byc} from \cite{LEPH}.
In the ratio $\xi^2=(g_{hZZ}/g^{SM}_{hZZ})^2$, $g_{hZZ}$ designates the non-standard 
$hZZ$ coupling and $g^{SM}_{hZZ}$ the same coupling in the Standard Model.
Whereas in Fig.~\ref{fig:1byc}, the Higgs boson is assumed to decay into fermions and bosons according to the Standard Model, when $BR(h\to b\bar{b})$ differs from the Standard Model one, the parameter in Fig.~\ref{fig:1byc}, $\xi^2$, must be replaced by $\xi^2\;BR(h\to b\bar{b})/BR_{SM}(h \rightarrow b \bar{b})$.
% , in parts $(b)$ and $(c)$ it is assumed that the Higgs boson decays exclusively into $b \bar{b}$
% or $\tau^+ \tau^-$ pairs.

For the $\mu \nu$SSM for each Higgs we can define the couplings $\xi_\alpha$, with $\alpha=1,...,8$, given by
\bea
 \xi_\alpha=[ v_u S(u,\alpha)+ v_d S(d,\alpha)+ \nu_i S(L_i,\alpha)]/ v\ ,
\label{coupling}
\eea
where $S(u,\alpha), \; S(d,\alpha), \;  S(L_i,\alpha)$ are the fraction composition of up-type Higgs doublet, down-type Higgs doublet and left-handed sneutrinos of the $h_\alpha$ neutral scalar mass eigenstate. A sum over $i=1,2,3$ is assumed in the last term, and $v^2=v^2_u+v^2_d+ \nu_i \nu_i $.

If more than one Higgs with mass below $114$~GeV are present but they are degenerated, we could define  $\xi^2=\xi_\alpha \xi_\alpha $, where the sum is over all Higgses below 114~GeV, and still use Fig. \ref{fig:1byc} for $\xi^2\; BR(h\to b\bar{b})/BR_{\text{SM}}(h\to b\bar{b})$.

% {\bf Very light Higgs,  with sneutrino left-handed like composition in such a way to scape LEP constraints.} Similar to the NMSSM,  but in this case with the possibility of three insted of only one sneutrino right-handed like Higgs since the sneutrino right-handed.

Also with more than one Higgs below $114$~GeV with arbitrary masses, for each Higgs these constraints can be used for the coupling $\xi_\alpha^2BR(h_\alpha\to b\bar{b})/BR_{\text{SM}}(h\to b\bar{b})$. Notice however that, given a value of $\xi_\alpha^2BR(h_\alpha\to b\bar{b})/BR_{\text{SM}}(h\to b\bar{b})$, the corresponding lower bound on the Higgs mass is a necessary but not sufficient condition to fulfil the LEP bounds.

Obviously, if the Higgs is mostly $\snuc$-like the coupling goes to zero, and we could have three very light Higgses 
%with $\xi_i \approx 0$ 
avoiding the LEP constraints. From the above discussion we can see that another way  to avoid them would be to make $BR(h \to b \bar{b})$  small.
% (the same for $Br(h \to \tau^+ \tau)$).

However, in the general case a more involved analysis is necessary, since for example more than $2b$ in the final state are possible. Let us remember that searches for $h\to \Phi \Phi$ and $\Phi \to b \bar{b}$ (where $\Phi$ is a CP-odd or CP-even Higgs) by  OPAL \cite{OPAL} and DELPHI \cite{DELPHI} impose a  strong constraint on the parameter space of the Standar Model. Once combined these analyses, one obtains $M_H > 110$~GeV for $\xi \sim 1$.
Nevertheless, in models with more scalars and pseudoscalars it is possible to obtain a larger number of  $b \bar{b}$, e.g. $ h_3 \to 2 h_2  \to 4 P_1 \to 4b 4\bar{b}$. 
%We should discuss this for the model.
It seems therefore that a re-analysis of the LEP data, to take into account this well motivated and complex phenomenology, would be interesting. Specially interesting would be to re-analyse the well-known $2.3\sigma$ excess in the $e^+ e^- \to Z + b\bar{b}$ channel in the LEP data around $100$~GeV. In the context of the NMSSM, the consistency of the excess with $h \rightarrow PP$ decays was discussed in \cite{Gunionexcess}.

Searches for $e^+ e^- \to hZ$ independent of the decay mode of the Higss by OPAL \cite{OPAL2}, could also be important to exclude some regions of the parameter space.

Searches for $h \to \Phi \Phi$ and $\Phi \to gg$, $\Phi \to c\bar{c}$, $\Phi \to \tau^+ \tau^-$ by OPAL \cite{OPAL3}, and  the recent analysis of the Higgs decaying into four taus carried out in \cite{exp4taus}, must also be taken into account. Nevertheless, the
$\mu \nu$SSM requires a more detailed analysis than the one available in the literature, since for instance a larger number of $\tau's$ in the final states is possible. % As we have already said, in this model scalar or pseudoscalar Higgses can decay to neutrinos at the tree-level, see Figs. \ref{fig: BolaCharginos and HiggsDecayToNeutrinos}b and \ref{fig: BolaCharginos and HiggsDecayToNeutrinos pseudo}b.  This could be important especially in the case of a very pure $\snuc$-like Higgs. In this case this decay could be in competition with the chargino loop  process, pseudoscalar to two photons, see Figs. \ref{fig: BolaCharginos and HiggsDecayToNeutrinos}a. and \ref{fig: BolaCharginos and HiggsDecayToNeutrinos pseudo}a.

It is also worth mentioning that an on-shell or off-shell Z could decay into neutralinos, with the three lightest neutralinos being very light and mainly composed by left-handed neutrinos. The decay of the neutralinos $\neutralino_a$ with $a=4,...,10$ was discussed in \cite{Ghosh:2008yh, Hirsch0}. Invisible Z width constraints \cite{zwidth} must be applied.

Let us finally discuss the sum rules. For the $\xi_\alpha$ defined in Eq. (\ref{coupling}), one can obtain the following sum rule:
\bea
\sum_{\alpha=1}^{8}\xi_\alpha^2 =1\ .
\label{sum1}
\eea
% Assuming for simplicity that the three $\snu$-like Higgses are the heaviest ones (it is straightforward to generalize the formula for the case when the $\snu$-like Higgses are in an arbitrary position), for the first five we have
% \bea
%  \xi_a \approx [ \sin\beta\ S(u,a)+ \cos\beta\ S(d,a)]\ ,
% \eea
% where $a=1, ..5$, and we have defined as usual $\tan\beta=\frac{v_u}{v_d}$ since
% the $\nu_i$ are very small as discussed above.
%$\sin(\beta)$ and $\cos(\beta)$
% Neglecting the left-handed sneutrino components the composition vectors are not orthonormals but not far from that and  we can write:
% \bea
% \sum_{a=1}^{5}\xi_a^2 \approx 1\ .
% \eea
Notice that for the three $\snu$-like Higgses the corresponding $\xi_\alpha$ can be neglected, and therefore one can write 
\bea
\sum_{\phi=1}^{5}\xi_\phi^2 \approx 1\ ,
\eea
where
\bea
 \xi_\phi \approx [ \sin\beta\ S(u,\phi)+ \cos\beta\ S(d,\phi)]\ ,
\eea
with $\tan\beta=\frac{v_u}{v_d}$ defined as usual, since
the $\nu_i$ are very small as discussed above.
%$\sin(\beta)$ and $\cos(\beta)$
% Neglecting the left-handed sneutrino components the composition vectors are not orthonormals but not far from that and  we can write:
% \bea
% \sum_{a=1}^{5}\xi_a^2 \approx 1\ .
% \eea

Also another important sum rule, in analogy with the one discussed in \cite{sumrule}, is  valid:
%More difficlt is to se how the sum rule
\bea
\sum^8_{\alpha=1}\xi^2_\alpha M^2_{h_\alpha} = M^2_{max}\ ,
\label{sum2}
\eea
% \Lnote{Esta \'ultima sum rule vale por analog\'ia con el NMSSM pero no la he demostrado .... Lo de abajo es correcto, lo que falta demostrar es si  $M^2_{max}$ es realmente la cota a la masa del Higgs en general. O sea a tree-level y despreciando los $Y_\nu$ y $\nu$ tenemos una formula, la de abajo, pero en general ?`  que es $M^2_{max}$ ?}
where, neglecting terms with $Y_\nu$ and $\nu$,  $M_{\max} $ is the upper bound on the lightest Higgs mass studied in \cite{MuNuSSM2}
% \bea
% M^2_{max}= M^2_z(cos^2{2 \beta} + \frac{\lambda_i \lambda_i}{g^2}\sin^2{2\beta}) + rad. \;  corrs.+ O(Y_\nu,\nu)
% \eea
\bea
M^2_{max}= M^2_Z\left( \cos^2{2 \beta} + \frac{2\lambda_i \lambda_i \cos^2\theta_W}{g_2^2}\sin^2{2\beta}\right) + rad. \;  corr.
\eea
Using
Eqs. (\ref{sum1}) and (\ref{sum2}) one can deduce, as in the case of the NMSSM \cite{Ham}, that
\bea
M^2_{h_2} \leq  \frac{1}{1 - \xi^2_1}(M^2_{\max}-\xi^2_1 M_{h_1}^2)\ ,
\eea
where $h_1$ and $h_2$ are the lightest and next-to-lightest Higgses.

Finally let us mention that a simple way to avoid current collider constraints is to make the new Higgses very heavy, in such a way that the constraints apply only to the first one, as we will see in benchmark point 6 presented in Section \ref{numericalresults}. Then very interesting signals could be expected from the Higgs cascade decays in experiments like LHC.

%%%%%%%%%%%%

% \begin{figure}[h!]
%  \begin{center}
% \hspace*{-8mm}
%     \begin{tabular}{cc}
%       \epsfig{file=figures/BolaCharginoA.eps,height=2.5cm,angle=-0}
%         \hspace*{0mm}&\hspace*{-3mm}
% \epsfig{file=figures/HiggsDecayToNeutrinosA.eps,height=2.5cm,angle=-0}
%       \\ & \\
%        \hspace*{-1.1cm} (a) & \hspace*{-1cm} (b)
%     \end{tabular}
% \captions{
% (a) Di-photon CP-odd Higgs decay. 
% We can observe the 
% proportionality between them. 
% (b) Tree-level CP-odd Higgs decay to neutrinos.}
%     \label{fig: BolaCharginos and HiggsDecayToNeutrinos pseudo}
%  \end{center}
% \end{figure}

\section{Signals at colliders} \label{numericalresults}

In the previous section we have tried to provide a general overview of the decays of the Higgses of the $\mu \nu$SSM. In this section we will concentrate in those decays
that are genuine of this model, and could therefore serve to 
distinguish it from other SUSY models. For that, we will focus first our attention on the decays of a MSSM-like Higgs with a mass about $114$ GeV (for being detectable in the near future), and with a sizeable branching ratio into the two lightest neutralinos.
These neutralinos could decay inside the detector leading to displaced vertices. This fact can be used to distinguish the $\mu\nu$SSM from R-parity conserving models such as the NMSSM. 
For example, as mentioned in subsection \ref{Decays}, the lightest 
neutralino\footnote{In our convention, when we refer to 'neutralino', we are excluding the three light left-handed neutrinos  $\tilde \chi_{1,2,3}^0$.
} 
$\tilde \chi_{4}^0$'s 
%(and $\tilde \chi_{5}^0$'s in the cases where they are degenerated in mass with the $\tilde \chi_{4}^0$'s).
can decay to an on-shell light singlet pseudoscalar (that subsequently decays to $b \bar b$) and a neutrino, and therefore the decay $h_{MSSM} \to \tilde \chi_4^0 \tilde \chi_4^0 \to 2 
P 
%P_{\tilde \nu^c} 
2\nu \to 2b 2 \bar b  2\nu$ is genuine of the $\mu \nu$SSM.
In other R-parity breaking models such as the BRpV, there are no singlet Higgses and a lightest neutralino lighter than gauge bosons could decay only through three-body decay processes. However, we have to point out that since the final decay products could be the same in both models, they may be difficult to distinguish experimentally.
% We will also see that the final state 8 $b$-jets plus missing energy is possible in situations where singlet-like scalars are produced by the decay of the neutralino, and they decay to pseudoscalars.

We will also discuss an example where the Higg-to-Higgs cascade decays studied in 
subsection \ref{Decays}
are relevant to distinguish the $\mu\nu$SSM from other SUSY models. 
% In particular, the $h_{\text{MSSM}}$ can decay into two CP-even Higgses, and they decay to pseudoscalars leading to the final state 8 $b$-jets.

Following the above strategy, in this section we will present a sample of numerical examples of viable benchmark points of interest for LHC searches.
The study of the heavier doublet-like Higgs, where the cascades described in subsection \ref{Decays} could also be relevant, is left for a future work.

% \textbf{In this work we will focus our attention in the decays of the SM like Higgs with a mass closed to $114 \ GeV$ and therefore it could be discovered soon in collider experiments. Notice that the study of the heavier doublet-like Higgs is left for future works and that the cascades discribed in subsection \ref{Decays} could be relevant.}

Let us mention that for the computation we have used a spectrum generator for the $\mu\nu$SSM (see \cite{MuNuSSM2} for a description\footnote{In this version we have included one-loop corrections to neutrino masses (in general to neutralinos). These corrections have been computed in \cite{Ghosh2010}.}), linked with modified subroutines for the model, based on the codes NMHdecay \cite{NMHdecay} and Spheno \cite{Spheno}. In particular, the modified subroutines based on the code NMHdecay are used to compute the two-body decays of all Higgses present in the $\mu \nu$SSM. We have also built a subroutine to compute the two-body decays of neutralinos. The modified subroutines based on the code Spheno are used to compute the three-body decays of neutralinos.

We have searched for points of the parameter space that are safe from exclusion by current collider constraints but that could be detected in the near future at LHC. Nevertheless, a full analysis of these points in the light of LEP and TEVATRON is beyond the scope of this paper and then it is not possible to totally guarantee that all of them satisfy all experimental constraints. In any case, if any of the benchmark points provided here is not completely safe from experimental constraints, it would be in the border and with small variations in the values of the parameters could be driven to the allowed experimental region. 
% and for this reason 
%In this respect, we would like to suggest a reanalysis of the LEP data for the case of the $\mu \nu$SSM.
% a variation of this Benchmark points is neccessary to satisfy all the collider constraints. 

Below we give a list with all the constraints that we are imposing on the points analyzed. Some of them have already been discussed in Subsection
\ref{Coupling with $Z$ bosons and summation rules.}.
%  Most of this constrait can not be really applied in the case of R-parity violating model, unless the live time of the LSP (defined in the limit where R-parity is conserve, $Y_\nu=0$ ) is very long and is escaping detection. But imposing this we are making a conservative approach trying to search benchmark points as safe as possible from collider constraints.

% {\bf Charged sector}
% 
% 1) We check that the Chargino is not to light imposing $m_{80}$~GeV. \cite{?}
% 
% 2) We check that the charged Higgses are not to light as also the charged  sleptons. Since in the model they mix in an unic mass matrix we have impose to sleptons and charged higgses the one that is more constraint. \cite{}

First, all points are true minima of the neutral scalar potential. We have checked that tachyons do not appear and that the couplings fulfil Landau pole constraints at the GUT scale. 

We have verified that all points satisfy the following $3\sigma$ neutrino sector constraints \cite{NeutrinoConstraints}, 
%(\textbf{see \cite{NeutrinoConstraintsValle} for the last update}):
\begin{align}\label{Synopsis}
7.14<\Delta m_{sol}^2/10^{-5}\mathrm{\ eV}^2<8.19 \ , \ 2.06<\Delta m_{atm}^2/10^{-3}\mathrm{\ eV}^2<2.81 \nonumber\\
 0.263<\sin^2\theta_{12}<0.375 \ , \ \sin^2\theta_{13}<0.046 \ , \  0.331<\sin^2\theta_{23}<0.644 
%\nonumber
\end{align}
% \begin{table}[t]
%=========================================================================%\begin{table}[t]
%\centering
%\resizebox{\textwidth}{!}{
%\begin{ruledtabular}
% \centering
% \begin{tabular}{|c|c|c|c|c|}
%\hline%---------------------------------------------------------------------
%$\Delta m_{sol}^2/10^{-5}\mathrm{\ eV}^2$ & $\sin^2\theta_{12}$ &
%            $\sin^2\theta_{13}$ & $\sin^2\theta_{23}$ &
%            $\Delta m_{atm}^2/10^{-3}\mathrm{\ eV}^2$ \\[5pt]
%\hline%---------------------------------------------------------------------
%7.14-8.19 & 0.263-0.375  & $<0.046$        &
%0.331-0.644  & 2.06-2.81 \\
%\hline%---------------------------------------------------------------------
%\end{tabular}
%\caption{\label{Synopsis} Allowed $3 \sigma$ ranges for the
%neutrino masses and mixings .}
%as discussed in \cite{ConstraintsFogli}
%\end{ruledtabular}
%}%end of resizebox
%\vspace*{.2cm}
%\end{table}
%We have also guaranteed that, being conservative, current limits on sparticle masses with R-parity conserved are satisfied, excluding points with charginos, squarks, gluinos and charged Higgs/sleptons too light \cite{LEPConstraintsSUSYMasses,TevatronConstraintsSUSYMasses}.
%%as well as tests on neutralinos (Z width into neutralinos). OJO CON LAS 2 citas anteriores !!!!!!

We have guaranteed that current limits on sparticle masses with R-parity conserved are satisfied, excluding points with charged Higgs/sleptons, charginos, squarks and gluinos too light \cite{LEPConstraintsSUSYMasses,TevatronConstraintsSUSYMasses}. We are being conservative, since strictly speaking these limits apply only to R-parity conserving models. %OJO CON LAS 2 citas anteriores !!!!!!

%We have also take into account constraints coming from processes $\tilde t \ \rightarrow \ b \ l \ \tilde \nu$, $\tilde t \ \rightarrow \ \tilde \chi^0 \ c$, $\tilde b \ \rightarrow \ \tilde \chi^0 \ b$ \textbf{CITAR????}.

% In principle, in expcon.f also the constraints from CLEO and from BABAR are taken into account.
% Should we put it in the draft?

In the neutral Higgs sector we have checked the constraints on the reduced couplings $\times$ branching ratios in terms of the masses, for all the CP-even and CP-odd scalars, in the following channels analysed at LEP:

1) For $e^+ e^- \to h Z$ with the following decays of h,
\begin{itemize}
\item $h \to \text{invisible}$ \cite{LEPHaleph, LEPHinvisible}. Here we are assuming as invisible the light neutrinos. 
%and the long-lived particles that are decaying clearly outside the detector (that is, the extremely pure CP-even or CP odd Higgses).  
A more elaborated analysis requires a re-analysis of LEP data, taking into account for instance that neutralinos could partially contribute to the missing energy when the decay distance is comparable to the size of the detector. We have checked that in the points where the decay length of the lightest neutralino is considerably greater than $\mathcal O(1 \ \text{m})$, considering also the LSP as invisible, the constraint is satisfied.
% However we have also checked that including the LSP the constraint is satisfy.
\item $h \to \gamma \gamma$, from LEP Higgs working group results  \cite{gamma}.
\item $h \to b\bar{b}$, from the LEP Higgs working group \cite{LEPH}.
\item $h$ to two jets, from OPAL and the LEP Higgs working group, both at LEP2 \cite{OPAL0312, LEP2}.
\item $h \to \tau^+ \tau^-$, from the LEP Higgs working group \cite{LEPH}.
\item $h \to PP$ with  $PP$ decaying to  4 jets, 2 jets + cc, 2 jets + $\tau^+ \tau^-$, 4 $\tau' s$, cccc, $\tau \tau + cc$, from OPAL results \cite{OPAL3}.
% 4 $b$, 4 $\tau$, 2 $b$ 2 $\tau$ , light pairs,
%

\end{itemize}
 
2) For $e^+ e^- \to h P$   with $hP$ decaying into $4 \, b$, $4 \, \tau$,
% $ 2 \, b \, 2 \tau$, , \ 6\tau$
   and $PPP \to 6b $ studied by DELPHI \cite{DELPHI}.

3) For $e^+ e^- \to h Z \to PPZ \to 4b + 2jets$  the DELPHI constraints  \cite{DELPHI}.

4) For $e^+ e^- \to h Z$ independent of $h$ decay mode, combining the results of ALEPH and OPAL collaborations \cite{LEPHaleph, LEPH}.

\vspace{0.5cm}

%We have chosen all our input parameters at EW scale and
On the other hand, as discussed in detail in \cite{MuNuSSM2}, using the eight minimization conditions for the neutral scalar potential we have solved the soft masses $m_{H_u}$, $m_{H_d}$, $m_{\tilde{L_i}}$ and $m_{\tilde{\nu^c_i}}$ in terms of $\tan\beta$, $\nu^c_i$, $\nu_i$, and we have used the fact that $\nu_i$ are very small in order to define $\tan\beta\approx \frac{v_u}{v_d}$ and  $v^2 \approx v_u^2 + v_d^2$, as usual.
% and used insted of  $v_u$, $v_d$.
%in favour of $v_u$, $v_d$, $\nu^c_i$, $\nu_i$. Also, using the fact that $\nu_i$ are very small  insted of use $v_u$ and $v_d$ we define $\tan(\beta)\approx \frac{v_u}{v_d}$ and  $v=v_u^2 + v_d^2$ and use this parametrs as inputs.
For simplicity, to perform the numerical analysis we have assumed a diagonal structure of the parameters in flavour space. 
We have also assumed universality for most of the parameters. In the case of the neutrino parameters this is not possible, since we need at least two generations with different $Y_{\nu_i}$ and $\nu_i$ in order to guarantee the correct hierarchy of neutrino masses. Besides, an exact universality of the other parameters would produce degenerations in the spectrum.
Since we are working with low-energy parameters, the presence of exact universality after the running from higher scales seems to be extremely unlikely. To avoid this artificial situation, but still maintaining the simplicity of using universal parameters in the computation, we have slightly broken the universality in the diagonal entries of the $\kappa$ tensor.
% except the ones for neutrinos. 
On the other hand, in the case of the trilinear terms we take all of them proportional to the corresponding Yukawa couplings.
% for example $(A_\kappa \kappa)_{ijk}=\kappa_{ijk}A_\kappa$.
%In addition we have assumed that all the soft masses are diaginal.

To summarize, the independent low-energy free parameters  that we are varying in our analysis are,
% \footnote{The only difference in the set of parameters  respect to the ones used in \cite{MuNuSSM2} is that in this paper we choose $\nu_1 \neq \nu_2=\nu_3$ instead of $\nu_1=\nu_2 \neq \nu_3$. Since this makes easier to find solutions with masses and mixing angles compatible with experiments, as was explained in \cite{MuNuSSM SCPV}.},
 \be
\lambda_i=\lambda, \; \; \tan\beta, \; \; \kappa_{iii}, \; \; \nu^c_i=\nu^c, \; \; \nu_1, \; \; \nu_2=\nu_3, \; \; Y_{\nu_1},  \; \; Y_{\nu_2}=Y_{\nu_3}, \;  \; A_{\lambda}, \; \; A_{\kappa}, \; \; M_2,
\label{parameters} \ee
where for $M_1$ and $M_3$ we are assuming a relation that mimics the one coming from unification at the GUT scale, $M_1=\frac{\alpha_1^2}{\alpha^2_2} \, M_2$, $M_3=\frac{\alpha_3^2}{\alpha_2^2 } \, M_2$, impliying $M_1 \approx 0.5 M_2$, $M_3=2.7 \, M_2$.
%in addition with a diagonal and degenerate form of the soft masses and the trilinear terms associated with the charged sector  $m_{\tilde{u}}$, $m_{\tilde{d}}$, $m_{\tilde{e}}$,  $A_E$, $A_T$, $A_D$.
%
% %m_{\tilde{Q}}, \; m_{\tilde{u}}  \; m_{\tilde{d}} \; m_{\tilde{e}}., 
% 
% % Also in  order to simplify our anaylisis we have imposed  the following unifications,
% % 
% % \bea &\lambda_i=\lambda, \; \; \kappa_{iii}=\kappa, \; \; \nu^c_i=\nu^c, \; \; \nu_1, \; \; \nu_2=\nu_3, \; \; Y_{\nu_1},  \; \; Y_{\nu_2}=Y_{\nu_3}, \;  \; \\ \noindent& A_{\lambda_i}=A_{\lambda}, \; \; A_{\kappa_{iii}}=A_{\kappa}, \; \; A_{\nu_{iii}}=A_{\nu}.  
% \eea
% 
% % $\lambda_i=\lambda$, $\kappa_{iii}=k$, $\nu^c_i\nu^c$, $\nu_1$, $\nu_2=\nu_3$, $Y_{\nu_1}$, $Y_{\nu_2}=Y_{\nu_3}$, $A_{\lambda_i}=A_{\lambda}$, $A_{\kappa_{iii}}=A_{\kappa}$, $A_{\nu_{iii}}=A_{\nu}$.
In addition we have fixed the following soft parameters as, $m_{\tilde{Q}}~=~1000\gev$,  $m_{\tilde{u}}=1000\gev$, $m_{\tilde{d}}=1000\gev$, $m_{\tilde{e}}=1000\gev$, $A_e=1000\gev$, $A_u=2400\gev$, $A_d=1000\gev$, $A_{\nu}=-1000\gev$. Let us remark, nevertheless, that we have varied the value of $A_u$ for certain points, since it is relevant for the 1-loop corrections to the mass of the Standard Model Higgs.

% \be
%  \lambda_i=\lambda  \; \kappa_iii=k \; \nu^c_i\nu^c \; \nu_1 \; \nu_2=\nu_3 \; Y_{\nu_1} \; Y_{\nu_2}=Y_{\nu_3} \;  A_{\lambda_i}=A_{\lambda}\; A_{\kappa_{iii}}=A_{\kappa} \; A_{\nu_{iii}}=A_{\nu}
% \ee

% 

For the values of the parameters that we will use in the benchmark points below, it is possible to 
show \cite{MuNuSSM2} using 
Appendix \ref{Appendix Mass matrices} that the mixing between the Higgses and the right-handed sneutrinos is
of the order of $a_{\lambda_i} v_u=A_\lambda \lambda v_u$, and therefore small compared with the relevant diagonal terms $\lambda_i \lambda_j \nu_i^c \nu_j^c=9 \lambda^2 (\nu^c)^2$.
Thus the Higgs doublets are basically decoupled from the right-handed sneutrinos.
%Because of the universality assumption, the masses of the latter are essentially degenerated up to 
%small contributions due to the non-universal values of $Y_\nu$ and $\nu$.
Note also that the right-handed neutrino masses are given by a value that can be approximated as $2 \kappa_{iii} \nu^c$ \cite{MuNuSSM2}.

% \textbf{As was obtained in \cite{MuNuSSM2}, using Appendix \ref{Appendix Mass matrices} one can check that in regions of the parameter space where the off-diagonal mixing terms of the neutral scalar mass matrices are smaller than the diagonal terms, Higgs doublets and right-handed sneutrinos are almost decoupled. For universal parameters as in our case, the mixing is of the order $A_\lambda \lambda v_u$ and it is smaller in our points to the relevant diagonal term $\lambda_i \lambda_j \nu_i^c \nu_j^c=9 \lambda^2 (\nu^c)^2$. In addition, the fact of assuming universality for most of the parameters implies that the masses of the quite pure singlets are essentially degenerated up to small contributions due to the non-universal values of $Y_\nu$ and $\nu$.
% Also in \cite{MuNuSSM2} was obtained that the right-handed neutrino masses are degenerated, with a value that can be approximated as $2 \kappa \nu^c$.}

%

Taking all the above into account, let us discuss now eight interesting benchmark points for collider physics. 
%It is also worth noticing that in all the cases that we will describe, once produced, the heavier (but almost degenerated) mainly right-handed neutrino neutralinos, are decaying in cascades into the lightest one (mainly right-handed neutrino or Higgsino/gaugino-like), plus neutrinos, through three-body decays with very short decay lengths.
For the first three points that we will consider, the lightest neutralino $\neutralino_4$ is mainly a right-handed neutrino, since 
we take the value of $2 \kappa_{iii} \nu^c$ small compared to the 
soft gaugino mass $M_2$ and Higgsino masses $\mu=\lambda_i \nu_i^c$.
% \textbf{and with the effective $\mu$ term that accounts for Higgsino masses of order $\mu_{eff} > 200 \ \text{GeV}$}.
This composition of the LSP is genuine of the $\mu\nu$SSM and hence, very interesting to study. 
%It is worth noticing here that due to the universality assumption, $\tilde \chi_5^0$ is basically degenerated in mass with $\tilde \chi_4^0$ and has a similar behaviour.
%and is also right-handed neutrino-like with a very pure composition and behaves as the LSP, similarly to $\tilde \chi_4^0$. 
The other right-handed neutrino-like neutralinos $\tilde \chi_{5,6}^0$ are slightly heavier than $\neutralino_{4}$, and once produced in the decay of a Higgs, they decay rapidly to
$\neutralino_{4}$ 
% $\tilde \chi_4^0$ and $\tilde \chi_5^0$ 
through 3-body processes such as $\tilde \chi_{5,6}^0 \to \tilde \chi_{4}^0 q \bar q$ or $\tilde \chi_{5,6}^0 \to \tilde \chi_{4}^0 l \bar l$.
%The other right-handed neutrino-like neutralinos $\neutralino_{5,6}$ are almost degenerated with the lightest one and, once produced, they decay mainly to it through three-body decay processes such as $\neutralino_5 \to \neutralino_4 \nu\nu$ (with an off-shell Z or an on-shell/off-shell Higgs),
%with very short lifetimes ({\bf ?????????????????????????????????}).

On the other hand, for benchmark points 4, 5 and 6,  
the lightest neutralino $\neutralino_4$ is MSSM-like.
For example, taking small enough values for $M_2$ one can have a MSSM lightest neutralino almost bino-like.
The right-handed neutrino-like neutralinos $\neutralino_{5,6,7}$ 
also decay through three-body processes to the lightest one and quarks/leptons very promptly.

Thus, additional quarks or leptons are present in the cascades due to the decays of the right-handed neutrino-like neutralinos into the lightest one.

Finally, in benchmark points 7 and 8 we work again with the lightest neutralino as a right-handed neutrino, although for benchmark point 7 it does not play an important role in the Higgs cascades and only Higgs-to-Higgs cascade decays are relevant.

Let us also remark that for all the eight benchmark points, 
$A_\kappa$ is chosen small for having light pseudoscalars, since its contribution 
is the dominant one in the diagonal element of the mass matrix.
In this way the neutralino can decay into a light pseudoscalar and a neutrino through two-body processes.
%and thus the model could easily be distinguished from the BRpV, as discussed above.
Since we have light singlets, we are also choosing for simplicity small values of $\tan \beta$ in order to be able to fulfill LEP constraints more easily. 
%  since only two $b's$ are expected from the decay of the pseudoscalar. 
%On the contrary, the lightest neutralino in the BRpV model can only decay through three-body decay processes when its mass is smaller than the gauge bosons mass. 

Benchmark point 1 is presented in Table \ref{Bench1}. 
There we only show the relevant masses and branching ratios for our discussion. The masses of the heavier doublet-like Higgs and left-handed sneutrinos (both scalars or pseudoscalars) are larger than the ones shown, and we do not study the decays of such Higgses. Neither the heavier MSSM-like neutralinos $\neutralino_{7,8,9,10}$ play any role on our discussion.
In this benchmark point
a doublet-like Higgs  with mass $m_{h_4} = 118.8$ GeV can decay into two neutralinos with masses $\mneutralino \approx 34-42 \gev$, and with a branching ratio of 4\%.
% leading after the cascades to the lightest one, to two lightest neutralinos and 
%neutrinos.
The lightest neutralino can decay through a two-body decay process to a scalar/pseudoscalar and a neutrino. Note that the branching ratios of the decays of neutralinos are referred only to two-body processes, while the decay lengths shown in the tables take into account two- and three-body processes.
The decay into
%\textbf{through a two-body decay process} 
a pseudoscalar $P_{1,2,3}$ and a neutrino takes place in 67\% of the cases.
%\textbf{as can be obtained with the help of Fig. \ref{fig:fermions}c}. 
These pseudoscalars are mainly decaying into $b \bar{b}$ and a displaced vertex could be detected since the decay length of the lightest neutralino is 23  cm. 
Besides the cascade $h_{4} \to \tilde \chi^0 \tilde \chi^0 \to 2 P
%P_{\tilde \nu^c} 
2\nu \to 2b2\bar b  2\nu$, the lightest neutralino could also decay to a CP-even singlet and a neutrino in 33\% of the cases, with the CP-even Higgs decaying into two pseudoscalars. Then, the following cascade is also relevant: $h_{4} \to \tilde \chi^0 \tilde \chi^0 \to 2 h 2\nu \to 4P2\nu \to 4b4\bar b 2\nu$, leading to 8 $b$-jets plus missing energy with a displaced vertex.
%The lightest neutralinos could also decay to long-lived pseudoscalars ({\bf AND NEUTRINOS}
 
%  in 17\% of the cases.
%\textbf{as can be obtained with the help of Fig. \ref{fig:fermions}c}. 
% These pseudoescalars are mainly decaying into $b \bar{b}$ and a displaced vertex could be detected since the decay length of the lightest neutralino is 45 cm. The lightest neutralinos could also decay to long-lived pseudoscalars ({\bf AND NEUTRINOS}), directly or through the decay into one scalar (going to long-lived pseudoescalars ({\bf Y QUE MAS?})) and a neutrino, giving rise to missing energy since the lifetime of these pseudoescalars is $ \mathcal O (10^ 4 \ s)$ and they decay mainly to neutrinos outside the detector.
% ({\bf POR QUE LOS BR A P1 Y P2 SON TAN DISTINTOS?})

% \textbf{Let us remark that in all the tables presented we are only showing the masses and branching ratios relevant for our discussion. The masses of the heavier Higgs doublet-like and of the left-handed sneutrinos, both scalars or pseudoscalars, are heavier than the ones we show and we do not study there the decays of such Higgses. The heavier MSSM-like neutralinos $\neutralino_{8,9,10}$ are neither playing any role on our discussion.}

Benchmark point 2 is given in Table \ref{Bench2}. In this case the decay of the Standard Model Higgs with a mass $m_{h_4} = 116.2$ GeV into neutralinos is enhanced to 12\%, since neutralino masses are smaller than in benchmark point 1 due to the smaller value of $2 \kappa_{iii} \nu^c$. Besides, the decay of the lightest neutralino into CP-even Higgses is kinematically forbidden. Notice also that in this case the decay of the pseudoscalars into two $b$'s is kinematically forbidden and then they decay into $\tau^+ \tau^-$. 
%The lightest pseudoscalar $P_1$ is the one that decays to $\tau^+ \tau^-$ and the lightest neutralino is decaying to it in a 55\% of the cases.
Summarizing, the following cascade leading to a displaced vertex takes place: $h_{4} \to \tilde \chi^0 \tilde \chi^0 \to 2 P 2\nu \to 2 \tau^+ 2 \tau^- 2\nu$.
The decay length of the neutralino $\tilde \chi_4^0$  is 1.89  m.
%and we have checked that the invisible Higgs constraint is satisfied even if we consider the (three) lightest neutralino(s) as invisible.
% since the heaviest ones decay rapidly to the lightest one).
% ({\bf PERO NO SE VE NADA DE LOS DECAYS DE LOS PSEUDOSCALARES PORQUE EL NEUTRALINO SE VA DEL DETECTOR???}).

% Benchmark point 2 is given in Table \ref{Bench2}. In this case the decay of the Standard Model Higgs with a mass $m_{h_4} = 112.2$ GeV to neutralinos is enhanced to 12\% \textbf{since the right-handed neutrino masses are smaller due to a smaller value of $2 \kappa \nu^c$ (??????)}. Besides, the decay of the lightest neutralino to CP-even Higgses is kinematically forbidden and the decay to a pseudoscalar $P_1$, that is mainly decaying to two Standard Model particles, is favored with a branching ratio of 55\% . Notice that in this case the decay of the pseudoescalar to two b's is kinematically forbidden and then it decays to $\tau^+ \tau^-$. 
% The lightest pseudoscalar $P_1$ is the one that decays to $\tau^+ \tau^-$ and the lightest neutralino is decaying to it in a 55\% of the cases. 
% The decay length of the neutralino is ({\bf AND NEUTRINOS}), and we have checked that the invisible Higgs constraint is satisfied considering also the lightest neutralino as invisible (to be precise, the three right-handed neutrinos since the heaviest ones decay rapidly to the lightest one ({\bf ???????})).
%  ({\bf PERO NO SE VE NADA DE LOS DECAYS DE LOS PSEUDOSCALARES PORQUE EL NEUTRALINO SE VA DEL DETECTOR???}).

Benchmark point 3 is given in Table \ref{Bench3}. A doublet-like Higgs  with mass $m_{h_4} = 116.6 \gev$ can decay into two neutralinos with masses $\mneutralino \approx 47-50 \gev$ in 0.5\% of the cases,
with the interesting cascade $h_{4} \to \tilde \chi^0 \tilde \chi^0 \to 2 P 2\nu \to 2 b 2 \bar b 2\nu$.
%The decay length of the lightest neutralino is 
%$\mathcal O (14 \ cm)$. The lightest neutralino decay 60\% of the times to a pseudoscalar with mass $m_{P_3} \approx 17 \gev$ plus a neutrino, with the pseudoscalar decaying 93 \% of the cases to two b's. 
% Benchmark point 3 is given in Table \ref{Bench3}. A doublet-like Higgs  with mass $m_{h_4} = 116.6 \gev$ can decay to two neutralinos with masses $\mneutralino \approx 48 \gev$ in 0.8\% of the cases. 
The decay length of the lightest neutralino $\tilde \chi_4^0$  is 
12 cm. 
% The lightest neutralino decay 60\% of the times to a pseudoscalar with mass $m_{P_3} \approx 17 \gev$ plus a neutrino, with the pseudoscalar decaying 93 \% of the cases to two b's. 

Let us finally remark that, as expected, we have observed that increasing the mass of the lightest neutralino, its decay length is reduced. On the other hand, reducing the mass of the light pseudoscalars a few \gev,  the decay into two $b$'s can be kinematically forbidden, producing a dominant decay to leptons. Also it is possible to decrease the mass of the Higgs to values about 100 \gev, and then have a Higgs scenario in the line of the work \cite{Gunionexcess}, escaping the large fine-tuning and little hierarchy problems.
% One simple way to do this is to change the value of $A_u$  that essentially controls the one loop contribution to the Higgs mass in such a way to have a lighter Higgs as we will explain for benchmark point 5. 
%({\bf DE QUE VA ESTO??})
We would also like to point out that, as was shown in \cite{MuNuSSM2}, modifying the value of $\lambda$, it is possible to increase the mass of the MSSM-like Higgs up to about $140 \ \text{GeV}$.

The input parameters of the benchmark point 4, presented in Table \ref{Bench4}, are similar to those of the benchmark point 3, except for the fact that we are
%({\bf ??????}),
%but in this case, 
decreasing the soft gaugino mass $M_2$, and therefore generating a MSSM-like lightest neutralino (almost bino-like). Thus the production through the Standard Model-like Higgs decay is increased to 42\%. 
Notice that while in the previous benchmark points only three neutralinos $\neutralino_{4,5,6}$ have masses below half of the mass of the Standard Model Higgs $h_4$, here four neutralinos  $\neutralino_{4,5,6,7}$ fulfill that condition.
The lightest neutralino has a decay length of 1.65 m and decays into a pseudoscalar $P_{1,2,3}$ and a neutrino, with the pseudoscalar decaying 93\% of the cases into two $b$'s. In this case, 
the production of $b$'s described through the cascade decays of the Standard Model Higgs, leading to displaced vertices, $h_{4} \to \tilde \chi^0 \tilde \chi^0 \to 2 P 2\nu \to 2 b 2 \bar b 2\nu$, is very enhanced and competes with a similar branching ratio for the direct decay of the Standard Model Higgs 
into two $b$'s.

% Benchmark point 4, presented in Table \ref{Bench4}, is similar to benchmark point 3 
% ({\bf ??????}),
% but in this case the lightest neutralino is MSSM-like as mentioned above, and therefore the production through the Standard Model-like Higgs decay is increased to 42\%. 
% Notice that while in the previous benchmark points only three neutralinos $\neutralino_{4,5,6}$ have masses below half of the mass of the Standard Model Higgs $h_4$, here four neutralinos  $\neutralino_{4,5,6,7}$ fulfill that condition.
% The lightest neutralino has a decay length of $\mathcal O (160 \ cm)$ and decays $87$\% of the cases to a pseudoescalar $P_3$ and a neutrino, with the pseudoescalar decaying 93\% of the cases to two b's. In this case, the production of b's described through the cascade decays of the Standard Model Higgs, leading to displaced vertices, is very enhanced and competes with a similar branching ratio for the direct decay of the Standard Model Higgs to two b's. 

Benchmark point 5 is given in Table \ref{Bench5}. It is very similar to benchmark point 4, but reducing the trilinear soft term $A_u$, that is important for the 1-loop corrections to the mass of the Higgs, we can decrease the Standard Model Higgs mass 
to $m_{h_4} \sim 112.8 \ \text{GeV}$.
%, below $114 \ GeV$. 
LEP constraints are still satisfied since the branching ratio of $h_4$ into two $b$'s is dramatically reduced in favour of the branching ratio to neutralinos. 
We have checked that in this case, the process $h_4 \to  \neutralino \neutralino \to 2P2\nu \to 2b2\bar b  2\nu$ satisfies the 4$b$'s LEP constraint. We have also checked that the invisible Higgs constraint is satisfied even if we consider the lightest neutralino as invisible. Nevertheless, a more involved analysis of LEP data would be necessary regarding this point, to take into account the missing energy carried by the neutrinos.
% that would decay outside the LEP detector. 

Benchmark point 6 is presented in Table \ref{Bench6}. In this case, the spectrum is heavier, with all CP-even singlet scalars above $114 \ GeV$, and with $h_1$ being the Standard Model Higgs. The pseudoscalars are also considerably heavier than in the other benchmark points. 
This case is similar to the usual ones of the MSSM. 
The small difference comes from the fact that the Standard Model Higgs would decay in a significant ratio of 2\% to neutralinos leading to displaced vertices. The lightest neutralino, MSSM-like, will have two-body decays kinematically forbidden and will decay only through three-body processes with a decay length of
5.33 m.
%Even considering the lightest neutralino as invisible, the invisible Higgs constraint is satisfied.
In Table \ref{Bench6} we show the branching ratios to the following decay products (with a notation neglecting the mixings): $\nu l l$, $l q \bar q$, $\nu q \bar q$, $3 \nu$.

Benchmark point 7 is presented in Table \ref{Bench7}. In this case, the universality assumption has been broken also for the $\lambda_i$ parameters in order to favour the decay of $h_4$ to two singlet-like scalars $h_1$. Now the neutralino does not play an important role in the cascade decays of the Higgs, since the branching ratio of $h_4$ into two neutralinos is very suppressed.
This is due to the fact that the only kinematically-allowed decay of Higgs to neutralinos is $h_4 \to \tilde \chi_4^0 \chi_4^0$, and $\tilde \chi_4^0$ is a quite pure right-handed neutrino-like.
As a consequence, displaced vertices are not expected for this benchmark point. The MSSM-like Higgs with a mass $m_{h_4}=119.6$ GeV will have the typical decay of the MSSM into $b \bar b$ or the typical cascades of the NMSSM, 
$h_{4} \to 2P \to 2b 2\bar b$, in most of the cases.
%The novelty that the extended Higgs sector of the $\mu \nu$SSM introduces in this benchmark point is the decay
Besides, the decay of the Higgs $h_{4}$ into two CP-even singlet-like Higgses with a branching ratio of 4\% is also possible.
Thus the following cascade is relevant
$h_{4} \to 2h_1 \to 4P \to 4b 4 \bar b$. 
These cascades serve to distinguish the $\mu \nu$SSM from other R-parity violating models.
Besides, once a SUSY particle is produced at the collider, decaying into the LSP, the displaced vertex could allow to distinguish the $\mu \nu$SSM from the NMSSM.
%This signal is also genuine of the $\mu \nu$SSM, no other SUSY model could lead to such kind of signals (except the (M+2)SSM, whose phenomenology has not been analyzed in the literature) and could also serve to distinguish it from other R-parity violating models.

Finally, let us discuss benchmark point 8 shown in 
Table \ref{Bench8}, where we work again with a right-handed neutrino-like lightest neutralino. The main feature of this case is that, whereas for the singlet-like pseudoscalars $P_{1,2}$ the decay into $b \bar b$ is kinematically forbidden, for $P_3$ it is allowed. Then, several cascade decays are expected. The MSSM-like Higgs, $h_4$, has a mass of $120.2$ GeV. Apart from the typical decay of the MSSM, $h_4 \to b \bar b$, it 
can also decay without leading to displaced vertices with the following relevant cascades: $h_4 \to 2 h_1 \to 4P_{1,2}\to 4\tau^+4\tau^-$, $h_4 \to 2P_3 \to 2b 2 \bar b$. This signal could be considered as genuine of the $\mu \nu$SSM. The MSSM-like Higgs can also decay into neutralinos in 6\% of the cases leading to the following relevant cascades, where displaced vertices and missing energy are expected: $h_4 \to \tilde \chi_4^0 \tilde \chi_4^0 \to 2  P_{1,2} 2\nu \to 2\tau^+2\tau^- 2\nu $, $h_4 \to \tilde \chi_4^0 \tilde \chi_4^0 \to 2 h_{1,2,3} 2\nu \to 4P_{1,2}2 \nu \to 4\tau^+4\tau^- 2 \nu$, $h_4 \to \tilde \chi_4^0 \tilde \chi_4^0 \to 2 P_{3} 2\nu \to 2 b 2\bar b 2  \nu$. This benchmark point shows how extremely characteristic signals could be expected in certain regions of the parameter space of the $\mu \nu$SSM.

%%%
\begin{table}[t] 
%\end{center}
\centering 
\resizebox{\textwidth}{!}{%
\begin{tabular}{|c|c | c  |c | c | c | }
\hline
\hline

$\lambda$  & $\kappa_{111}$ & $\kappa_{222}$ & $\kappa_{333}$ & $\akappa$ (GeV) & $M_2$ (GeV)  \\
\hline
$ 1.0 \times 10^{-1}$ &  $2.1 \times 10^{-2}$& $1.9 \times 10^{-2}$ & $1.7 \times 10^{-2}$ & -5.0  &  $-1.7 \times 10^{3}$ \\
\hline
\hline
$\tan \beta$ & $A_\lambda$ (GeV) &~~~~$\nu_1$ (GeV)~~~~ &~~~~$\nu_{2,3}$(GeV)~~~~& $Y_{\nu_1}$ & $Y_{\nu_{2,3}}$  \\
\hline
$3.9$ & $1.0 \times 10^{3}$ &  $2.61 \times 10^{-5}$  & $1.31 \times 10^{-4}$  & $5.56 \times 10^{-8}$ & $2.66 \times 10^{-7}$ \\
\hline
\hline
~~~~ $\nu^c$ (GeV) ~~~~&~~~~  $m_{h_1}$ (GeV) ~~~~ & ~~~~ $m_{h_{2}}$ (GeV) ~~~~ & ~~~~  $m_{h_{3}}$ (GeV) ~~~~ &~~~~  $m_{h_{4}}$ (GeV) ~~~~& $m_{P_1}$ (GeV) \\
\hline
$1.0 \times 10^3$ & 27.9  & 33.3 & 37.9 & 118.8 & 12.2\\
\hline
\hline
$m_{P_2}$ (GeV) ~~~~   &~~~~ $m_{P_3}$ (GeV) ~~~~ & $m_{\tilde \chi_4^0}$ (GeV) & $m_{\tilde \chi_5^0}$ (GeV) & $m_{\tilde \chi_6^0}$ (GeV)  &  --- \\
\hline
13.8 & 20.3 & 34.4 & 38.4 & 42.5 & ---  \\
\hline
\hline
$BR(h_4 \to \sum_{i,j=4}^6 \neutralino_i \neutralino_j)$ &  $BR(\neutralino_{4} \to \sum_{i=1}^3 P_i \nu)$  & $BR(\neutralino_{4} \to  h_1 \nu)$
&    $BR(h_{1} \to \sum_{i,j=1}^3 P_i P_j)$   & $BR(P_{1,2,3}) \to b \bar b  $ & $l_{\neutralino_4 \to}$ (cm)  \\
\hline
0.04  & 0.67 & 0.33 & 0.89 & 0.93 & 23
  \\ 

\hline
\hline

\end{tabular}}
%\end{center}

\caption[Values of interest.]{
Relevant input parameters, masses and branching ratios of benchmark point 1.
}
\label{Bench1}
\end{table}

%%%
\begin{table}[t] 
%\end{center}
\centering 
\resizebox{\textwidth}{!}{%
\begin{tabular}{|c|c | c  |c | c | c | }
\hline
\hline

$\lambda$  & $\kappa_{111}$ & $\kappa_{222}$ & $\kappa_{333}$ & $\akappa$ (GeV) & $M_2$ (GeV)  \\
\hline
$ 1.0 \times 10^{-1}$ &  $7.7 \times 10^{-3}$& $7.5 \times 10^{-3}$ & $7.3 \times 10^{-3}$ & -1.0  &  $-1.7 \times 10^{3}$ \\
\hline
\hline
$\tan \beta$ & $A_\lambda$ (GeV) &~~~~$\nu_1$ (GeV)~~~~ &~~~~$\nu_{2,3}$(GeV)~~~~& $Y_{\nu_1}$ & $Y_{\nu_{2,3}}$  \\
\hline
$3.7$ & $1.0 \times 10^{3}$ &  $2.92 \times 10^{-5}$  & $1.46 \times 10^{-4}$  & $2.70 \times 10^{-8}$ & $1.51 \times 10^{-7}$ \\
\hline
\hline
~~~~ $\nu^c$ (GeV) ~~~~&~~~~  $m_{h_1}$ (GeV) ~~~~ & ~~~~ $m_{h_{2}}$ (GeV) ~~~~ & ~~~~  $m_{h_{3}}$ (GeV) ~~~~ &~~~~  $m_{h_{4}}$ (GeV) ~~~~& $m_{P_1}$ (GeV) \\
\hline
$8.0 \times 10^2$ & 13.6  & 13.9 & 17.0 & 116.2 & 8.4\\
\hline
\hline
$m_{P_2}$ (GeV) ~~~~   &~~~~ $m_{P_3}$ (GeV) ~~~~ & $m_{\tilde \chi_4^0}$ (GeV) & $m_{\tilde \chi_5^0}$ (GeV) & $m_{\tilde \chi_6^0}$ (GeV)  &  --- \\
\hline
9.5 & 9.6 & 11.8 & 12.2 & 14.0 & ---  \\
\hline
\hline
$BR(h_4 \to \sum_{i,j=4}^6 \neutralino_i \neutralino_j)$ &  $BR(\neutralino_{4} \to \sum_{i=1}^3 P_i \nu)$  &    $BR(P_{1} \to \tau^+ \tau^-)$   & $BR(P_{2} \to \tau^+ \tau^-)$ & $BR(P_{3} \to \tau^+ \tau^-)$ & $l_{\neutralino_4 \to}$ (cm)  \\
\hline
0.12  & 1.0 & 0.89 & 0.83 & 0.82 & 189
  \\ 

\hline
\hline
\end{tabular}}
%\end{center}
\caption[Values of interest.]{
Relevant input parameters, masses and branching ratios of benchmark point 2.
}
\label{Bench2}
\end{table}

%%%
\begin{table}[t] 
%\end{center}
\centering 
\resizebox{\textwidth}{!}{%
\begin{tabular}{|c|c | c  |c | c | c | }
\hline
\hline

$\lambda$  & $\kappa_{111}$ & $\kappa_{222}$ & $\kappa_{333}$ & $\akappa$ (GeV) & $M_2$ (GeV)  \\
\hline
$ 1.0 \times 10^{-1}$ &  $3.1 \times 10^{-2}$& $3.0 \times 10^{-2}$ & $2.9 \times 10^{-2}$ & -1.0  &  $-1.7 \times 10^{3}$ \\
\hline
\hline
$\tan \beta$ & $A_\lambda$ (GeV) &~~~~$\nu_1$ (GeV)~~~~ &~~~~$\nu_{2,3}$(GeV)~~~~& $Y_{\nu_1}$ & $Y_{\nu_{2,3}}$  \\
\hline
$3.7$ & $1.0 \times 10^{3}$ &  $3.04 \times 10^{-5}$  & $1.18 \times 10^{-4}$  & $5.10 \times 10^{-8}$ & $2.95 \times 10^{-7}$ \\
\hline
\hline
~~~~ $\nu^c$ (GeV) ~~~~&~~~~  $m_{h_1}$ (GeV) ~~~~ & ~~~~ $m_{h_{2}}$ (GeV) ~~~~ & ~~~~  $m_{h_{3}}$ (GeV) ~~~~ &~~~~  $m_{h_{4}}$ (GeV) ~~~~& $m_{P_1}$ (GeV) \\
\hline
$8.0 \times 10^2$ & 46.0  & 47.9 & 49.5 & 116.6 & 14.6\\
\hline
\hline
$m_{P_2}$ (GeV) ~~~~   &~~~~ $m_{P_3}$ (GeV) ~~~~ & $m_{\tilde \chi_4^0}$ (GeV) & $m_{\tilde \chi_5^0}$ (GeV) & $m_{\tilde \chi_6^0}$ (GeV)  &  --- \\
\hline
14.8 & 16.6 & 46.7 & 48.4 & 50.3 & ---  \\
\hline
\hline
$BR(h_4 \to \sum_{i,j=4}^6 \neutralino_i \neutralino_j)$ &  $BR(\neutralino_{4} \to \sum_{i=1}^3 P_i \nu)$  &    $BR(P_{1,2,3} \to b \bar b)$    & $l_{\neutralino_4 \to}$ (cm) & --- & ---  \\
\hline
0.005  & 1.0 & 0.93 & 12 & --- & ---
  \\ 

\hline
\hline

\end{tabular}}
%\end{center}
\caption[Values of interest.]{
Relevant input parameters, masses and branching ratios of benchmark point 3.
}
\label{Bench3}
\end{table}

%%%
\begin{table}[t] 
%\end{center}
\centering 
\resizebox{\textwidth}{!}{%
\begin{tabular}{|c|c | c  |c | c | c | }
\hline
\hline

$\lambda$  & $\kappa_{111}$ & $\kappa_{222}$ & $\kappa_{333}$ & $\akappa$ (GeV) & $M_2$ (GeV)  \\
\hline
$ 1.0 \times 10^{-1}$ &  $3.6 \times 10^{-2}$& $3.5 \times 10^{-2}$ & $3.4 \times 10^{-2}$ & -1.0  &  $-1.0 \times 10^{2}$ \\
\hline
\hline
$\tan \beta$ & $A_\lambda$ (GeV) &~~~~$\nu_1$ (GeV)~~~~ &~~~~$\nu_{2,3}$(GeV)~~~~& $Y_{\nu_1}$ & $Y_{\nu_{2,3}}$  \\
\hline
$3.7$ & $1.0 \times 10^{3}$ &  $4.11 \times 10^{-6}$  & $1.59 \times 10^{-5}$  & $4.89 \times 10^{-8}$ & $3.27 \times 10^{-7}$ \\
\hline
\hline
~~~~ $\nu^c$ (GeV) ~~~~&~~~~  $m_{h_1}$ (GeV) ~~~~ & ~~~~ $m_{h_{2}}$ (GeV) ~~~~ & ~~~~  $m_{h_{3}}$ (GeV) ~~~~ &~~~~  $m_{h_{4}}$ (GeV) ~~~~& $m_{P_1}$ (GeV) \\
\hline
$8.0 \times 10^2$ & 53.7  & 55.7 & 57.4 & 119.7 & 15.5\\
\hline
\hline
$m_{P_2}$ (GeV) ~~~~   &~~~~ $m_{P_3}$ (GeV) ~~~~ & $m_{\tilde \chi_4^0}$ (GeV) & $m_{\tilde \chi_5^0}$ (GeV) & $m_{\tilde \chi_6^0}$ (GeV)  &  $m_{\tilde \chi_7^0}$ (GeV) \\
\hline
15.7 & 17.9 & 51.8 & 54.8 & 56.6 & 58.9  \\
\hline
\hline
$BR(h_4 \to \sum_{i,j=4}^7 \neutralino_i \neutralino_j)$ &  $BR(\neutralino_{4} \to \sum_{i=1}^3 P_i \nu)$  &    $BR(P_{1,2,3} \to b \bar b)$    & $l_{\neutralino_4 \to}$ (cm) & --- & ---  \\
\hline
0.42  & 1.0 & 0.93 & 165 & --- & ---
  \\ 

\hline
\hline

\end{tabular}}
%\end{center}
\caption[Values of interest.]{
Relevant input parameters, masses and branching ratios of benchmark point 4.
}
\label{Bench4}
\end{table}

%%%
\begin{table}[t] 
%\end{center}
\centering 
\resizebox{\textwidth}{!}{%
\begin{tabular}{|c|c | c  |c | c | c | }
\hline
\hline

$\lambda$  & $\kappa_{111}$ & $\kappa_{222}$ & $\kappa_{333}$ & $\akappa$ (GeV) & $M_2$ (GeV)  \\
\hline
$ 1.0 \times 10^{-1}$ &  $3.6 \times 10^{-2}$& $3.5 \times 10^{-2}$ & $3.4 \times 10^{-2}$ & -1.0  &  $-1.0 \times 10^{2}$ \\
\hline
\hline
$\tan \beta$ & $A_\lambda$ (GeV) &~~~~$\nu_1$ (GeV)~~~~ &~~~~$\nu_{2,3}$(GeV)~~~~& $Y_{\nu_1}$ & $Y_{\nu_{2,3}}$  \\
\hline
$3.7$ & $1.0 \times 10^{3}$ &  $4.11 \times 10^{-6}$  & $1.59 \times 10^{-5}$  & $4.89 \times 10^{-8}$ & $3.27 \times 10^{-7}$ \\
\hline
\hline
~~~~ $\nu^c$ (GeV) ~~~~& $A_u$ (GeV) &~~~~  $m_{h_1}$ (GeV) ~~~~ & ~~~~ $m_{h_{2}}$ (GeV) ~~~~ & ~~~~  $m_{h_{3}}$ (GeV) ~~~~ &~~~~  $m_{h_{4}}$ (GeV)  \\
\hline
$8.0 \times 10^2$ & $1.2 \times 10^3$ & 53.3  & 55.6 & 57.4 & 112.8 \\
\hline
\hline
$m_{P_1}$ (GeV) ~~~~   &~~~~ $m_{P_2}$ (GeV) ~~~~& $m_{P_3}$ (GeV) & $m_{\tilde \chi_4^0}$ (GeV) & $m_{\tilde \chi_5^0}$ (GeV) & $m_{\tilde \chi_6^0}$ (GeV)   \\
\hline
15.4 & 15.6 & 17.8 & 51.7 & 54.8 & 56.5 \\
\hline
\hline
$m_{\tilde \chi_7^0}$ (GeV) & $BR(h_4 \to \sum_{i,j=4}^7 \neutralino_i \neutralino_j)$ &  $BR(\neutralino_{4} \to \sum_{i=1}^3 P_i \nu)$  &    $BR(P_{1,2,3} \to b \bar b)$    & $l_{\neutralino_4 \to}$ (cm) & ---   \\
\hline
58.9 & 0.30  & 1.0 & 0.93 & 164 & --- 
  \\ 

\hline
\hline

\end{tabular}}
%\end{center}
\caption[Values of interest.]{
Relevant input parameters, masses and branching ratios of benchmark point 5.
}
\label{Bench5}
\end{table}

% \section{Comments and remarks} \label{comments}

% In this section we want first to make a brief comment about the gravitino as dark matter in this model and its role on collider physics. Second, we want to discuss certain special limits of the $\mu \nu$SSM. 

\vspace{0.5cm}

Let us finally discuss in more detail the detectability of these signals at the LHC.
For that we need to study first the production cross section of the Higgs in the context of the 
$\mu \nu$SSM. 
%We estimate that the dominant production process at the LHC for the benchmark points discussed above is gluon fusion. 
It is well known that gluon fusion and $b$-quark fusion are the two main production processes of a Higgs at the LHC in the context of SUSY. 
%shown in Fig.  \ref{fig: GluonFusion},. 
Gluon fusion dominates over $b$-quark fusion in our benchmark points, as can be shown using the relevant 
equations  \cite{Formulas Xs}:
\begin{eqnarray}
\sigma(gg \rightarrow h_4)=\sigma(gg \rightarrow H_{\text{SM}})\frac{\Gamma(h_4 \rightarrow gg)}{\Gamma(H_{\text{SM}}\rightarrow gg)} \simeq \sigma(gg \rightarrow H_{\text{SM}})\ ,
\label{Cross-section gluon fusion}
%\\
\end{eqnarray}
\begin{eqnarray}
\sigma(b \bar b \rightarrow h_4)=\sigma(b \bar b \rightarrow H_{\text{SM}}) \left( \frac{Y_{bbh_4}}{Y_{bbH_{\text{SM}}}} \right)^2=\sigma(b \bar b \rightarrow H_{\text{SM}}) \frac{S^2(d,4)}{\cos^2 \beta}\ .
\label{Cross-section b-quark fusion}
\end{eqnarray}
%since, comparing Eq. (\ref{Cross-section gluon fusion}) for gluon fusion and Eq. (\ref{Cross-section b-quark fusion}) for b-quark fusion \cite{Formulas Xs}, 
We can see that for the case of $b$-quark fusion, the production cross section is reduced compared to the one of the Standard Model because in our benchmark points the value 
of $\tan \beta$ is low, and the main component of the Higgs is $H_u^0$. However, the production cross section for gluon fusion is very similar to the one of the Standard Model.
%, since the decay into gluons is comparable to the Standard Model case.
Note that in all benchmark points studied, we were interested in the production of a doublet-like Higgs 
%very pure doublet-Higgs 
($h_4$ in our notation, except for the benchmark point 6 where it is the lightest Higgs and therefore is denoted as $h_1$). 
%Then, the production cross section in gluon fusion process is expected to be very similar to that of the Standard Model Higgs boson. This can be checked from Eq. (\ref{Cross-section gluon fusion}) where, as $h_4$ is very pure doublet-Higgs 
In addition, our gluinos and squarks are heavy, and as a consequence the decay width into gluons is very similar to the one of the Standard Model.

We have used the code HIGLU \cite{HIGLU} to compute explicitly 
the production cross section of a Standard Model Higgs and the decay widths into gluons for our benchmark points, finding that $0.75 \ \sigma(gg \rightarrow H_{\text{SM}}) \lesssim \sigma(gg \rightarrow h_4) \lesssim \sigma(gg \rightarrow H_{\text{SM}})$. For a center of mass energy of $7 \ \text{TeV}$ we find that $\sigma(gg \rightarrow H_{\text{SM}})$ is about $17-19.5 \ \text{pb}$ and, as a consequence, we obtain production cross sections of about $\sigma(gg \rightarrow h_4) \simeq 15-19 \ \text{pb}$. Then, in principle we expect that the LHC could detect the signals described in this paper except maybe for cascades with a very small branching ratio (see Table \ref{Xs}). For example, the cascade described above with the largest value of the product of the cross section multiplied by the branching ratio is the one of the benchmark point 4, $h_4 \rightarrow \tilde \chi_4^0 \tilde \chi_4^0 \rightarrow 2P2\nu \rightarrow 2b2\bar b 2\nu$, with a result of $5860 \ \text{fb}$. The cascade with the smallest value of this product is the one of the benchmark point 8, $h_4 \rightarrow \chi_4^0 \tilde \chi_4^0 \rightarrow 2P_3 2 \nu \rightarrow 2b2\bar b 2\nu$, with a result of $20 \ \text{fb}$. The study of the detectability of these signals at the LHC with an event generator is beyond the scope of this paper and is left for a future work.

\section{Gravitino and colliders} \label{gravitino}

As was mentioned in the Introduction, since $R$-parity is broken in the $\mu \nu$SSM, neutralinos or sneutrinos, with very short lifetimes, are no longer candidates for the dark matter of the Universe.
Nevertheless, if the gravitino $\Psi_{3/2}$ is the LSP, it was shown in 
\cite{gravitino} that it could be a good candidate for dark matter, with a lifetime much longer than the age of the Universe. There, it was also shown that because the gravitino decays producing a monochromatic photon, the indirect detection of gravitinos 
in the {\it Fermi} satellite  \cite{fermi} with a mass range between 0.1-10 GeV is possible. Larger masses are disfavored by current {\it Fermi}  measurements.

In this case of gravitino LSP, one should check whether or not the collider signals studied in the previous section, are altered. In particular, the neutralino partial decay length into gravitino and photon must be computed.
For this computation we can use 
% The gravitino, the superpartner of the graviton, present in any local version of supersymmetry, is a good candidate for dark matter since its coupling with matter (and its decay) is suppressed by the gravitational interaction \cite{yamaguchi}.
% As was shown in \cite{gravitino} the gravitino is not only a viable dark matter candidate but also its indirect detection is possible in this model. Regarding this, in order to be detected by the Fermi satellite \cite{fermi}, the mass of the gravitino must be not far from $\mathcal O(1 \ GeV) $.
the expression of the decay length  $\neutralino_4 \rightarrow \Psi_{3/2}\ \gamma$ \cite{Ibarra}.
One obtains:

\begin{eqnarray}
c \ \tau^{3/2}_{\tilde \chi_4^0} \sim 80 \ \text{km} \ \left(  \frac{m_{3/2}}{10 \ \text{keV}} \right)^2 \ \left( \frac{m_{\tilde \chi_4^0}}{50 \ \text{GeV}} \right)^{-5}.
\end{eqnarray}
We can easily see that in order to have a significant decay to gravitinos, the mass of the gravitino must be very low, less than $10 \ \kev$. That is, for gravitino masses larger than $10 \ \kev$, the decay width of neutralino into gravitino and photon is much smaller than the decay widths into Standard Model particles.
Thus the collider signals studied in the previous section are not altered.

% Below we will show that for such masses, the branching ratios of neutralinos to gravitinos are irrelevant. On the other hand we want to point out that very light gravitinos, of $\mathcal O( 10 \ \kev)$, can  contribute to collider physics.

Summarizing, we want to emphasize that in the $\mu \nu$SSM the gravitino could be a viable dark matter candidate, accessible to indirect detection experiments, and without altering the collider phenomenology described along this paper.

%%%
\begin{table}[t] 
%\end{center}
\centering 
\resizebox{\textwidth}{!}{%
\begin{tabular}{|c|c | c  |c | c | c | }
\hline
\hline

$\lambda$  & $\kappa_{111}$ & $\kappa_{222}$ & $\kappa_{333}$ & $\akappa$ (GeV) & $M_2$ (GeV)  \\
\hline
$ 1.12 \times 10^{-1}$ &  $7.12 \times 10^{-2}$& $7.11 \times 10^{-2}$ & $7.10 \times 10^{-2}$ & -18  &  $-1.0 \times 10^{2}$ \\
\hline
\hline
$\tan \beta$ & $A_\lambda$ (GeV) &~~~~$\nu_1$ (GeV)~~~~ &~~~~$\nu_{2,3}$(GeV)~~~~& $Y_{\nu_1}$ & $Y_{\nu_{2,3}}$  \\
\hline
$3.7$ & $1.0 \times 10^{3}$ &  $7.21 \times 10^{-7}$  & $1.04 \times 10^{-7}$  & $6.66 \times 10^{-8}$ & $4.53 \times 10^{-7}$ \\
\hline
\hline
~~~~ $\nu^c$ (GeV) ~~~~&~~~~  $m_{h_1}$ (GeV) ~~~~ & ~~~~ $m_{h_{2}}$ (GeV) ~~~~ & ~~~~  $m_{h_{3}}$ (GeV) ~~~~ &~~~~  $m_{h_{4}}$ (GeV) & $m_{P_1}$ (GeV) ~~~~ \\
\hline
$8.47 \times 10^2$ & 113.7  & 115.1 & 115.3 & 118.9 & 57.8\\
\hline
\hline
$m_{P_2}$ (GeV) ~~~~& $m_{P_3}$ (GeV) & $m_{\tilde \chi_4^0}$ (GeV) & $m_{\tilde \chi_5^0}$ (GeV) & $m_{\tilde \chi_6^0}$ (GeV) & $m_{\tilde \chi_7^0}$ (GeV)  \\
\hline
57.9 & 61.3 & 52.1 & 114.2 & 120.2 & 120.4  \\
\hline
\hline
$BR(h_1 \to \neutralino_4 \neutralino_4)$    & $BR(\tilde \chi_4^0 \to l q \bar q)$  & $BR(\tilde \chi_4^0 \to \nu l \bar l)$ & $BR(\tilde \chi_4^0 \to \nu q \bar q)$ & $BR(\tilde \chi_4^0 \to 3 \nu)$ & $l_{\neutralino_4 \to}$ (cm) \\
\hline
 0.02    & 0.52 & 0.28 & 0.15 & 0.05 & 533
  \\ 

\hline
\hline

\end{tabular}}
%\end{center}
\caption[Values of interest.]{
Relevant input parameters, masses and branching ratios of benchmark point 6.
}
\label{Bench6}
\end{table}

%%%
\begin{table}[t] 
%\end{center}
\centering 
\resizebox{\textwidth}{!}{%
\begin{tabular}{|c|c | c  |c | c | c | }
\hline
\hline

$\lambda_{1,2}$  & $\lambda_3$ & $\tan \beta$ & $A_\lambda$ (GeV) &$\akappa$ (GeV) & $M_2$ (GeV)  \\
\hline
$ 1.0 \times 10^{-2}$ &  $2.8 \times 10^{-1}$& $3.7$ & $1.0 \times 10^{3}$ & -1  &  $-5.88 \times 10^{3}$ \\
\hline
\hline
$\kappa_{111}$  & $\kappa_{222}$ & $\kappa_{333}$ & $Y_{\nu{_1}}$ & $Y_{\nu_2}$ &  $Y_{\nu_3}$ \\
\hline
$ 7.12 \times 10^{-2}$ &  $6.95 \times 10^{-2}$& $3.15 \times 10^{-2}$ & $8.58 \times 10^{-8}$ & $2.42 \times 10^{-7}$  &  $2.13 \times 10^{-6}$ \\
\hline
\hline
~~~~~~~~$\nu_1$ (GeV) &~~~~~~~~$\nu_2$ (GeV) &~~~~~~~~$\nu_3$ (GeV) &~~~~$\nu^c$ (GeV)~~~~& ~~~~~~~~& ---  \\
\hline
$1.19 \times 10^{-4}$ & $1.71 \times 10^{-4}$  & $4.72 \times 10^{-7}$   & $8.0 \times 10^{2}$  &  & --- \\
\hline
\hline
~~~~ $m_{h_1}$ (GeV) ~~~~&~~~~  $m_{h_2}$ (GeV) & $m_{h_3}$ (GeV) & ~~~~ $m_{h_{4}}$ (GeV) ~~~~ & ~~~~  $m_{P_{1,2}}$ (GeV) ~~~~ &~~~~  $m_{P_{3}}$ (GeV) ~~~~ \\
\hline
47.9 & 110.9  & 113.6 & 119.6 & 14.0 & 25.7\\
\hline
\hline
$m_{\neutralino_4}$ (GeV) ~~~~   &~~~~ $m_{\neutralino_5}$ (GeV) ~~~~ & $m_{\neutralino_{6,7}}$ (GeV) & --- &  --- &  --- \\
\hline
53.9 & 111.2 & 113.9 & --- & --- & ---  \\
\hline
\hline
$BR(h_4 \to h_1 h_1)$    & $BR(h_1 \to \sum_{i,j=1}^3 P_i P_j)$  & $BR(P_i \to b \bar b)$ & $BR(h_4 \to \sum_{i,j=1}^3 P_i P_j)$ & $BR(h_4 \to b \bar b)$ &  --- \\
\hline
 0.04    & 0.97 & 0.93 & 0.39 & 0.40 & ---
  \\ 

\hline
\hline

\end{tabular}}
%\end{center}
\caption[Values of interest.]{
Relevant input parameters, masses and branching ratios of benchmark point 7.
}
\label{Bench7}
\end{table}

%%%
\begin{table}[t] 
%\end{center}
\centering 
\resizebox{\textwidth}{!}{%
\begin{tabular}{|c|c | c  |c | c | c | }
\hline
\hline
$\lambda$  & $\kappa_{111}$ & $\kappa_{222}$ & $\kappa_{333}$ & $\akappa$ (GeV) & $M_2$ (GeV)  \\
\hline
$ 1.0 \times 10^{-1}$ &  $1.66 \times 10^{-2}$& $1.65 \times 10^{-2}$ & $1.64 \times 10^{-2}$ & -5.0  &  $-1.7 \times 10^{3}$ \\
\hline
\hline
$\tan \beta$ & $A_\lambda$ (GeV) &~~~~$\nu_1$ (GeV)~~~~ &~~~~$\nu_{2,3}$(GeV)~~~~& $Y_{\nu_1}$ & $Y_{\nu_{2,3}}$  \\
\hline
$4.9$ & $1.0 \times 10^{3}$ &  $5.84 \times 10^{-5}$  & $2.25 \times 10^{-4}$  & $1.25 \times 10^{-7}$ & $2.26 \times 10^{-7}$ \\
\hline
\hline
~~~~ $\nu^c$ (GeV) ~~~~&~~~~  $m_{h_1}$ (GeV) ~~~~ & ~~~~ $m_{h_{2}}$ (GeV) ~~~~ & ~~~~  $m_{h_{3}}$ (GeV) ~~~~ &~~~~  $m_{h_{4}}$ (GeV) ~~~~& $m_{P_1}$ (GeV) \\
\hline
$8.0 \times 10^2$ & 19.8  & 21.6 & 21.8 & 120.2 & 8.8\\
\hline
\hline
$m_{P_2}$ (GeV) ~~~~   &~~~~ $m_{P_3}$ (GeV) ~~~~ & $m_{\tilde \chi_4^0}$ (GeV) & $m_{\tilde \chi_5^0}$ (GeV) & $m_{\tilde \chi_6^0}$ (GeV)  &  --- \\
\hline
8.9 & 16.9 & 26.3 & 26.5 & 27.8 & ---  \\
\hline
\hline
$BR(h_4 \to h_1 h_1)$ &  $BR(h_4 \to P_3 P_3)$  &    $BR(h_4 \to \sum_{i,j=4}^6 \neutralino_i \neutralino_j)$ & $BR(h_4 \to b \bar b)$ & $BR(h_{1} \to \sum_{i,j=1}^2 P_i P_j)$  & $BR(h_{2,3} \to \sum_{i,j=1}^2 P_i P_j)$    \\
\hline
0.05  & 0.12 & 0.06 & 0.55 & 0.98 & 1.0   \\ 
\hline
\hline
$BR(P_{1,2} \to \tau^+ \tau^-)$ & $BR(P_{3} \to b \bar b)$ &  $BR(\tilde \chi_4^0 \to \sum_{i=1}^3 h_i \nu)$ & $BR(\tilde \chi_4^0 \to P_{1,2} \nu)$ & $BR(\tilde \chi_4^0 \to P_3 \nu)$    & $l_{\neutralino_4}$ (cm)  \\
\hline
0.88 & 0.93  & 0.51 & 0.33 & 0.16   & 15 \\ 
\hline
\hline

\end{tabular}}
%\end{center}
\caption[Values of interest.]{
Relevant input parameters, masses and branching ratios of benchmark point 8.
}
\label{Bench8}
\end{table}

%%%
\begin{table}[t] 
%\end{center}
\centering 
\resizebox{\textwidth}{!}{%
\begin{tabular}{|c|c | c  | }
\hline
\hline
\text{Benchmark point}  & \text{Cascade} & $\sigma(gg \rightarrow h_4) \times BR_{\text{cascade}} \ (\text{fb})$    \\
\hline
1 & $h_4 \rightarrow \tilde \chi_4^0 \tilde \chi_4^0 \rightarrow 2P 2\nu \rightarrow 2b2\bar b 2\nu$ & 270  \\
\hline
  & $h_4 \rightarrow \tilde \chi_4^0 \tilde \chi_4^0 \rightarrow 2h 2\nu \rightarrow 4P2\nu \rightarrow 4b4\bar b 2\nu$ & 44 \\
\hline
\hline
2 & $h_4 \rightarrow \tilde \chi_4^0 \tilde \chi_4^0 \rightarrow 2P 2\nu \rightarrow 2\tau^+2\tau^- 2\nu$ & 1620 \\
\hline
\hline
3 & $h_4 \rightarrow \tilde \chi_4^0 \tilde \chi_4^0 \rightarrow 2P 2\nu \rightarrow 2b2\bar b 2\nu$ &   70    \\
\hline
\hline
4 & $h_4 \rightarrow \tilde \chi_4^0 \tilde \chi_4^0 \rightarrow 2P 2\nu \rightarrow 2b2\bar b 2\nu$ & 5860  \\
\hline
\hline
5 & $h_4 \rightarrow \tilde \chi_4^0 \tilde \chi_4^0 \rightarrow 2P 2\nu \rightarrow 2b2\bar b 2\nu$  & 4870 \\
\hline
\hline
6  & $h_1 \rightarrow \tilde \chi_4^0 \tilde \chi_4^0 \rightarrow 2l 2q 2\bar q$ & 150  \\
\hline
   & $h_1 \rightarrow \tilde \chi_4^0 \tilde \chi_4^0 \rightarrow 2\nu 2l 2\bar l$ & 80 \\
\hline
 & $h_1 \rightarrow \tilde \chi_4^0 \tilde \chi_4^0 \rightarrow 2\nu 2q 2\bar q$ &  40  \\
\hline
 & $h_1 \rightarrow \tilde \chi_4^0 \tilde \chi_4^0 \rightarrow 6 \nu$ & 15 \\ 
\hline
\hline
7 & $h_4 \rightarrow 2 P \rightarrow 2b2\bar b$   & 5450        \\
\hline
  & $h_4 \rightarrow 2h_1 \rightarrow 4P \rightarrow 4b4\bar b$ & 460  \\ 
\hline
\hline
8 & $h_4 \rightarrow 2 P_3 \rightarrow 2b2\bar b$ & 1660    \\
\hline
 & $h_4 \rightarrow h_1h_1 \rightarrow 4 P_{1,2} \rightarrow 4\tau^+ 4\tau^-$  & 460  \\
\hline
 & $h_4 \rightarrow \tilde \chi_4^0 \tilde \chi_4^0 \rightarrow 2P_{1,2}2\nu \rightarrow 2\tau^+ 2\tau^- 2 \nu$ & 80 \\
\hline
 & $h_4 \rightarrow \tilde \chi_4^0 \tilde \chi_4^0 \rightarrow 2h 2\nu \rightarrow 4 P_{1,2} 2\nu \rightarrow 4 \tau^+ 4\tau^- 2\nu $ & 150 \\
\hline
 & $h_4 \rightarrow \tilde \chi_4^0 \tilde \chi_4^0 \rightarrow 2P_3 2\nu  \rightarrow 2 b 2\bar b 2\nu $ & 20 \\
\hline
\hline

\end{tabular}}
%\end{center}
\caption[Values of interest.]{
Production cross section multiplied by branching ratios of the cascades, for the benchmark points discussed in the text.}
\label{Xs}
\end{table}

\section{Conclusions}
\label{Conclusions}

In this work we have studied the Higgs sector of the $\mu \nu$SSM focusing our attention on collider physics. In certain regions of the parameter space, the phenomenology of the Higgs sector in this model is very rich and different from other SUSY models. On the one hand, the Higgs sector is extended due to the presence of left- and right-handed sneutrinos mixing with the MSSM Higgses.
% The singlets are three-times replicated and the breaking of R-parity gives rise to an small mixing of left-handed sneutrinos with the Higgses.
On the other hand the breaking of R-parity, could lead to signatures different from the usual missing energy.
%%In this work we have studied in detail the Higgs sector of the $\mu \nu$SSM focusing our attention on collider physics. The phenomenology of the Higgs sector of this model is very rich and different from other SUSY models. On one hand, the Higgs sector is extended in comparison to other models. The singlets are three-times replicated and the breaking of R-parity gives rise to an small mixing of left-handed sneutrinos with the Higgses. On the other hand the breaking of R-parity could lead to signatures different from the usual missing energy.

First, we have analized the mixings in the Higgs sector of the $\mu \nu$SSM.
Assuming three families of right-handed neutrino superfields, one obtains eight CP-even and seven CP-odd Higgses in the model.
Although the three left-handed sneutrinos are basically decoupled from the rest of the Higgses, the mixing between Higgs doublets and right-handed sneutrinos is not necessarily small.
In this work we have deduced general conditions to suppress the latter. 
This can be useful to obtain very light singlets avoiding collider constraints, but also to have a doublet-like Higgs as the lightest one being as heavy as possible.

Then, we have provided an overview of new decays in the Higgs sector with respect to other SUSY models with extra singlets like the NMSSM. Due to the extended Higgs sector, Higgs-to-Higgs cascade decays could be more complicated, as shown in subsection \ref{Decays}. In addition, the breaking of R-parity gives rise to new decays. 
%as for example, the decay of a neutral Higgs to two neutrinos at the tree-level that could be important in the case of very light pseudoescalars ({\bf ????????????}).
LEP constraints have also been discussed in the context of the $\mu \nu$SSM. For this, we have computed the couplings of the Higgses with Z bosons and the sum rules.
% Also the production mechanisms of Higgses at lepton and hadron colliders in this model have been briefly reviewed.

Finally, in Section \ref{numericalresults} we have concentrated on Higgs decays that are genuine of 
the $\mu \nu$SSM, and could serve to distinguish it from other SUSY models.
We have provided benchmark points that should pass current constraints and are interesting for LHC. 
In particular, we have focused first our attention on the decays of a MSSM-like light Higgs $h_{MSSM}$
%with mass closed to $114$ GeV (for being detectable in the near future), and 
with a sizeable branching ratio to two lightest neutralinos.
% \footnote{In our convention, when we refer to 'neutralino', we are excluding the three light left-handed neutrinos  $\tilde \chi_{1,2,3}^0$.
% } $\tilde \chi_4^0$'s.
These neutralinos could decay inside the detector leading to displaced vertices. This fact can be used to distinguish the $\mu\nu$SSM from $R$-parity conserving models such as the NMSSM/MSSM. Let us remark, however, that in models of gauge mediated SUSY breaking, where the gravitino is the LSP, a displaced vertex can also be obtained depending on the lifetime of the next-to-LSP. 
%(see \cite{GMSB} for a review).

Besides, 
%As mentioned in subsection \ref{Decays}, the lightest neutralino can 
the decays can be into a neutrino and an on-shell light singlet pseudoscalar $P$, that subsequently decays into $b \bar b$ (or if kinematically forbidden into $\tau^+ \tau^-$), and therefore the decay $h_{MSSM} \to \tilde \chi^0 \tilde \chi^0 \to 2 
P 
%P_{\tilde \nu^c} 
2\nu \to 2b2\bar b 2\nu$ is genuine of the $\mu \nu$SSM.
For example, in other R-parity breaking models such as the BRpV, there are no singlet Higgses and a lightest neutralino lighter than gauge bosons could decay only through three-body decay processes. However, as the final products of the cascades can be the same in both models, it may be difficult to distinguish them experimentally.
%, while in the $\mu \nu$SSM the lightest neutralino can decay through a two-body process to a pseudoscalar and a neutrino, as mentioned above.
We have also seen that a final state with 8 $b$-jets plus missing energy is possible in situations where singlet-like scalars are produced first by the decay of the neutralino, and they decay into pseudoscalars,
%as shown in (\ref{decayeq}).
$h_{MSSM} \to \tilde \chi^0 \tilde \chi^0 \to 2 h 2\nu \to 4P2\nu \to 4b4\bar b 2\nu$. 
% Note that cascade decays of a Higgs leading to 8 $b$-jets plus missing energy is a novel feature of the $\mu\nu$SSM.

We have also studied a case with an spectrum similar to the one of the MSSM, where
all CP-even singlet scalars are above  $114 \ GeV$, and the pseudoscalars are heavier than the neutralinos. Then, 
the $h_{MSSM}$ will decay in a significant ratio to neutralinos, and these will decay only through three-body processes leading to displaced vertices.

In another case the neutralino does not play an important role and only Higg-to-Higgs cascade decays are relevant. Although displaced vertices are not expected, the decays $h_{MSSM} \to 2P \to 2b 2\bar b$,
$h_{MSSM} \to 2 h \to 4P \to 4b 4\bar b$ are possible, allowing to distinguish the $\mu \nu$SSM from other R-parity violating models. Besides, once a SUSY particle is produced at the collider, decaying into the LSP, the displaced vertex would allow to distinguish the $\mu \nu$SSM from the NMSSM.

Finally, we have studied a case where 
for singlet-like pseudoscalars $P_{1,2}$ the decay into $b \bar b$ is kinematically forbidden, but for $P_3$ is allowed. Then, several interesting cascade decays are expected without leading to displaced vertices: $h_{MSSM} \to 2 h_1 \to 4P_{1,2}\to 4\tau^+4\tau^-$,
$h_{MSSM} \to 2P_3 \to 2b 2 \bar b$. This is a genuine feature of
the $\mu \nu$SSM. In addition, the following cascades are possible, with displaced vertices and missing energy: $h_{MSSM}\to \tilde \chi_4^0 \tilde \chi_4^0 \to 2  P_{1,2} 2\nu \to 2\tau^+2\tau^- 2\nu $, $h_{MSSM} \to \tilde \chi_4^0 \tilde \chi_4^0 \to 2 h_{1,2,3} 2\nu \to 4P_{1,2}2 \nu \to 4\tau^+4\tau^- 2 \nu$, $h_{MSSM} \to \tilde \chi_4^0 \tilde \chi_4^0 \to 2 P_{3} 2\nu \to 2 b 2\bar b 2  \nu$.

% \begin{figure}[h!]
%  \begin{center}
% \hspace*{-8mm}
%     \begin{tabular}{cc}
% \epsfig{file=figures/AssociatedProductionWithHeavyQuarks.eps,height=3.0cm,angle=-0}
%          \hspace*{0mm}&\hspace*{-3mm}
%  \epsfig{file=figures/AssociatedProductionWithHeavyQuarks2.eps,height=3.0cm,angle=-0}
%        \\ & \\
%         \hspace*{-1.1cm} (a) & \hspace*{-1cm} (b)
%     \end{tabular}
% \captions{
%(a) 
% Examples of associated production of a Higgs with heavy quarks where $\phi$ can be either a CP-even or a CP-odd Higgs.
% }
%      \label{fig: AssociatedProductionWithHeavyQuarks}
%  \end{center}
% \end{figure}

In conclusion, the above discussion gives us the idea that extremely characteristic cascades can be expected in certain regions of the parameter space of the $\mu \nu$SSM.

We have also emphasized that in the $\mu \nu$SSM the gravitino could be a viable dark matter candidate, accessible to indirect detection experiments, and without altering the collider phenomenology described along this paper. In particular, the 
branching ratio of neutralino to gravitino-photon turns out to be negligible.
%We have also emphasized that, being the gravitino a viable dark matter candidate in this model, and its %indirect detection being possible, it should not alter the collider phenomenology studied along this paper.

Let us finally remark that the collider phenomenology of the $\mu \nu$SSM is very rich and peculiar, as shown here using several benchmark points, and, as a consequence, we still need to carry out much work in the future to cover all interesting aspects of the model. 
For example, although we have discussed in Section \ref{numericalresults} is some detail the detectability of the signals at the LHC, computing the production cross section multiplied by the branching ratios for the different cascades, the analysis with an event generator is beyond the scope of this paper and is left for a future work.

\vspace{2cm}

%\pagebreak
\noindent {\bf Acknowledgments} 

D.E. L\'opez-Fogliani would like to thank Ulrich Ellwanger for extremely helpful conversations. 
The work of J. Fidalgo and C. Mu\~noz
was supported in part by MICINN under grants FPA2009-08958 and FPA2009-09017, 
by
the Comunidad de Madrid under grant HEPHACOS S2009/ESP-1473, and by the
European Union under the Marie Curie-ITN program PITN-GA-2009-237920.
The work of D.E. L\'opez-Fogliani was supported by the French ANR TAPDMS ANR-09-JCJC-0146. The work of R. Ruiz de Austri was supported by MICINN under the project PARSIFAL FPA2007-60323. 
The authors also acknowledge the support of the Consolider-Ingenio 2010 
Programme under grant MultiDark CSD2009-00064.

\noindent 
%  
% 
% The work of J. Fidalgo and C. Mu\~noz was
% supported in part by the MICINN under grants FPA2009-08958 and FPA2009-09017, by
% the Comunidad de Madrid under grant HEPHACOS S2009/ESP-1473, and by the
% European Union under the Marie Curie-ITN program PITN-GA-2009-237920. 
% D. E. L\'opez-Fogliani thanks CNRS for financial support. 
% The work of R. Ruiz de Austri was
% supported in part by MICINN under grant FPA2007-60323, and by the Generalitat Valenciana
% under grant PROMETEO/2008/069. 
% The authors also thank the support of the Spanish 
% MICINN's Consolider-Ingenio 2010 Programme under
% grants MultiDark CSD2009-00064 and PAU CSD2007-00060.

\vspace{0.5cm}

%\newpage
\newpage
%%%%%%%%%%%%%%%%%%%%%%%%%%%%%%%%%%%%%%%%%%%%%%%%%%%%%%%%%%%%%%%%%%%%%
\appendix
%%%%%%%%%%%%%%%%%%%%%%%%%%%%%%%%%%%%%%%%%%%%%%%%%%%%%%%%%%%%%%%%%%%%%%
%\newpage
\section{Neutral scalar mass matrices} \label{Appendix Mass matrices}
\subsection{CP-even neutral scalars}
\label{appendixA.1.1}

The quadratic potential includes
\begin{align}
V_{\text{quadratic}}= \frac{1}{2} \mathbf{h'}_{\alpha} M^2_{h_{\alpha\beta}}  \mathbf{h'}_{\beta} +... \ ,
\label{matrix1}
\end{align}
where  $\mathbf{h}'_\alpha=(h_d, h_u,(\widetilde{\nu}^c_i)^R, (\widetilde\nu_i)^R)$ is in the unrotated basis, and
% \begin{align}
% V_{\text{quadratic}}=(h_d,h_u,,(\widetilde{\nu}^c_1)^R,(\nu_1)^R,(\widetilde{\nu}^c_2)^R,(\nu_2)^R,(\widetilde{\nu}^c_3)^R,(\nu_3)^R)M^2_s \left( \begin{array}{c} h_d \\h_u\\ (\widetilde{\nu}^c_1)^R \\ (\nu_1)^R\\ (\widetilde{\nu}^c_2)^R\\ (\nu_2)^R\\ (\widetilde{\nu}^c_3)^R\\ (\nu_3)^R  \end{array}\right)+...
% \end{align}
below we give the expressions for the independent coefficients of $M^2_{h_{\alpha\beta}}$\\
% The independent $m_{hh}$ terms:
\begin{align}
M_{h_{d}h_{d}}^{2}=m_{H_d}^{2}+\frac{G^2}{4}\{3v_{d}^{2}-v_{u}^{2}+\nu_{i}\nu_{i}\}+\lambda_{i}\lambda_{j}\nu_i^c \nu_j^c 
+\lambda_{i}\lambda_{i}v_{u}^{2}\ ,
\end{align}
\begin{align}
M_{h_{u}h_{u}}^{2}=m_{H_u}^{2}+\frac{G^2}{4}(-v^2_{d}+3v_{u}^{2}-\nu_{i}\nu_{i})+ \lambda_{i}\lambda_{j} \nu_i^c \nu_j^c +\lambda_{i}\lambda_{i}v_{d}^2 \nonumber \\ 
-2Y_{\nu_{ij}} \lambda_{j}v_{d}\nu_{i}+ Y_{\nu_{ik}} Y_{\nu_{ij}}\nu_j^c \nu_k^c+Y_{\nu_{ik}}Y_{\nu_{jk}} \nu_{i}\nu_{j}\ ,
\end{align}
\begin{align}
M_{h_{d}h_{u}}^{2}=-a_{\lambda_{i}}\nu_i^c-\frac{G^2}{2}v_{d}v_{u}+2v_{d}v_{u}\lambda_{i}\lambda_{i}-(\lambda_{k}\kappa_{ijk} \nu_i^c \nu_j^c + 2Y_{\nu_{ij}}\lambda_{{j}}v_{u}\nu_{i})\ , 
\end{align}
% The independent $m_{h\nu^c}^{2}$ terms:
\begin{align}
M_{h_{d}(\widetilde{\nu}_i^c)^R }^{2}=-a_{\lambda_{i}}v_{u}+2\lambda_{i}\lambda_{j}v_{d}\nu_j^c -2\lambda_{k} \kappa_{ijk}v_{u}\nu_j^c-Y_{\nu_{ji}}\lambda_{k}\nu_{j}\nu_k^c -Y_{\nu_{jk}}\lambda_{i}\nu_{j}\nu_k^c\ ,
\label{Adr}
\end{align}
\begin{align}
M_{h_{u}(\widetilde{\nu}_i^c)^R }^{2}=-a_{\lambda_{i}}v_{d}+a_{\nu_{ji}}\nu_{j}+2\lambda_{i}\lambda_{j}v_{u}\nu^c_{j}-2\lambda_{k}\kappa_{ilk}v_{d}\nu^c_{l}+2Y_{\nu_{jk}}\kappa_{ilk}\nu_{j}\nu^c_{l}
+2Y_{\nu_{jk}}Y_{\nu_{ji}}v_{u}\nu^c_{k}\ ,
\label{Aur}
\end{align}
\begin{align}
M_{h_{d}(\widetilde{\nu}_i)^R}^{2}=\frac{1}{2} G^2v_{d}\nu_{i}-(Y_{\nu_{ij}}\lambda_{j}v_{u}^{2}+Y_{\nu_{ij}}\lambda_{k}\nu^c_{k}\nu^c_{j})\ ,
\label{Adl}
\end{align}
\begin{align}
M_{h_{u}(\widetilde{\nu}_i)^R}^{2}=a_{\nu_{ij}}\nu^c_{j}-\frac{G^2}{2}v_{u}\nu_{i}-2Y_{\nu_{ij}}\lambda_{j}v_{d}v_{u}+Y_{\nu_{ik}}\kappa_{ljk}\nu^c_{l}\nu^c_{j}+2Y_{\nu_{ij}}Y_{\nu_{kj}}v_{u}\nu_{k}\ ,
\label{Aul}
\end{align}
% The independent $m_{\widetilde{\nu}^{c}\widetilde{\nu}}^{2}$terms:
\begin{align}
M_{(\widetilde{\nu}_{i})^R (\widetilde{\nu}_{j})^R}^{2}=m_{\tilde{L}_{ij}}^{2}+\frac{G^2}{2}\nu_{i}\nu_{j}+\frac{1}{4}
G^2(\nu_{k}\nu_{k}+v_{d}^{2}-v_{u}^{2})\delta_{ij}+Y_{\nu_{ik}}Y_{\nu_{jk}}v^2_{u}+Y_{\nu_{ik}}Y_{\nu_{jl}}\nu^c_{k}{\nu}^{c}_l\ ,
\end{align}
\begin{align}
M_{(\widetilde{\nu}_{i})^R (\widetilde{\nu}^{c}_{j})^R}^{2}=a_{\nu_{ij}}v_{u}-Y_{\nu_{ij}}\lambda_{k} v_{d}{\nu}^{c}_k-Y_{\nu_{ik}}\lambda_{j}v_{d}{\nu}^{c}_k
+2Y_{\nu_{ik}}\kappa_{jlk}v_{u}{\nu}^{c}_l\nonumber\\
+Y_{\nu_{ij}}Y_{\nu_{kl}}\nu_{k}{\nu}^{c}_l
+Y_{\nu_{il}}Y_{\nu_{kj}}\nu_{k}{\nu}^{c}_l\ , 
\label{Alr}
\end{align}
\begin{align}
M_{(\widetilde{{\nu}}^{c}_{i})^R (\widetilde{{\nu}}^{c}_{j})^R}^{2}=m_{\widetilde{{\nu}}^{c}_{ij}}^{2}+2a_{\kappa_{ijk}}{\nu^c_{k}}-2\lambda_{k}\kappa_{ijk}v_{d}v_{u}+ 2\kappa_{ijk}\kappa_{lmk}\nu^{c}_l\nu^{c}_m+4\kappa_{ilk}\kappa_{jmk}\nu^{c}_l\nu^{c}_m\nonumber \\
+\lambda_{i}\lambda_{j}(v_{d}^{2}+v_{u}^{2})+2Y_{\nu_{lk}}\kappa_{ijk}v_{u}\nu_{l}-(Y_{\nu_{kj}}\lambda_{i}+Y_{\nu_{ki}}\lambda_{j})v_{d}\nu_{k}+Y_{\nu_{ki}}Y_{\nu_{kj}}v_{u}^{2}
+Y_{\nu_{ki}}Y_{\nu_{lj}}\nu_{k}\nu_{l}\ .
\label{evenrr}
\end{align}

Then the mass eigenvectors are
\bea
 \mathbf{h}_{\alpha}=R^h_{\alpha\beta} \mathbf{h'}_\beta\ ,
\eea
with the diagonal mass matrix 
\bea
(M^{\text{diag}}_{h_{\alpha \beta}})^2=R^h_{\alpha \gamma} M^2_{h_{\gamma \delta}} R^h_{\beta \delta}\ .
\eea

\subsection{CP-odd neutral scalars}
\label{A.1.2}

In the unrotated basis $\mathbf{P'}_{\alpha}=\left( P_d,P_u,(\widetilde{\nu}^c_i)^I,(\widetilde{\nu}_i)^I \right)$ we have

% \begin{align}
% V_{\text{quadratic}}=\left(P_d,P_u,(\widetilde{\nu}^c_1)^I,(\nu_1)^I,(\widetilde{\nu}^c_2)^I,(\nu_2)^I,(\widetilde{\nu}^c_3)^I,(\nu_3)^I \right)M^2_s \left( \begin{array}{c} h_d \\h_u\\ (\widetilde{\nu}^c_1)^I \\ (\nu_1)^I\\ (\widetilde{\nu}^c_2)^I\\ (\nu_2)^I\\ (\widetilde{\nu}^c_3)^I\\ (\nu_3)^I  \end{array}\right)+...
% \end{align}
\begin{align}
V_{\text{quadratic}}= \frac{1}{2} \mathbf{P'}_{\alpha} M^2_{P_{\alpha \beta}} \mathbf{P'}_{\beta}+...
\label{matrix2}
\end{align}
Below we give the expressions for the independent coefficients of $M^2_{P_{\alpha \beta}}$\\
% 
% Below we give the expresion for the independent cofficient\\
% The independent $m_{PP}$terms (For simplify the notation we don't
% writte the supraindex I in $\widetilde{\nu}_{i}^{I}$and $(\nu^c_{i})^{I}$):
\begin{align}
M_{P_{d}P_{d}}^{2}=m_{H_{d}}^{2}+\frac{G^2}{4}(v_{d}^{2}-v_{u}^{2}+\nu_{i}\nu_{i})+\lambda_{i}\lambda_{j}\nu^c_i \nu^c_{j}
+\lambda_{i}\lambda_{i}v_{u}^{2}\ ,
\end{align}
\begin{align}
M_{P_{u}P_{u}}^{2}=m_{H_{u}}^{2}+\frac{G^2}{4}(v_{u}^{2}-v_{d}^{2}-\nu_{i}\nu_{i})+\lambda_{i}\lambda_{j}\nu^c_i\nu^c_{j}+\lambda_{i}\lambda_{i}v_{d}^2\nonumber \\
-2Y_{\nu_{ij}}\lambda_{j}v_{d}\nu_{i}+Y_{\nu_{ik}}Y_{\nu_{ij}}\nu^c_k\nu^c_{j}+Y_{\nu_{ik}}Y_{\nu_{jk}}\nu_{i}\nu_{j}\ ,
\end{align}
\begin{align}
M_{P_{d}P_{u}}^{2}=a_{\lambda_{i}}\nu^c_{i}+\lambda_{k}\kappa_{ijk}\nu^c_{i}\nu^c_{j}\ ,
\end{align}
%\\
% The independent $m_{h\widetilde{\nu}^c}^{2}$ terms:
\begin{align}
M_{P_{d}(\widetilde{\nu}^c_{i})^{I}}^{2}=a_{\lambda_{i}}v_{u}-2\lambda_{k}\kappa_{ijk}v_{u}\nu^c_{j}-Y_{\nu_{ji}}\lambda_{k} \nu^c_{k}\nu_{j}+Y_{\nu_{jk}}\lambda_{i}\nu^c_{k}\nu_{j}\ ,
\end{align}
\begin{align}
M_{P_{d}(\widetilde{\nu}_{i})^{I}}^{2}=-Y_{\nu_{ij}}\lambda_{j}v_{u}^{2}-Y_{\nu_{ij}}\lambda_{k}\nu^c_{k}\nu^c_{j}\ ,
\end{align}
\begin{align}
M_{P_{u}(\widetilde{\nu}^c_{i})^{I}}^{2}=&a_{\lambda_{i}}v_{d}-a_{\nu_{ji}} \nu_{j}-2\lambda_{k}\kappa_{ilk}v_{d}\nu^c_{l}+2Y_{\nu_{jk}}\kappa_{ilk} \nu_{j}\nu^c_{l}\ ,
\end{align}
\begin{align}
M_{P_{u}(\widetilde{\nu}_{i})^{I}}^{2}=-a_{\nu_{ij}}\nu^c_{j}-Y_{\nu_{ik}}\kappa_{ljk}\nu^c_{l}\nu^c_{j}\ ,
\end{align}
% The independent $m_{\widetilde{\nu}^c\widetilde{\nu}}^{2}$terms :
\begin{align}
M_{(\widetilde{\nu}_{i})^{I}(\widetilde{\nu}_{j})^{I}}^{2}=m_{\widetilde{L}_{ij}}^{2}+\frac{1}{4}G^2(\nu_{k}\nu_{k}+v_{d}^{2}-v_{u}^{2})\delta_{ij}+Y_{\nu_{ik}}Y_{\nu_{jk}}v^2_{u}+Y_{\nu_{ik}}Y_{\nu_{jl}}\nu^c_{k}\nu^c_{l}\ ,
\end{align}
\begin{align}
M_{(\widetilde{\nu}_{i})^{I}(\widetilde{\nu}^c_{j})^{I}}^{2}=&-a_{\nu_{ij}}v_{u}-Y_{\nu_{ik}}\lambda_{j}v_{d}\nu^c_{k}-Y_{\nu_{ij}}Y_{\nu_{lk}}\nu_{l}\nu^c_{k}+Y_{\nu_{ik}}Y_{\nu_{lj}}\nu_{l}\nu^c_{k}+Y_{\nu_{ij}}\lambda_{k}v_{d}\nu^c_{k}+2Y_{\nu_{il}}\kappa_{jlk}v_{u}\nu^c_{k}\ ,
\end{align}
\begin{align}
&&M_{(\widetilde{\nu}^c_{i})^{I}(\widetilde{\nu}^c_{j})^{I}}^{2}=m_{\widetilde{\nu}^c_{ij}}^{2}-2a_{\kappa_{ijk}}\nu^c_{k}+2\lambda_{k}\kappa_{ijk}v_{d}v_{u}
-2\kappa_{ijk}\kappa_{lmk}\nu^c_{l}\nu^c_{m}
+4\kappa_{imk}\kappa_{ljk}\nu^c_{l}\nu^c_{m}\nonumber\\ &+&\lambda_{i}\lambda_{j}(v_{d}^{2}+v_{u}^{2})-(Y_{\nu_{ki}}\lambda_{j}+Y_{\nu_{kj}}\lambda_{i})v_{d}\nu_{k}-2Y_{\nu_{lk}}\kappa_{ijk}v_{u}\nu_{l}+Y_{\nu_{ki}}Y_{\nu_{kj}}v_{u}^{2}	
+Y_{\nu_{li}}Y_{\nu_{kj}}\nu_{k}\nu_{l}\ .
\label{oddrr}
\end{align}

Then the mass eigenvectors are 
\bea
\mathbf{P}_\alpha=R^P_{\alpha \beta} \mathbf{P'}_\beta\ ,
\eea
with the diagonal mass matrix 
\bea
(M^{\text{diag}}_{P_{\alpha \beta}})^2=R^P_{\alpha \gamma} M^2_{P_{\gamma \delta} }R^P_{\beta \delta}\ .
\eea

%%%%%%%%%%%%%%%%%%%%%%%%%%%%%%%%%%%%%%%%%%%%%%%%%%%%%%%%%%%%%%%%
\section{Higgs sector couplings \label{Appendix Higgs sector couplings}}
%%%%%%%%%%%%%%%%%%%%%%%%%%%%%%%%%%%%%%%%%%%%%%%%%%%%%%%%%%%%%%%%
\subsection{Coupling between three CP-even Higgses} \label{Appendix Three CP-even Higgses coupling}

$h_\delta h_\epsilon h_\eta:$
%\\
\begin{eqnarray}
&& \dfrac{\lambda_i \lambda_j}{\sqrt{2}}[\nu_i^c(\Pi_{\delta \epsilon \eta}^{11(j+2)}+\Pi_{\delta \epsilon \eta}^{22(j+2)})+v_d \Pi_{\delta \epsilon \eta}^{1(i+2)(j+2)}+v_u \Pi_{\delta \epsilon \eta}^{2(i+2)(j+2)}] \nonumber \\
&& + \dfrac{1}{\sqrt{2}} \lambda_l \lambda_l [v_d \Pi_{\delta \epsilon \eta}^{122}+v_u \Pi_{\delta \epsilon \eta}^{211}]-\dfrac{1}{\sqrt{2}}\lambda_l \kappa_{ljk}[v_d \Pi_{\delta \epsilon \eta}^{2(j+2)(k+2)}+v_u \Pi_{\delta \epsilon \eta}^{1(j+2)(k+2)}+2 \nu_j^c \Pi_{\delta \epsilon \eta}^{12(k+2)}] \nonumber \\
&& + \sqrt{2} \kappa_{ljk}\kappa_{lbd}[\nu_j^c \Pi_{\delta \epsilon \eta}^{(k+2)(b+2)(d+2)}]+\dfrac{Y_{\nu_{ij}}Y_{\nu_{kl}}}{\sqrt{2}}[\nu_i \Pi_{\delta \epsilon \eta}^{(j+2)(l+2)(k+5)}+\nu_j^c \Pi_{\delta \epsilon \eta}^{(l+2)(i+5)(k+5)}] \nonumber \\
&& - \dfrac{1}{\sqrt{2}}Y_{\nu_{ij}}\lambda_k[\nu_i \Pi_{\delta \epsilon \eta}^{1(j+2)(k+2)}+\nu_j^c \Pi_{\delta \epsilon \eta}^{1(k+2)(i+5)}+\nu_k^c \Pi_{\delta \epsilon \eta}^{1(j+2)(i+5)}+v_d \Pi_{\delta \epsilon \eta}^{(j+2)(k+2)(i+5)}] \nonumber \\
&& +\dfrac{1}{\sqrt{2}}Y_{\nu_{lj}}Y_{\nu_{lm}}[v_u \Pi_{\delta \epsilon \eta}^{2(j+2)(m+2)}+\nu_j^c \Pi_{\delta \epsilon \eta}^{22(m+2)}]+\dfrac{1}{\sqrt{2}}Y_{\nu_{il}}Y_{\nu_{jl}}[v_u \Pi_{\delta \epsilon \eta}^{2(i+5)(j+5)}+\nu_i \Pi_{\delta \epsilon \eta}^{22(j+5)}] \nonumber \\
&& -\dfrac{1}{\sqrt{2}}\lambda_l Y_{\nu_{il}}[2 v_u \Pi_{\delta \epsilon \eta}^{12(i+5)}+v_d \Pi_{\delta \epsilon \eta}^{22(i+5)}+\nu_i \Pi_{\delta \epsilon \eta}^{122}] \nonumber \\
&& +\dfrac{1}{\sqrt{2}} \kappa_{ljk} Y_{\nu_{il}}[2 \nu_j^c \Pi_{\delta \epsilon \eta}^{2(k+2)(i+5)}+v_u \Pi_{\delta \epsilon \eta}^{(j+2)(k+2)(i+5)}+\nu_i \Pi_{\delta \epsilon \eta}^{2(j+2)(k+2)}] \nonumber \\
&& -\dfrac{1}{\sqrt{2}}(A_\lambda \lambda)_i \Pi_{\delta \epsilon \eta}^{12(i+2)}+\dfrac{1}{\sqrt{2}}(A_\nu Y_\nu)_{ij}\Pi_{\delta \epsilon \eta}^{2(i+2)(j+5)}+\dfrac{1}{3\sqrt{2}}(A_\kappa \kappa)_{ijk}\Pi_{\delta \epsilon \eta}^{(i+2)(j+2)(k+2)} \nonumber \\
&& +\dfrac{g_1^2+g_2^2}{4\sqrt{2}}[\nu_i \Pi_{\delta \epsilon \eta}^{(i+5)(j+5)(j+5)}+\nu_i\Pi_{\delta \epsilon \eta}^{11(i+5)}-\nu_i \Pi_{\delta \epsilon \eta}^{22(i+5)} \nonumber \\
&& +v_d \Pi_{\delta \epsilon \eta}^{1(i+5)(i+5)}+v_d \Pi_{\delta \epsilon \eta}^{111}-v_d \Pi_{\delta \epsilon \eta}^{122}-v_u \Pi_{\delta \epsilon \eta}^{2(i+5)(i+5)}+v_u \Pi_{\delta \epsilon \eta}^{222}-v_u \Pi_{\delta \epsilon \eta}^{112}]\ ,
\end{eqnarray}
where $b,d,i,j,k,l,m=1,2,3$; $\alpha,\beta,\gamma,\delta,\epsilon,\eta=1,...,8$, and
\begin{eqnarray}
\Pi_{\delta \epsilon \eta}^{\alpha \beta \gamma}=R^h_{\delta \alpha}R^h_{\epsilon \beta}R^h_{\eta \gamma}+R^h_{\delta \alpha}R^h_{\eta \beta}R^h_{\epsilon \gamma}+R^h_{\epsilon \alpha}R^h_{\delta \beta}R^h_{\eta \gamma}+R^h_{ \epsilon \alpha}R^h_{\eta \beta}R^h_{\delta \gamma}+R^h_{\eta \alpha}R^h_{\delta \beta}R^h_{\epsilon \gamma}+R^h_{\eta \alpha}R^h_{\epsilon \beta}R^h_{\delta \gamma}\ . \nonumber \\
\end{eqnarray}

\subsection{Coupling between one CP-even and two CP-odd Higgses} \label{Appendix One CP-even and two CP-odd Higgses coupling}

$h_\delta P_\epsilon P_\eta:$
\\
\begin{eqnarray}
&& \dfrac{\lambda_i \lambda_j}{\sqrt{2}}[\nu_i^c(\Pi_{\delta \epsilon \eta}^{(j+2)11}+\Pi_{\delta \epsilon \eta}^{(j+2)22})+v_d \Pi_{\delta \epsilon \eta}^{1(i+2)(j+2)}+v_u \Pi_{\delta \epsilon \eta}^{2(i+2)(j+2)}] + \dfrac{1}{\sqrt{2}} \lambda_l \lambda_l [v_d \Pi_{\delta \epsilon \eta}^{122}+v_u \Pi_{\delta \epsilon \eta}^{211}] \nonumber \\
&& +\dfrac{1}{\sqrt{2}}\lambda_l \kappa_{ljk}[v_d( \Pi_{\delta \epsilon \eta}^{2(j+2)(k+2)}-2 \Pi_{\delta \epsilon \eta}^{(j+2)2(k+2)})+v_u (\Pi_{\delta \epsilon \eta}^{1(j+2)(k+2)}-2 \Pi_{\delta \epsilon \eta}^{(j+2)1(k+2)}) \nonumber \\
&& +2 \nu_j^c (\Pi_{\delta \epsilon \eta}^{(k+2)12}-\Pi_{\delta \epsilon \eta}^{12(k+2)}-\Pi_{\delta \epsilon \eta}^{21(k+2)})] \nonumber \\
&& + \sqrt{2} \kappa_{ljk}\kappa_{lbd}[-\nu_j^c \Pi_{\delta \epsilon \eta}^{(k+2)(b+2)(d+2)}+2 \nu_j^c \Pi_{\delta \epsilon \eta}^{(b+2)(k+2)(d+2)}] \nonumber \\
&& -\dfrac{\lambda_l Y_{\nu_{il}}}{\sqrt{2}}[v_d \Pi_{\delta \epsilon \eta}^{(i+5)22}+\nu_i \Pi_{\delta \epsilon \eta}^{122}+2 v_u \Pi_{\delta \epsilon \eta}^{21(i+5)}] \nonumber \\
&& +\dfrac{\kappa_{ljk}Y_{\nu_{il}}}{\sqrt{2}}[2 \nu_j^c \Pi_{\delta \epsilon \eta}^{2(k+2)(i+5)}+2 \nu_j^c \Pi_{\delta \epsilon \eta}^{(i+5)2(k+2)}-2 \nu_j^c \Pi_{\delta \epsilon \eta}^{(k+2)2(i+5)}+2 v_u \Pi_{\delta \epsilon \eta}^{(j+2)(i+5)(k+2)} \nonumber \\
&& -v_u \Pi_{\delta \epsilon \eta}^{(i+5)(j+2)(k+2)}+2 \nu_i \Pi_{\delta \epsilon \eta}^{(j+2)2(k+2)}-\nu_i \Pi_{\delta \epsilon \eta}^{2(j+2)(k+2)}] \nonumber \\
&& -\dfrac{Y_{\nu_{ij}\lambda_k}}{\sqrt{2}}[-\nu_i \Pi_{\delta \epsilon \eta}^{(j+2)(k+2)1}+\nu_i \Pi_{\delta \epsilon \eta}^{(k+2)(j+2)1}+\nu_i \Pi_{\delta \epsilon \eta}^{1(k+2)(j+2)}-\nu_j^c \Pi_{\delta \epsilon \eta}^{(i+5)(k+2)1}-\nu_j^c \Pi_{\delta \epsilon \eta}^{(k+2)(i+5)1} \nonumber \\
&& +\nu_j^c \Pi_{\delta \epsilon \eta}^{1(k+2)(i+5)}-\nu_k^c \Pi_{\delta \epsilon \eta}^{1(i+5)(j+2)}+\nu_k^c \Pi_{\delta \epsilon \eta}^{(i+5)1(j+2)}+\nu_k^c \Pi_{\delta \epsilon \eta}^{(j+2)1(i+5)}-v_d \Pi_{\delta \epsilon \eta}^{(k+2)(i+5)(j+2)} \nonumber \\
&& +v_d \Pi_{\delta \epsilon \eta}^{(j+2)(k+2)(i+5)}+v_d \Pi_{\delta \epsilon \eta}^{(i+5)(j+2)(k+2)}] + \dfrac{Y_{\nu_{ij}}Y_{\nu_{kl}}}{\sqrt{2}}[-\nu_i \Pi_{\delta \epsilon \eta}^{(j+2)(k+5)(l+2)}+\nu_i \Pi_{\delta \epsilon \eta}^{(l+2)(j+2)(k+5)} \nonumber \\
&& +\nu_i \Pi_{\delta \epsilon \eta}^{(k+5)(l+2)(j+2)}-\nu_j^c \Pi_{\delta \epsilon \eta}^{(i+5)(k+5)(l+2)}+\nu_j^c \Pi_{\delta \epsilon \eta}^{(k+5)(i+5)(l+2)}+\nu_j^c \Pi_{\delta \epsilon \eta}^{(l+2)(k+5)(i+5)}] \nonumber \\
&& +\dfrac{Y_{\nu_{lj}}Y_{\nu_{lm}}}{\sqrt{2}}[v_u \Pi_{\delta \epsilon \eta}^{2(j+2)(m+2)}+\nu_j^c \Pi_{\delta \epsilon \eta}^{(m+2)22}]+\dfrac{Y_{\nu_{il}}Y_{\nu_{jl}}}{\sqrt{2}}[v_u \Pi_{\delta \epsilon \eta}^{2(i+5)(j+5)}+\nu_i \Pi_{\delta \epsilon \eta}^{(j+5)22}] \nonumber \\
&&+\dfrac{(A_{\lambda}\lambda)_i}{\sqrt{2}}[\Pi_{\delta \epsilon \eta}^{(12(i+2))}+\Pi_{\delta \epsilon \eta}^{21(i+2)}+\Pi_{\delta \epsilon \eta}^{(i+2)12}]-\dfrac{(A_{\nu}Y_{\nu})_{ij}}{\sqrt{2}}[\Pi_{\delta \epsilon \eta}^{2(i+5)(j+2)}+\Pi_{\delta \epsilon \eta}^{(i+5)2(j+2)} \nonumber \\
&& +\Pi_{\delta \epsilon \eta}^{(i+2)2(j+5)}]-\dfrac{(A_\kappa \kappa)_{ijk}}{\sqrt{2}} \Pi_{\delta \epsilon \eta}^{(i+2)(j+2)(k+2)}+\dfrac{g_1^2+g_2^2}{4\sqrt{2}}[\nu_i \Pi_{\delta \epsilon \eta}^{(i+5)(j+5)(j+5)}+\nu_i \Pi_{\delta \epsilon \eta}^{(i+5)11} \nonumber \\
&&-\nu_i \Pi_{\delta \epsilon \eta}^{(i+5)22}+v_d(\Pi_{\delta \epsilon \eta}^{1(i+5)(i+5)}+\Pi_{\delta \epsilon \eta}^{111}-\Pi_{\delta \epsilon \eta}^{122})+v_u(\Pi_{\delta \epsilon \eta}^{222}-\Pi_{\delta \epsilon \eta}^{211}-\Pi_{\delta \epsilon \eta}^{2(i+5)(i+5)})]\ ,
\end{eqnarray}
where $b,d,i,j,k,l,m=1,2,3$; $\alpha,\beta,\gamma,\delta=1,...,8$; 
$\epsilon,\eta=1,...,7$, 
and
\begin{eqnarray}
\Pi_{\delta \epsilon \eta}^{\alpha \beta \gamma}=R^h_{\delta \alpha}(R^P_{\epsilon \beta}R^P_{\eta \gamma}+R^P_{\eta \beta}R^P_{\epsilon \gamma})\ . \nonumber \\
\end{eqnarray}

\end{document}